\documentclass[english,twocolumn,floatfix,prb,aps,showpacs,superscriptaddress,nofootinbib,bibliography]{revtex4}
\usepackage[english]{babel}                  
 \usepackage{xcolor}   
\usepackage[utf8]{inputenc}                 
\usepackage[T1]{fontenc}
\usepackage{epsfig,mathtools}
\usepackage{amsmath,dsfont,amsfonts}
\usepackage{hyphenat}
\usepackage{hyperref}
\usepackage{placeins}
\usepackage{caption}
\usepackage{subcaption}
\usepackage{accents}

\renewcommand{\i}{\imath}

  \let\D=\Delta

\def\be{\begin{equation}}
	\def\ee{\end{equation}}
\def\bea{\begin{eqnarray}}
	\def\eea{\end{eqnarray}}
\def\ba{\begin{array}}
	\def\ea{\end{array}}

\begin{document}
	\title{Hydrodynamics of charged two-dimensional Dirac systems II:\\the role of collective modes}
	
	\author{Kitinan Pongsangangan}
		\affiliation{Institute for Theoretical Physics and Center for Extreme Matter and Emergent Phenomena,
		Utrecht University, Princetonplein 5, 3584 CC Utrecht, The Netherlands}
	\author{Tim Ludwig}
		\affiliation{Institute for Theoretical Physics and Center for Extreme Matter and Emergent Phenomena,
		Utrecht University, Princetonplein 5, 3584 CC Utrecht, The Netherlands}
	\author{Henk T.C. Stoof}
		\affiliation{Institute for Theoretical Physics and Center for Extreme Matter and Emergent Phenomena,
		Utrecht University, Princetonplein 5, 3584 CC Utrecht, The Netherlands}
	\author{Lars Fritz}
	\affiliation{Institute for Theoretical Physics and Center for Extreme Matter and Emergent Phenomena,
		Utrecht University, Princetonplein 5, 3584 CC Utrecht, The Netherlands}


	\begin{abstract}
		\noindent 
		 We study the hydrodynamic  properties of ultraclean interacting two-dimensional Dirac electrons with Keldysh quantum field theory. We study it from a weak-coupling and a strong-coupling perspective. We demonstrate that long-range Coulomb interactions play two independent roles: (i) they provide the inelastic and momentum-conserving scattering mechanism that leads to fast local equilibration; (ii) they facilitate the emergence of collective excitations, for instance plasmons, that contribute to transport properties on equal footing with electrons. Our approach is based on an effective field theory of the collective field coupled to electrons. Within a conserving approximation for the coupled system we derive a set of coupled quantum-kinetic equations.  This builds the foundation of the derivation of the Boltzmann equations for the interacting system of electrons and plasmons. From this, we explicitly derive all the conservation laws and identify the extra contributions of energy density and pressure from the plasmons. We demonstrate that plasmons show up in thermo-electric transport properties as well as in quantities that enter the energy-momentum tensor, such as the viscosity. In a parallel paper we discuss some of the phenomenology of the corresponding hydrodynamic equations with an eye on thermo-electric transport properties. 
	\end{abstract}
	\maketitle

\section{introduction}

The conventional theory of electronic transport in a solid-state setting describes the motion of electrons in the following way: individual electrons diffuse on the background of a disordered lattice, primarily scattering from impurities and/or lattice vibrations~\cite{Drude1900,Ashcroft&Mermin1976,Ziman2000}.  
However, recent years have seen tremendous progress in realizing electronic transport that follows a different paradigm: hydrodynamic electrons, meaning electrons flowing collectively like a viscous liquid, such as water or honey. This idea has first been discussed in the 1960s by Gurzhi~\cite{Gurzhi1963}. However, the subject has only recently picked up a lot of pace. This is mostly due to recent advances in the preparation of ultrapure mono- and bilayer graphene samples with sufficiently strong interactions~\cite{Novoselov2005,CastroNeto2009,Bandurin2018,Bandurin2016,Sulpizio2019,Molenkamp1995}. The prerequisite for the experimental observation of electron hydrodynamics is that microscopic momentum-conserving electron-electron collisions due to Coulomb interactions must be considerably faster than momentum-relaxing scatterings of electrons from impurities and/or phonons. This allows to locally establish equilibrium. In that situation, it is justified to speak of approximate conservation laws, sufficient to open the door for the observation of electron hydrodynamics~\cite{Fritz2008a,Fritz2008b,Landau1965,Lucas2016,Lucas2018,Narozhny2019}. The theoretical method used to investigate this hydrodynamic flow is usually based on the traditional Navier-Stokes equation which expresses conservation of momentum, energy and electric charge~\cite{LandauLifshitz1987,Lucas2018,Narozhny2015}. 
	One way to derive these hydrodynamic equations from microscopics starts from the Boltzmann equation~\cite{Lifshitz2981,Narozhny2019}. There, usually, the effect of the Coulomb interactions enters only through electron-electron collisions leading to local equilibrium. This effect is usually calculated from a weak-coupling perspective within second-order perturbation theory in the interaction strength, referred to as the Born approximation for the cross-section~\cite{Landau1965,Fritz2008a,Kashuba2008,KadanoffBaym1962}. 
	However, hydrodynamics of a charged liquid must be expected to behave differently in many respects from the archetypal example of fluid like water flow. The reason is that for a charged electronic system, in addition to said inelastic scatterings, Coulomb interactions provide also long-range mean-field forces between electrons. This facilitates the emergence of collective excitations, for instance plasma oscillations or plasmons~\cite{Pines&Bohm1952,Bohm&Pines1953,Vlasov1968}. Under the correct circumstances, as present in the situation studied here, these plasmons behave as `bona fide' quasi-particles that have their own dynamics. This effect is usually not considered in the context of electron hydrodynamics.

{\it{Background and main idea:}	} Strongly correlated many-particle systems can often be regarded as a collection of weakly interacting excitations. One of the prime examples of this kind is the Landau Fermi liquid~\cite{Abrikosov1959}. Its excitations behave as well-defined entities, called quasi-particles. This means, the following conditions must be fulfilled: (i) an excitation with momentum $\vec{p}$ possesses a well-defined complex energy spectrum, say $\omega(\vec{p}) = \epsilon(\vec{p})-i\gamma(\vec{p})$, where the imaginary part of the energy $\gamma(\vec{p})$ describes the decay rate of the particle, inversely proportional to the lifetime of the excitation; (ii) an excitation must be long lived which means that the decay must be at least underdamped $\gamma(\vec{p}) \ll \epsilon(\vec{p})$. These requirements usually limit one to the low-frequency and/or long-wavelength behavior of the system. Under such circumstances, it is justified to consider the complicated interacting system as a collection of independent elementary quasi-particles~\cite{Pines1999}.  Once this paradigm is adopted for a given many-body system, there are two main questions in need of answer: (i) what are the quasi-particles involved in physical phenomena of interest at the relevant energy scale and (ii) what is the population of these quasi-particles in each momentum state especially when the system is exposed to external disturbances? The latter is quantified by a distribution function $f(\vec{x},t,\vec{p})$ which gives the probability density that the particle in a momentum state $\vec{p}$ is found in a neighborhood of a spatial position $\vec{x}$ at time $t$. For a given non-equilibrium situation, the time evolution of the distribution function is described by the Boltzmann equation.
 	 The knowledge of the distribution function and the energy spectrum is vital to determine all thermodynamic as well as transport properties of the system in a straightforward fashion. 
	 The main question we address in this paper are: (i) what are the quasi-particles involved in hydrodynamical transport phenomena of ultraclean two-dimensional Dirac electrons at an accessible temperature and (ii) how do they equilibrate? 
The conventional theory for hydrodynamic behavior in two-dimensional Dirac systems is a version of a two-component hydrodynamics, consisting of electrons and holes~\cite{Lucas2016,Narozhny2015,Fritz2008a}. This scenario is very popular and can formally be derived from a weak-coupling analysis using a Hartree-Fock-Born approximation. In that framework, Coulomb interaction plays two different roles: (i) it is seen by an electron as an internal potential produced by all the other electrons in the system through the Hartree term; (ii) Coulomb interaction manifests itself as an inelastic and momentum conserving scattering mechanism~\cite{Fritz2008a,Fritz2008b,Kashuba2008,KadanoffBaym1962,Landau1965} which locally equilibrates the system. This process is an important requirement for observing the electron hydrodynamical regime with its fascinating transport properties~\cite{Lucas2016,Lucas2018,Narozhny2019}. 

Perturbation theory up to, in this case, second order in the coupling constant is unfortunately unable to describe many important physical phenomena, such as the emergence of collective modes. Collective modes, however, are the hallmark of interacting electronic systems. One such mode are plasma oscillations, also called plasmons. In conventional three-dimensional metals, plasmons are a gapped degree of freedom with a large energy gap, larger than the Fermi energy. This implies that thermal plasmons cannot be excited at realistic experimental temperatures and hence are largely irrelevant, both for thermodynamic as well as transport properties. As an example, aluminium at room temperature has a ratio of the plasmon gap to the thermal energy $\hbar\omega_p/k_B T \approx 16\;\text{eV}/0.25\;\text{eV} = 64$. Consequently, the plasmon occupation number is negligibly small, $n_B(\omega_p) \approx 10^{-28}$, where $n_B$ is the equilibrium Bose-Einstein distribution function~\cite{Pines1953}. 
In contrast, in two dimensions, plasmons are massless and have a square-root dispersion, {\it i.e.}, $\omega \propto \sqrt{q}$~\cite{Sarma2013,Wunsch2006,Stern1967}. This is not only true for Dirac fermions but for a generic two-dimensional electronic system. Consequently, at accessible temperatures, plasmons can be excited and thus constitute proper low-energy elementary excitations. There might be various effects that potentially destabilize the plasmon and broaden their spectral function such as disorder, electron-phonon, and electron-electron collision~\cite{Glazman2004,Kivenson1969}. However, it turns out that the plasmons are typically remarkably stable. 	 
	 
The main result in this paper is that we offer a novel derivation of the equations of hydrodynamics from a strong-coupling perspective. This results in a combined description of electrons, holes, and plasmons, that are coupled to each other. In this description, plasmons enter on equal footing with the electronic degrees of freedom: we find that plasmons make a contribution of the same order of magnitude as the fermionic degrees of freedom to heat currents as well as the energy-momentum tensor and consequently should show up in measurements that measure thermal transport but also viscous effects. The results presented here provide a starting point for more phenomenological descriptions, including transport properties, which we present in a parallel paper.

{\it{Organization of the paper}}:
We start with Sec.~\ref{sec:model} where we introduce the model of interacting Dirac fermions and describe how it connects, for instance, to graphene. We then proceed to a technical section, Sec.~\ref{sec:formalism}, where we review the formalism of real-time quantum field theory. It is a summary of the most important steps that lead from a fully quantum mechanical treatment towards the semiclassical Boltzmann equation. This includes the partition function on the Schwinger-Keldysh time contour, Sec.~\ref{subsec:generatingfunction}, as well as the structure of the Green functions on the closed time contour in Sec.~\ref {subsec:Greenfunction}. We then discuss the Dyson equation, including the Keldysh or quantum-kinetic equation in Sec.~\ref{subsec:dysonequation}. We proceed to introduce the Wigner transform and the Moyal product in Sec.~\ref{subsec:wignertransform} which allows to perform the gradient expansion on the Keldysh equation in Sec.~\ref{subsec:gradientkeldysh}. More detailed accounts of the formalism can be found in Refs.~\cite{KadanoffBaym1962,Kamenev2011,Rammer&Smith1986,Keldysh1965}. This section can be skipped by a reader familiar with the Schwinger-Keldsyh or Kadanoff-Baym approach. In Sec.~\ref{sec:HartreeBornapproximation}, we discuss two-component hydrodynamics as currently used for the description of graphene. We derive all the equations from the Schwinger-Keldysh approach and show how the two-component fluid picture emerges within a weak-coupling approach to second order in the Coulomb interaction. We start with introducing the non-interacting Green functions in Sec.~\ref{subsec:noninteracting}. Within weak coupling, Coulomb interactions play two different roles: (i) It is seen by an electron as an internal potential produced by all the other electrons in the system. This is the result of perturbation theory to first order, called the Hartree-Fock scheme~\cite{Mahan2000}, where the Fock term leads to a renormalization of the Fermi velocity~\cite{vozmediano,Elias2011}. The potential energy, on the other hand, is referred to as the Hartree potential. In thermal equilibrium, it is canceled by a potential from the underlying positively charged background in which the particles move. However, it builds the basis for a derivation of the collective excitations of the fluid~\cite{Vlasov1968,Pines1999}. This aspect is discussed in Sec.~\ref{subsec:collisionless}; (ii) The Coulomb interaction manifests itself as an inelastic and momentum conserving scattering mechanism~\cite{Fritz2008a,Fritz2008b,Kashuba2008,KadanoffBaym1962,Landau1965}. This process is an important requirement for observing the electron hydrodynamical regime since it leads to local equilibration. In Sec.~\ref{subsec:born}  we discuss the scattering process in detail. We then derive the ensuing conservation laws starting from the Boltzmann equation in Sec.~\ref{subsec:conservationlawsborn} and finish the section with a discussion of collective modes in Sec.~\ref{subsec:collectivemodes}. We find a collective mode that we later identify with the plasmon. We show explicitly that it matches with the strong-coupling treatment if interpreted correctly. 

Low-order perturbation theory is unable to describe many important physical phenomena, for instance, collective modes, such as plasmons. We address this situation in Sec.~\ref{sec:strongcoupling}. We start with a formalized version of the random-phase approximation (RPA) in Sec.~\ref{subsec:rpa}. To this end, we introduce a new quantum field associated with a plasmon excitation by means of a Hubbard-Stratonovich transformation~\cite{Kleinert1978,Hubbard1959}, which is an exact rewriting of the theory. After integrating out the fermions, the plasmons acquire their own dynamics, which is discussed in Sec.~\ref{sec:plasmons}. The numerical details of this are discussed in Sec.~\ref{subsec:polarization} whereas in Sec.~\ref{subsec:analyticalplasmon} we discuss an analytical approximation that allows us to make further progress. An important side product is that we show that there is also a feedback effect, that renormalizes the fermions and provides scattering for them, despite having integrated them out at an earlier stage. Importantly, we can show explicitly that this is not an instance of double counting, as one might suspect, but is indeed required to preserve conservation laws. In Sec.~\ref{subsec:kineticplasmons} we find a set of coupled kinetic equations for electrons, holes, and the plasmons~\cite{WyldPines1962,Kitinan2020,PinesSchrieffer1962}. It is important that, within the conserving approximation, the electrons scatter from plasmons and vice versa. Using a series of approximations, we derive Boltzmann equations from this effective field theory. Based on this, we derive the conservation laws of the system in Sec.~\ref{subsec:conservationlawsrpa} which indeed show that the fermion dynamics is strongly influenced by the plasmons. Furthermore, we observe, that the plasmons make a sizable contribution to the heat current (this provides an alternative derivation of the heat-current operator starting from the quantum-kinetic equation) and energy density. Additionally, it makes similar contributions to the momentum flux and therefore shows up in quantities related to the viscosity. This section also confirms that RPA is indeed a conserving approximation. 
We conclude our results in Sec.~\ref{sec:conclusionandoutlook} as well as provide an outlook for future work. The most technical details of the calculations are usually presented in a number of appendices.

\section{The model}\label{sec:model}
	
We study a theory of  charged Dirac electrons in two spatial dimensions interacting via long-range Coulomb interactions. The Hamiltonian is given by
\begin{equation}
	\label{eq:hamiltonian}
	\hat{H}=\hat{H}_0+\hat{H}_{\text{ex}}+\hat{H}_{I}\;,
\end{equation}
with $\hat{H}_0$ being the non-interacting part. In coordinate space, it reads
\begin{equation}
	\label{eq:nonintHamiltonian}
	\hat{H}_0 =  \sum_{i=1}^N\sum_{\lambda,\lambda'=\pm}\int d\vec{x}\;\hat{\Psi}_{i,\lambda}^\dagger(\vec{x}) \hat{H}_{D,\lambda\lambda'}(\vec{x})\hat{\Psi}_{i,\lambda'}(\vec{x})\;, 
\end{equation}
Here $\hat{H}_{D,\lambda\lambda'}(\vec{x})=\left(-i\hbar v_F \vec{\sigma} \cdot \vec{\nabla}-\mu\right)_{\lambda\lambda'}$ is the Dirac Hamiltonian with a Fermi velocity $v_F$ and chemical potential $\mu$, and $\hat{\Psi}^\dagger_{i,\lambda}(\vec{x})$ $(\hat{\Psi}_{i,\lambda}(\vec{x}))$ creates (annihilates) an electron at a position $\vec{x}$. The flavor index denoted by $i$ range from $i=1, ..., N$. The symbols $\lambda,\lambda'\in \{+,-\}$ denote spinor indices.

 The static potential energy $V_{\rm{ex}}(\vec{x})$ is added in order to take into account the positively charge background in which the electrons move, {\it i.e.},
\begin{equation}
	V_{\rm{ex}}(\vec{x})= -n_0\int d\vec{x}' V(\vec{x}-\vec{x}')\;, 
	\label{eq:jelliumpotential}
\end{equation}
where $n_0$ is the average density of the background ions, identical to the electron density in thermal equilibrium. The interaction of the electrons with the inert positively charged background is explained by
\begin{equation}
	\hat{H}_{\text{ex}} = \sum_{i=1}^{N}\sum_{\lambda=\pm}\int d\vec{x}\; \hat{\Psi}^\dagger_{i,\lambda}(\vec{x}) V_{\text{ex}}(\vec{x}) \hat{\Psi}_{i,\lambda}(\vec{x}).
\end{equation}
 In addition, $\hat{H}_I$ is the interaction part of the Hamiltonian. The electrons interact via a long-range Coulomb interaction, which is included in our model by the term
\begin{eqnarray}
	\label{eq:interaction}
	\hat{H}_I &=&\frac{1}{2} \sum_{i=1}^N\sum_{\lambda,\lambda'=\pm} \int d\vec{x}d\vec{x}'\;  \hat{\Psi}_{i,\lambda}^\dagger(\vec{x})\hat{\Psi}_{i,\lambda'}^\dagger(\vec{x}')V(\vec{x}-\vec{x}')\nonumber\\&&\hspace{5cm}\hat{\Psi}_{i,\lambda'}(\vec{x}')\hat{\Psi}_{i,\lambda}(\vec{x})\;.\nonumber\\
\end{eqnarray}
Let us emphasize that we assume that the fermions of different flavours are decoupled. The interaction potential between two electrons of charge $e$ separated by a distance $|\vec{x}-\vec{x}'|$ is given by the instantaneous Coulomb interaction
\begin{equation}
	V(\vec{x}-\vec{x}')=\frac{e^2}{4\pi\epsilon|\vec{x}-\vec{x}'|}\;,
\end{equation} 
where $\epsilon$ is the average dielectric constant. After a Fourier transformation, the non-interacting Hamiltonian acquires the form
\begin{equation}
	\hat{H}_0 =  \sum_{i=1}^N \sum_{\lambda,\lambda'=\pm} \int \frac{d\vec{p}}{(2\pi)^2} \;\hat{\Psi}_{i,\lambda}^\dagger(\vec{p}) \hat{\mathcal{H}}_{D,\lambda\lambda'}(\vec{p})\hat{\Psi}_{i,\lambda'}(\vec{p})\;,
\end{equation} 
where $\hat{\mathcal{H}}_{D,\lambda\lambda'}(\vec{p})=\left(\hbar v_F \vec{\sigma} \cdot \vec{p}-\mu\right)_{\lambda\lambda'}$ and the interaction part of the Hamiltonian becomes
\begin{eqnarray}
	\hat{H}_I &=& \frac{1}{2} \sum_{i=1}^N\sum_{\lambda,\lambda'=\pm}  \int \frac{d\vec{p}}{(2\pi)^2}\frac{d\vec{k}}{(2\pi)^2}\frac{d\vec{q}}{(2\pi)^2}\; \nonumber\\&&\hat{\Psi}_{i,\lambda}^\dagger(\vec{k}-\vec{q})\hat{\Psi}_{i,\lambda'}^\dagger(\vec{p}+\vec{q})V(\vec{q})\hat{\Psi}_{i,\lambda'}(\vec{k})\hat{\Psi}_{i,\lambda}(\vec{p}).\nonumber\\
\end{eqnarray}
The Fourier transformation of the $1/r$ Coulomb interaction between electrons in two dimensions reads
\begin{equation}
	V(\vec{p}) = \frac{e^2}{2\epsilon p} = \frac{2\pi\alpha v_F}{p},
	\label{coulombpotential}
\end{equation} 
where $p = |\vec{p}|$ denotes the norm of the two-dimensional momentum vector.
 The strength of the Coulomb interaction is usually characterised by the ratio of the potential energy to the kinetic energy. For the Dirac fermion, this ratio boils down to the fine structure constant $\alpha=e^2/4\pi\epsilon \hbar v_F$.  The theory is identical for each flavor $i=1,..,N$. 

Let us mention that this model is applicable to low-energy interacting electrons in graphene where momenta are measured with respect to the $K$ ($K'$) point in the Brillouin zone. In this case, $\lambda$ and $\lambda'$ denote the pseudo-spin degree of freedom taking into account the $A$ and $B$ sub-lattices of the hexagonal lattice. The number of fermion is then $N=4$ and it corresponds to spin and valley degrees of freedom. It was found that $v_F \approx 10^6$ m/s, and  $\epsilon=\epsilon_0 \epsilon_r$ measures the average value of the dielectric constant of materials above and below it, {\it{e.g.}}, $\epsilon = 1$ for suspended graphene in vacuum and $\epsilon \approx 7$ for graphene sandwiched in hBN layers. Thus for these two cases, $\alpha = 2.2$ and $\alpha \approx 0.3$, respectively~\cite{Narozhny2019}.

\section{The formalism}\label{sec:formalism}

In this section we review the basics of non-equilibrium field theory. More detailed accounts can be found in excellent books, Refs. \cite{KadanoffBaym1962,Kamenev2011,Rammer&Smith1986,Keldysh1965,CHOU1985}. In Sec.~\ref{subsec:generatingfunction} we start with a discussion of the generating function of the Green functions on the Schwinger-Keldysh closed time contour. This builds the basis of the whole approach. We then move on to discuss the structure of the Green function and the Keldysh representation in Sec.~\ref{subsec:Greenfunction}. In Sec.~\ref{subsec:dysonequation} we discuss the Dyson equation for interacting problems, as well as the quantum-kinetic or Keldysh equation. In Sec.~\ref{subsec:wignertransform}, we discuss the Wigner transform, including the gradient expansion and the Moyal product. Finally, in Sec.~\ref{subsec:gradientkeldysh}, we discuss the gradient expansion of the Keldysh equation, which is essential in the derivation of the Boltzmann equation.

\begin{figure}[tbh!]
	\includegraphics[width=0.45\textwidth]{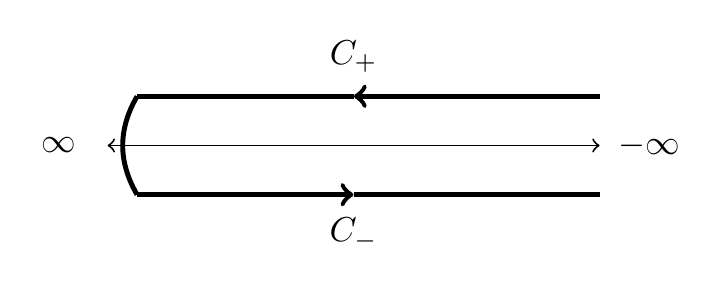}
	\caption{The closed time integration contour, $C=C_+ \cup C_-$. The upper branch $C_+$ goes
		forward from the initial time ($t =- \infty$) to the final time ($t = \infty$) and the lower branch $C_-$ goes backward
		from the final time to the initial time.
	}
	\label{fig:closetimecontour}
\end{figure}

 \subsection{The generating function}\label{subsec:generatingfunction}
The central object in the following Schwinger-Keldysh formalism is the generating function. It allows for the derivation of Green functions and associated physical observables by means of functional differentiation. In this section we review the construction of the partition function for the theory introduced in Eq.~\eqref{eq:hamiltonian}. The main idea is to assume that in the distant past ($t=-\infty$), the system was in thermal equilibrium at a specific temperature $T$. Its state is fully specified by the quantum-mechanical density matrix  $\hat{\rho} = e^{-\hat{H}_0/k_B T}$, where $\hat{H}_0$ is the non-interacting Hamiltonian given by Eq.~\eqref{eq:nonintHamiltonian}, and $k_B$ is the Boltzmann constant. Henceforth, we use $\hbar=k_B=1$, unless stated otherwise. The interaction will be switched on adiabatically to reach its actual strength before the observation. In addition, external perturbations might be subsequently established and drive the system away from equilibrium. The partition function is defined as $Z = \text{Tr}[\hat{U}_C \hat{\rho}]/\text{Tr}\left[\hat{\rho}\right]$ where $\hat{U}_C = \mathcal{T}_C \exp\left(-i\oint_C \hat{H}(t) dt\right)$ is the evolution operator along the closed time integration contour $C = C_+ \cup C_-$ depicted in Fig.~\ref{fig:closetimecontour}. The operator $\mathcal{T}_C$ orders the operators according to the position of their time arguments on the contour $C$. The evolution operator describes the evolution of the system from $t = -\infty$, where the system is non-interacting and in equilibrium, towards $t=\infty$, and back to the equilibrium state at $t=-\infty$, that is again non-interacting. During the evolution, the system may be exposed to external perturbations, and its response can be examined. 
 As in the conventional path-integral approach, one discretizes the closed time contour into $M$ infinitesimal time intervals and inserts the resolution of unity in the coherent-state basis at each discrete point in time along the contour. Subsequently, by taking the continuum limit, we obtain the generating function written as a functional integral according to
\begin{equation}
	Z = \int \mathcal{D}\psi^\dagger\mathcal{D}\psi \exp(i S[\psi^\dagger,\psi])\;.
	\label{eq:partitionfuntionelectrononlytheory}
\end{equation}
In writing down the partition function, we absorb an irrelevant normalization constant into the measure. The total action  reads $S[\psi^\dagger,\psi] = 
\oint_{C}  dt \mathcal{L}(\psi^\dagger,\psi)$, with the Lagrangian  
\begin{eqnarray}
&&\mathcal{L}(\psi^\dagger,\psi)=
\nonumber\\&&\int d\vec{x}\;  \sum_{i=1}^{N}\sum_{\lambda=\pm} \psi_{i,\lambda}^\dagger(\vec{x},t) i\partial_t \psi_{i,\lambda}(\vec{x},t) -\hat{H}(\psi^\dagger,\psi)\;.\nonumber \\
\end{eqnarray} 
It is convenient to split the action into two parts, $S[\psi^{\dagger},\psi] = \int_{C_+} dt\; \mathcal{L}(\psi^\dagger_+,\psi_+) + \int_{C_-} dt\; \mathcal{L}(\psi^\dagger_-,\psi_-)$. It is important to note that, while the two parts seem independent, this is not true due to the special boundary conditions when the contours meet, see discussion in Ref.~\cite{Kamenev2011}. The first term describes the fermions on the forward branch of the contour denoted by $C_+$ (the integration over the time variable extends from $-\infty$ to $\infty$) whereas the second term describes the fermions on the backward branch denoted by $C_-$. We introduced the subscript $\pm$ for the fermion fields on the different branches. Thus $\psi^{(\dagger)}_+$ and $\psi^{(\dagger)}_-$
denote the fermion fields on the forward and backward branches of the contour, respectively.  Moreover, we interchange the limits of the time integration on the backward branch, leading to an extra minus sign as the second integral then also goes from $-\infty$ to $\infty$.

Eventually, we find that the action reads 
 \begin{widetext}
 \begin{eqnarray}
 	S[\psi^\dagger,\psi]  &=&\sum_{i=1}^N\sum_{\lambda=\pm}\int dt d\vec{x}\;  \psi_{i,\lambda+}^\dagger(\vec{x},t) \Big(i\partial_t +i  v_F\vec{\nabla}\cdot\vec{\sigma}+\mu-V_{\rm{ex}}(\vec{x})\Big)\psi_{i,\lambda,+}(\vec{x},t)\nonumber\\&& - \frac{1}{2} \sum_{i=1}^N\sum_{\lambda,\lambda'=\pm}\int dt d\vec{x} d\vec{x}'\;  \psi_{i,\lambda,+}^\dagger(\vec{x},t)\psi_{i,\lambda',+}(\vec{x},t)V(\vec{x}-\vec{x}')\psi_{i,\lambda,+}(\vec{x}',t)\psi_{i,\lambda',+}^\dagger(\vec{x}',t)\nonumber\\&&- \sum_{i=1}^N\sum_{\lambda=\pm}\int d\vec{x}\;  \psi_{i,\lambda,-}^\dagger(\vec{x},t) \Big(i\partial_t +i  v_F\vec{\nabla}\cdot\vec{\sigma}+\mu-V_{\rm{ex}}(\vec{x})\Big)\psi_{i,\lambda,-}(\vec{x},t) \nonumber\\&& + \frac{1}{2}\sum_{i=1}^N\sum_{\lambda,\lambda'=\pm} \int d\vec{x} d\vec{x}'\;  \psi_{i,\lambda,-}^\dagger(\vec{x},t)\psi_{i,\lambda',-}(\vec{x},t)V(\vec{x}-\vec{x}')\psi_{i,\lambda,-}(\vec{x}',t)\psi_{i,\lambda',-}^\dagger(\vec{x}',t).\nonumber\\ 
 	\label{theoryonthecontour}
 \end{eqnarray}
 \end{widetext}

\subsection{Structure of the Green functions}\label{subsec:Greenfunction}
For the ensuing discussion we neglect all internal indices such as flavor and sublattice and only concentrate on the time arguments. The Green function on the closed time contour is defined as
\begin{equation}
	iG(\vec{x},\vec{x}',t,t') = \langle \mathcal{T}_C \hat{\Psi}(\vec{x},t)\hat{\Psi}^\dagger(\vec{x}',t') \rangle,
\end{equation}
where $C$ and $\mathcal{T}_C$ have been introduced above. The operator at the earliest time is arranged to the right-most position. There are in total four different cases:
\begin{eqnarray}
		iG_{-+}(\vec{x},\vec{x}',t,t') &=& \langle  \hat{\Psi}(\vec{x},t)\hat{\Psi}^\dagger(\vec{x}',t') \rangle\nonumber\\
		iG_{+-}(\vec{x},\vec{x}',t,t') &=& -\langle \hat{\Psi}^\dagger(\vec{x}',t') \hat{\Psi}(\vec{x},t) \rangle\nonumber\\
		iG_{++}(\vec{x},\vec{x}',t,t') &=& \langle \mathcal{T} \hat{\Psi}(\vec{x},t)\hat{\Psi}^\dagger(\vec{x}',t') \rangle\nonumber\\
	iG_{--}(\vec{x},\vec{x}',t,t') &=& \langle \tilde{\mathcal{T}} \hat{\Psi}(\vec{x},t)\hat{\Psi}^\dagger(\vec{x}',t') \rangle.		
\end{eqnarray}
 Here $iG_{-+}$ implies that the first time argument $t$ is on the branch $C_-$ while the second time argument $t'$ is on the branch $C_+$. In this case, the operators are already arranged in a correct order, so $\mathcal{T}_C$ can be dropped. Similarly, $iG_{+-}(\vec{x},\vec{x}',t,t')$ means that the first time argument $t$ lies on the forward branch whereas $t'$ lies on the backward branch. Since in this case $t'$ is always further on the contour than $t$, meaning $\mathcal{T}_C$ switches the annihilation and creation operators together with giving an extra minus sign due to their fermionic nature.
Then $iG_{++}(\vec{x},\vec{x}',t,t')$ means that both $t$ and $t'$ are on the forward branch of the contour $C$, and thus the time-contour ordering operator $\mathcal{T}_C$ becomes a normal time-ordering operator $\mathcal{T}$. This implies that
\begin{eqnarray}
	&&iG_{++}(\vec{x},\vec{x}',t,t') \nonumber\\&&=  \theta(t-t')iG_{-+}(\vec{x},\vec{x}',t,t')  + \theta(t'-t)iG_{+-}(\vec{x},\vec{x}',t,t')\;, \nonumber\\			
\end{eqnarray}
where $\theta(t-t')$ denotes the Heaviside theta function.
In contrast, for $iG_{--}(\vec{x},\vec{x}',t,t')$, both $t$ and $t'$ are on the backward branch. Since the direction of the backward branch is opposite to the direction of the time axis (the backward branch extends from $-\infty$ to $\infty$),  the time-contour ordering operator $\mathcal{T}_C$ becomes an anti-time-ordering operator, denoted by $\tilde{\mathcal{T}}$. This means that
\begin{eqnarray}
	&&iG_{--}(\vec{x},\vec{x}',t,t') \nonumber\\&&=  \theta(t-t')iG_{+-}(\vec{x},\vec{x}',t,t')  + \theta(t'-t)iG_{-+}(\vec{x},\vec{x}',t,t') \;.\nonumber\\			
\end{eqnarray}
These components satisfy the following relation:
\begin{eqnarray}
	&&iG_{++}(\vec{x},\vec{x}',t,t') + iG_{--}(\vec{x},\vec{x}',t,t') \nonumber\\&&\hspace{1cm}= iG_{+-}(\vec{x},\vec{x}',t,t') + iG_{-+}(\vec{x},\vec{x}',t,t')\;.
	\label{eq:Greenfunctionconstraint}
\end{eqnarray}
The Green functions can straightforwardly be calculated within the functional-integral formalism according to
\begin{eqnarray}
	&&iG_{AB}(\vec{x},\vec{x}',t,t')\nonumber\\&&=\frac{1}{Z}\int \mathcal{D}\psi^\dagger\mathcal{D}\psi \; \psi_A(\vec{x},t) \psi^\dagger_B(\vec{x}',t') \exp(i S[\psi^\dagger,\psi]),
\end{eqnarray}
where $A,B = \pm$ here label the branch index of the close-time contour. The generating function $Z$ is defined in Eq.~\eqref{eq:partitionfuntionelectrononlytheory} together with the action in Eq.~\eqref{theoryonthecontour}. It is convenient to rotate the fields using the so-called Larkin-Ovchinnikov transformation
\begin{widetext}
\begin{eqnarray}
	\begin{pmatrix}
		\psi_+(\vec{x},t) \\
		\psi_-(\vec{x},t)
	\end{pmatrix}
	=\frac{1}{\sqrt{2}}\begin{pmatrix}
		1&1\\1&-1
	\end{pmatrix} 
	\begin{pmatrix}
		\psi_1(\vec{x},t)\\
		\psi_2(\vec{x},t)
	\end{pmatrix}\;\quad \rm{and}\quad
	\begin{pmatrix}
		\psi^\dagger_+(\vec{x},t)&
		\psi^\dagger_-(\vec{x},t)
	\end{pmatrix}
	= \frac{1}{\sqrt{2}}
	\begin{pmatrix}
		\psi^\dagger_1(\vec{x},t)&
		\psi_2^\dagger(\vec{x},t)
	\end{pmatrix}\begin{pmatrix}
	1&-1\\1&1
\end{pmatrix}.\nonumber
	\label{eq:Larkinovchinnikovtransformation}
\end{eqnarray}
After this transformation, Eq.~\eqref{theoryonthecontour} reads
	\begin{eqnarray}
		S[\psi^\dagger,\psi] &=& \int_{-\infty}^{\infty} dtdt'\; \Big[ \int d\vec{x}d\vec{x}' \psi_{a,\lambda,i}^\dagger(\vec{x},t)G^{-1}_{0,ab;\lambda\lambda';ij}(\vec{x},\vec{x}',t,t') \psi_{b,\lambda',j}(\vec{x}',t')   \nonumber \\ &-& \frac{1}{2}\int d\vec{x}d\vec{x}' \rho^a(\vec{x},t)D_{0,ab}(\vec{x},\vec{x}',t,t')\rho^b(\vec{x}',t')\Big]\;.
		\label{eq:electrononlyaction}
	\end{eqnarray}	
\end{widetext}
Additionally, we have introduced a number of short-hand notations. The inverse of the non-interacting Green function for the fermion field reads 
\begin{eqnarray}
&&G^{-1}_{0,ab;\lambda\lambda';ij}(\vec{x},\vec{x}';t,t') = \delta_{ij}\delta_{ab}  \times \nonumber \\ &&\times \left(i\partial_t+iv_F\vec{\nabla}\cdot\vec{\sigma}+\mu-V_{\rm{ex}}(\vec{x})\right)_{\lambda\lambda'} \delta(\vec{x}-\vec{x}')\delta(t-t')\;.\nonumber \\
\label{eq:noninteractinginverseGreenfunction}
\end{eqnarray} 
The Latin letters ($a,b,\dots$) take on the values $1$ and $2$ labeling the Keldysh indices, the Greek letters ($\lambda,\lambda',\dots$) assume $\pm1$ denoting the spinor indices, whereas $i,j=1,..,N$ denotes the flavor indices. All double indices are summed over, unless stated otherwise.
 
 The main advantages of the Larkin-Ovchinnikov basis is that the condition, Eq.~\eqref{eq:Greenfunctionconstraint} is implemented and the remaining components are independent, meaning
\begin{eqnarray}
	&&\hspace{-1cm}iG_{ab;\lambda\lambda'}(\vec{x},\vec{x}',t,t') \nonumber\\ &=& \left\langle	
	\begin{pmatrix}
		\psi_1(\vec{x},t)\\
		\psi_2(\vec{x},t)
	\end{pmatrix} 	\begin{pmatrix}
	\psi^\dagger_1(\vec{x}',t')&
	\psi_2^\dagger(\vec{x}',t')
\end{pmatrix}\right\rangle
	\nonumber\\&=&  \begin{pmatrix}
		iG^{R}(\vec{x},\vec{x}',t,t')&iG^{K}(\vec{x},\vec{x}',t,t')\\0&iG^{A}(\vec{x},\vec{x}',t,t')
	\end{pmatrix}_{ab}.\nonumber\\
\end{eqnarray} 
The superscripts $R$, $A$ and $K$ stand for the
retarded, advanced and Keldysh components of the Green function, respectively. 
The Green functions are calculable within the functional integral formalism according to
\begin{eqnarray}
	&&iG_{ab}(\vec{x},\vec{x}',t,t')\nonumber\\&&=\int \mathcal{D}\psi^\dagger\mathcal{D}\psi \; \psi_a(\vec{x},t) \psi^\dagger_b(\vec{x}',t') \exp(i S[\psi^\dagger,\psi]),
\end{eqnarray} 
with the action given in Eq.~\eqref{eq:electrononlyaction}.
It can be worked out that the inverse Green function is given by
\begin{eqnarray}
	&&G^{-1}_{ab}(\vec{x},\vec{x}';t,t')\nonumber\\&&\hspace{1cm} = \begin{pmatrix}
		(G^{-1})^{R}(\vec{x},\vec{x}',t,t')&
		(G^{-1})^{K}(\vec{x},\vec{x}',t,t')\\0&(G^{-1})^{A}(\vec{x},\vec{x}',t,t')
	\end{pmatrix}_{ab},\nonumber\\
\end{eqnarray} 
where
\begin{equation}
	(G^{-1})^{R/A} = (G^{R/A})^{-1},
	\label{eq:generalinverseGreensunction}
\end{equation}
and
\begin{equation}\label{eq:KeldyshcomponentofinverseGreenfunction}
	G^R \circ (G^{-1})^K = -G^K \circ (G^{-1})^A.
\end{equation}
The convolution operator $\circ$ is short for the integration over space and time coordinates as well as the summation over the spinor indices. The Keldysh component ($G_{12;\lambda\lambda'} \equiv G^K_{\lambda\lambda'}$) is usually parametrized  in terms of the retarded and advanced components according to
\begin{eqnarray}\label{eq:KeldyshGreenfunction}
G^K_{\lambda\lambda'}(\vec{r},\vec{r'};t,t') \equiv G^R \circ F - F \circ G^A.
\end{eqnarray}
Here,  $F$ is a Hermitian two-point function.
By inserting this equation into  Eq.~\eqref{eq:KeldyshcomponentofinverseGreenfunction}, we find 
\begin{equation}\label{eq:inverseKeldyshGreenfunction}
	(G^{-1})^K = (G^R)^{-1} \circ F - F \circ (G^A)^{-1}.
\end{equation}
This shows that the Keldysh component of the Green function and its inverse can be parametrized in the same way.

Furthermore, we introduced the density operator $\rho^a(\vec{x},t)=\psi_{b;\lambda,i}^\dagger(\vec{x},t)\gamma^{a}_{bc}\psi_{c;\lambda,i}(\vec{x},t)$ (all double indices are summed over) and the Coulomb interaction in Keldysh space 
\begin{eqnarray}
D_{0,ab}(\vec{x},\vec{x}',t,t') =\sigma^x_{ab}\delta(t-t') \frac{V(\vec{x}-\vec{x}')}{2}\;.
\label{eq:coulombGreenfunction}
\end{eqnarray} 

\noindent The vertex operators $\gamma$ are third rank tensors operating on the Keldysh space of fermions as well as bosons. They are defined as $\gamma^1_{ab}=1_{ab}$ and $\gamma^2_{ab}=\sigma^x_{ab}$. Let us emphasize that our choice of convention gives an extra factor $1/2$ to the Coulomb interaction term, {\it i.e.}, $D_{0,ab} \propto \frac{V(\vec{x}-\vec{x}')}{2}$. 

\subsection{The Dyson and the quantum-kinetic equation}\label{subsec:dysonequation}

\begin{figure}[tbh!]
	\includegraphics[width=0.45\textwidth]{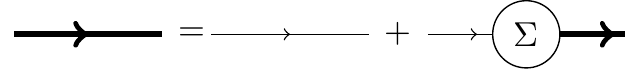}
	\caption{Diagrammatic representation of the Dyson equation for the fermion Green function.}
	\label{fig:Dysonequationforfermion}
\end{figure}

In the presence of interactions, the Green function obeys the Dyson equation (pictorially represented in Fig.~\ref{fig:Dysonequationforfermion})
\begin{equation}\label{eq:schwingerdysonequationforfermion}
	G = G_0 + G_0 \overset{\otimes}{} \Sigma\overset{\otimes}{} G.
\end{equation}

\noindent Here $\Sigma$ denotes the self-energy which is evaluated order by order from a perturbative expansion in the interaction.  It is represented diagrammatically by a set of  one-particle irreducible diagrams. The convolution $\otimes$ denotes integration over time and space, as well as summation of internal indices, including the Keldysh indices. One can rewrite Eq.~\eqref{eq:schwingerdysonequationforfermion}  as
\begin{equation}
	G^{-1} = G_0^{-1} - \Sigma.
\end{equation} 
The fermion self-energy inherits the Keldysh matrix structure of the inverse Green function as
\begin{eqnarray}\label{eq:fermionselfenergy}
	&&\Sigma_{ab;\lambda\lambda'}(\vec{x},\vec{x}',t,t')\nonumber\\&&\hspace{1cm} = \begin{pmatrix}
		\Sigma^{R}_{\lambda\lambda'}(\vec{x},\vec{x}',t,t')&
		\Sigma^{K}_{\lambda\lambda'}(\vec{x},\vec{x}',t,t')\\0&\Sigma^{A}_{\lambda\lambda'}(\vec{x},\vec{x}',t,t')
	\end{pmatrix}_{ab}.
\end{eqnarray} 

\noindent For the retarded and advanced components,  it consequently enters according to

\begin{equation}
	(G^{-1})^{R/A}= (G_0^{-1})^{R/A} - \Sigma^{R/A}\;.
	\label{eq:Dysonequationfor retardedandadvancedcomponents}
\end{equation}
The self-energy plays two major roles: It can affect the dispersion relation and lead to a finite lifetime. This is most easily seen in frequency-momentum space. The modified pole $\omega = \omega(\vec{p}) - i \gamma(\vec{p})$ is the solution of $(G_0^{-1})^{R}(\vec{p},\omega) - \Sigma^{R}(\vec{p},\omega)=0$. It gives a new dispersion relation, $\omega(\vec{p})$, as well as a decay rate, $\gamma(\vec{p})$, of the excitation. When the decay rate is sufficiently small, this excitation is called a quasi-particle.

The Keldysh component of the Dyson equation leads to a kinetic equation that governs the time evolution of the fermion distribution function. As a result, we have
\begin{eqnarray}\label{eq:DysonequationforKeldyshcomponent}
	- \Sigma^{K}&=&(G^{-1})^{K} \nonumber\\
	&=&
	(G^R)^{-1} \circ F - F \circ (G^A)^{-1} \;,
\end{eqnarray}
where we used the fact that $(G^{-1}_0)^K $ is a pure regularization and can be neglected in the presence of interactions, see Eq.~\eqref{eq:noninteractinginverseGreenfunction}. To arrive at the second equality, we use Eq.~\eqref{eq:inverseKeldyshGreenfunction} and parametrize the Keldysh component of the inverse Green function in terms of the Hermitian function $F$. Subsequently, by substituting Eq.~\eqref{eq:Dysonequationfor retardedandadvancedcomponents} into  Eq.~\eqref{eq:DysonequationforKeldyshcomponent}, one finds
\begin{equation}\label{eq:Keldyshequationfermion}
	G_0^{-1}\circ F - F \circ G_0^{-1} = -\Sigma^K + \Sigma^R \circ F - F \circ \Sigma^A.
\end{equation}
The regularization $\pm i\delta$ can be omitted from the retarded and advanced components in the presence of a non-zero imaginary part of the self-energy. The above equation is called the quantum-kinetic equation for the distribution matrix $F$. The solution of the full quantum kinetic equation, Eq.~\eqref{eq:Keldyshequationfermion} is usually exceedingly difficult. However, after some approximations, discussed below, the quantum kinetic equation reduces to a Boltzmann equation. The latter can be solved by, for instance, a variational method \cite{Ziman2000}.

\begin{figure}[tbh!]
	\includegraphics[width=0.45\textwidth]{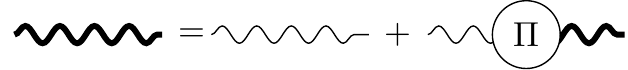}
	\caption{Diagrammatic representation of the Dyson equation for the boson Green function.}
	\label{fig:Dysonequationforboson}
\end{figure}

Later in this paper, we show that the Coulomb interactions play a role in facilitating the emergence of a boson associated with the electron density fluctuations. To this end, we also summarize the salient features of the Keldysh technique for a bosonic field. Similar to the fermion Green function, the boson Green function and its inverse have three non-vanishing components expressed in the following matrix structure
\begin{eqnarray}
	&&D_{ab}(\vec{x},\vec{x}',t,t')\nonumber\\&&\hspace{1cm} = \begin{pmatrix}
		D^{K}(\vec{x},\vec{x}',t,t')&D^{R}(\vec{x},\vec{x}',t,t')\\D^{A}(\vec{x},\vec{x}' ,t,t')&0
	\end{pmatrix}_{ab},
	\label{eq:bosongrennfunctionstructure}
\end{eqnarray} 
and
\begin{eqnarray}
	&&D^{-1}_{ab}(\vec{x},\vec{x}',t,t')\nonumber\\&&\hspace{1cm} = \begin{pmatrix}
		0&(D^{-1})^{A}(\vec{x},\vec{x}',t,t')\\(D^{-1})^{R}(\vec{x},\vec{x}',t,t')&(D^{-1})^{K}(\vec{x},\vec{x}',t,t')
	\end{pmatrix}_{ab},\nonumber\\
\end{eqnarray} 
together with the following relations
\begin{equation}
	(D^{-1})^{R/A} = (D^{R/A})^{-1},
\end{equation}
and
\begin{equation}\label{eq:KeldyshcomponentofinverseGreenfunctionboson}
	D^R \circ (D^{-1})^K = -D^K \circ (D^{-1})^A.
\end{equation} 
The interacting Green function is again determined from a Dyson equation (pictorially represented in Fig.~\ref{fig:Dysonequationforboson})
\begin{equation}\label{eq:schwingerdysonequationforboson}
	D = D_0 + D_0 \overset{\otimes}{} \Pi\overset{\otimes}{} D,
\end{equation}
with a self-energy $\Pi$, which is, in general, approximated in a perturbative series. The self-energy assumes the same Keldysh matrix structure as the inverse Green function to preserve causality, namely
\begin{eqnarray}
	&&\Pi_{ab}(\vec{x},\vec{x}',t,t')\nonumber\\&&\hspace{1cm} = \begin{pmatrix}
		0&\Pi^{A}(\vec{x},\vec{x}',t,t')\\\Pi^{R}(\vec{x},\vec{x}',t,t')&\Pi^{K}(\vec{x},\vec{x}',t,t')
	\end{pmatrix}_{ab}.
	\label{eq:selfenergyboson}
\end{eqnarray} 
The poles of the retarded and advanced components are again shifted by interaction effects leading to a new energy dispersion and a finite life time for the dressed particles by means of
\begin{equation}\label{eq:Dysonequationfor retardedandadvancedcomponentsboson}
	(D^{-1})^{R/A} = (D_0^{-1})^{R/A} - \Pi^{R/A}.
\end{equation}
The Keldysh component of the Dyson equation leads to a kinetic equation for the boson distribution function $B$.
\begin{equation}\label{eq:Keldyshequationboson}
	D_0^{-1}\circ B - B \circ D_0^{-1} = -\Pi^K + \Pi^R \circ B - B \circ \Pi^A.
\end{equation}
The left-hand side again describes a streaming term, whereas the right-hand side accounts for collision events. The Hermitian function $B$ is employed to parametrize the Keldysh Green function according to
\begin{eqnarray}
	D^K=D^R \circ B - B \circ D^A.
	\label{eq:KeldyshbosonGreenfunction}
\end{eqnarray}

Before we continue, it is convenient to rewrite the Keldysh equations, Eqs.~\eqref{eq:Keldyshequationfermion} and~\eqref{eq:Keldyshequationboson}. The rewriting seems arbitrary at this point, but later it will allow us to identify the left-hand sides with the streaming terms of a Boltzmann equation, whereas the right-hand sides will be the collision integrals. We can use the fact that we can decompose the self-energies according to $\Sigma^{R/A} = \Re \Sigma^R \pm i\Im \Sigma^R$ and $\Pi^{R/A} = \Re \Pi^R \pm i\Im \Pi^R$. This allows to rewrite Eq.~\eqref{eq:Keldyshequationfermion} as 
\begin{equation}
	\left[G_0^{-1}-\Re\Sigma^R\overset{\circ}, F\right] = -\Sigma^K +i\{\Im \Sigma^R \overset{\circ}, F\}.
	\label{eq:Keldysh0}
\end{equation}
and Eq.~\eqref{eq:Keldyshequationboson} as 
	\begin{equation}
		\left[D_0^{-1}-\Re\Pi^R\overset{\circ}, B\right] = -\Pi^K +i\{\Im \Pi^R \overset{\circ}, B\}.
	\end{equation}
Here, we introduced the notation $[V\overset{\circ},W] = V \circ W - W \circ V$  and $\{V\overset{\circ},W\}  = V \circ W + W \circ V$, defining the commutator and anticommutator of the functions $V$ and $W$ with $\circ$ as defined before.

\subsection{The Wigner transform and the gradient expansion}\label{subsec:wignertransform}

In equilibrium quantum field theory, diagrammatic approaches are, because of homogeneity, usually carried out in momentum and energy space instead of coordinate-space and time, meaning it is a simple Fourier transform. The semiclassical limit, however, is most conveniently accessed using the Wigner transform, which is a mixed representation. 
We briefly summarize it here for convenience. A generic space-time function, $g(\vec{x}_1,\vec{x}_2,t_1,t_2)$, can be rewritten in terms of center-of-mass ($\vec{r},t$) = ($(\vec{x}_1+\vec{x}_2)/2,(t_1+t_2)/2$) and relative coordinates ($\vec{x},\tau$) = ($\vec{x}_1-\vec{x}_2,t_1-t_2$). The Wigner transform is now a Fourier transform over the relative coordinates while the centre-of-mass coordinates are kept intact. Consequently, one obtains a function of center-of-mass spacetime, momentum, and frequency, {\it i.e.}, 
$\tilde{g}(\vec{r},t,\vec{p},\omega)= \int d\vec{x} d \tau \; g(\vec{r},t,\vec{x},\tau) \; e^{-i\vec{p}\cdot \vec{x}+i\omega \tau}$. 
There are two important  Wigner transforms that will be needed later for our derivation of the Boltzmann equation. 

(i) For a two-point function which can be decompose into an algebraic product of other two-point functions, {\it i.e.},
$C(\vec{r}_1,\vec{r}_2;t_1,t_2)=A(\vec{r}_1,\vec{r}_2;t_1,t_2)B(\vec{r}_1,\vec{r}_2;t_1,t_2)$, one can show that its Wigner transform is given by the momentum-frequency convolution
\begin{widetext}
\begin{eqnarray}
	C(\vec{r},t,\vec{p},\omega) = \int \frac{d\vec{p}_1}{(2\pi)^d}\frac{d\omega_1}{2\pi} \; A(\vec{r},t,\vec{p}_1,\omega_1) B(\vec{r},t,\vec{p}-\vec{p}_1,\omega-\omega_1).\nonumber\\
\end{eqnarray}
\end{widetext}

(ii) The Wigner transform of the space-time convolution of two two-point functions {\it i.e.}, 
$D(\vec{r}_1,\vec{r}_2;t_1,t_2)=(A \circ B) (\vec{r}_1,\vec{r}_2;t_1,t_2)$ is given by their Moyal product as
\begin{widetext}
\begin{eqnarray}
	D(\vec{r},t,\vec{p},\omega) &=& A(\vec{r},t,\vec{p},\omega) \star B(\vec{r},t,\vec{p},\omega)=A(\vec{r},t,\vec{p},\omega)\exp \left(\frac{i}{2}\left( \overleftarrow{\partial}_{\vec{r}} \overrightarrow{\partial}_{\vec{p}}-\overleftarrow{\partial}_{\vec{p}} \overrightarrow{\partial}_{\vec{r}}  - \overleftarrow{\partial}_{t} \overrightarrow{\partial}_{\omega} +\overleftarrow{\partial}_{\omega} \overrightarrow{\partial}_{t} \right)\right)B(\vec{r},t,\vec{p},\omega)\;. \nonumber \\
\end{eqnarray}
\end{widetext}
In the cases we are interested in, the function varies slowly with the center-of-mass coordinates.  Consequently, it is legitimate to keep only the lowest-order gradient terms according to
\begin{widetext}
 \begin{eqnarray}
	D(\vec{r},t,\vec{p},\omega) &\approx& A(\vec{r},t,\vec{p},\omega)B(\vec{r},t,\vec{p},\omega) -\frac{i}{2} \Big[\partial_{\vec{p}}A(\vec{r},t,\vec{p},\omega) \partial_{\vec{r}}B(\vec{r},t,\vec{p},\omega)-\partial_{\vec{r}}A(\vec{r},t,\vec{p},\omega) \partial_{\vec{p}}B(\vec{r},t,\vec{p},\omega)\Big]
	\nonumber\\ &&+\frac{i}{2} \Big[\partial_{\omega}A(\vec{r},t,\vec{p},\omega) \partial_{t}B(\vec{r},t,\vec{p},\omega)-\partial_{t}A(\vec{r},t,\vec{p},\omega) \partial_{\omega}B(\vec{r},t,\vec{p},\omega)\Big]+ ... \;.
	\label{eq:gradientexpansion}
\end{eqnarray}
\end{widetext}

\subsection{Gradient expansion of the Keldysh equation}\label{subsec:gradientkeldysh}

The Wigner transform of the Keldysh equation in Eq.~\eqref{eq:Keldyshequationfermion}, reads
\begin{eqnarray}
		\left[G_0^{-1}-\Re \Sigma^R_\star \overset{\star},  F\right]_-&=&-\Sigma^K_\star+i\{  \Im \Sigma^R_\star \overset{\star},F \}_+ \;.
	\end{eqnarray}
In writing the above equation, we briefly introduced the notation $\Sigma_\star$. It accounts for the fact that the self-energy itself has a gradient expansion according to
\begin{eqnarray}
	\Sigma_\star\approx \Sigma+\Sigma_\times+...
	\end{eqnarray}
where $\Sigma$ involves no gradients, while $\Sigma_\times$ involves both one spatial and momentum gradient (or, equivalently, frequency and time)~\cite{Kral1997,knoll2001}.
To leading order in non-vanishing gradients we find
\begin{eqnarray}
	&&\left[G_0^{-1}-\Re\Sigma^R, F\right]+i\Big\{\partial_{\vec{x}}\left(G_0^{-1}-\Re\Sigma^R\right) \cdot \partial_{\vec{p}} F \nonumber\\&& - \partial_{\vec{p}}\left(G_0^{-1}-\Re\Sigma^R\right) \cdot \partial_{\vec{x}} F-\partial_{t}\left(G_0^{-1}-\Re\Sigma^R\right) \partial_{\epsilon} F \nonumber\\ &&+ \partial_{\epsilon}\left(G_0^{-1}-\Re\Sigma^R\right) \partial_{t} F\Big\}  = -\Sigma^K + i\{\Im\Sigma^R,  F\}\nonumber 
\\&&- \Sigma^K_\times +i \{\Im \Sigma^R_\times,F\} .\nonumber\\
	\label{eq:gradientexpansionkeldyshequation}
\end{eqnarray}
It turns out that the contributions due to $\Sigma_\times$ are of the non-quasi-particle type  and vanish once we perform the quasi-particle approximation. Consequently, we drop them form our following discussion. 

In order to set up a formalism that accommodates for a two- or even multi-band scenario we make the following assumption: There is a transformation $\mathcal{U}_{\vec{p}}$ (in the case of a Dirac-type theory we specify it later) that projects the Green function into a diagonal basis according to
\begin{eqnarray}
g^{R/A}_{0}(\vec{x},\vec{p},\omega) &=& \mathcal{U}^\dagger_{\vec{p}} G^{R/A}_0(\vec{x},\vec{p},\omega) \mathcal{U}_{\vec{p}}
\end{eqnarray}
with $g^{R/A}_{0}(\vec{x},\vec{p},\omega)$ being a diagonal matrix. While in the following we present matrix equations, we only concentrate on the diagonal elements. Equivalently, we project the self-energies into the quasi-particle basis according to
\begin{eqnarray}
\sigma^{R/A}(\vec{x},\vec{p},\omega) &=& \mathcal{U}^\dagger_{\vec{p}} \Sigma^{R/A}(\vec{x},\vec{p},\omega) \mathcal{U}_{\vec{p}}\;.
\end{eqnarray}
This leads to
\begin{eqnarray}
	&&i\partial_{\vec{x}}\left(g_0^{-1}-\Re\sigma^R\right) \cdot \left(\partial_{\vec{p}} F+i\left[\mathcal{A}_{\vec{p}},F\right]\right)  \nonumber\\&& - \left(\partial_{\vec{p}}\left(g_0^{-1}-\Re\sigma^R\right)+i\left[\mathcal{A}_{\vec{p}},g_0^{-1}-\Re\sigma^R\right]\right) \cdot \partial_{\vec{x}} F\nonumber\\ &&-\partial_{t}\left(g_0^{-1}-\Re\sigma^R\right) \partial_{\epsilon} F + \partial_{\epsilon}\left(g_0^{-1}-\Re\sigma^R\right) \partial_{t} F \nonumber\\&& \hspace{4.5cm}= -\sigma^K + 2i\Im\sigma^R  F\;.\nonumber\\
	\label{eq:gradientexpansionkeldyshequationinquasiparticlebasis}
\end{eqnarray}
We observe that two additional terms that involve the Berry connection  
 $\mathcal{A}_{\vec{p}} = -i \mathcal{U}^\dagger_{\vec{p}}\partial_{\vec{p}}\mathcal{U}_{\vec{p}}$ are obtained. We are interested here in the solution to zeroth order in $\mathcal{A}_{\vec{p}}$ that is
  \begin{equation}
  	F_{\lambda\lambda'} = F_{\lambda\lambda'}^{(0)} + \mathcal{O}(\mathcal{A}_{\vec{p}}) = (1-2f_{\lambda}(\vec{r},\vec{p},t))\delta_{\lambda\lambda'}+ \mathcal{O}(\mathcal{A}_{\vec{p}}),
  	\label{eq:distributionfunctionzeroorderinA}
  \end{equation}
  containing only the diagonal elements of the distribution function. The function $f$ introduced above will later play the role of the fermionic distribution function and in equilibrium, it reduces to the Fermi-Dirac distribution. Within the quasi-particle approximation, $f_{\lambda}$ is independent of the frequency variable. Moreover, it is assumed to be diagonal in the spinor space. Consequently, only the main diagonal elements of the self-energies are important. In total, this leads to
 \begin{eqnarray}
 		&&\partial_{\epsilon}\left(g_{0,\lambda\lambda}^{-1}(\vec{x},t,\vec{p},\epsilon)-\Re\sigma^R_{\lambda\lambda}(\vec{x},t,\vec{p},\epsilon)\right)\partial_t f_\lambda(\vec{x},t,\vec{p}) \nonumber\\&&+\;\partial_{\vec{x}}\left(g_{0,\lambda\lambda}^{-1}(\vec{x},t,\vec{p},\epsilon)-\Re\sigma^R_{\lambda\lambda}(\vec{x},t,\vec{p},\epsilon)\right) \cdot \partial_{\vec{p}} f_\lambda(\vec{x},t,\vec{p}) \nonumber\\&&-\; \partial_{\vec{p}}\left(g_{0,\lambda\lambda}^{-1}(\vec{x},t,\vec{p},\epsilon)-\Re\sigma^R_{\lambda\lambda}(\vec{x},t,\vec{p},\epsilon)\right) \cdot \partial_{\vec{x}} f_\lambda(\vec{x},t,\vec{p})\nonumber\\&& = -\frac{i}{2}\sigma^K_{\lambda\lambda}(\vec{x},t,\vec{p},\epsilon) - \Im\sigma^R(\vec{x},\vec{p},t,\epsilon)  (1-2f(\vec{x},t,\vec{p}))\;.\nonumber\\
 		\label{eq:Keldyshfermion}
 \end{eqnarray}

The left-hand side will contain the so-called streaming terms, consisting of three contributions. The first term describes the time derivative of the distribution function with the quasi-particle weight
$\partial_{\epsilon}\left(g_0^{-1}-\Re\sigma^R\right) = 1-\partial_{\epsilon}\Re\sigma^R$.
The second term accounts for the change of the distribution function due to a force $\partial_{\vec{x}}\left(g_0^{-1}-\Re\sigma^R\right)$, whereas the last term tracks the change of the distribution function due to the diffusion of excitations with the velocity  $\partial_{\vec{p}}\left(g_0^{-1}-\Re\sigma^R\right)$. The right-hand side describes collisions that drive the system towards equilibrium.
 
 We now consider the Keldysh equation in Eq. (\ref{eq:Keldyshequationboson}) of the plasmon field and we drop the terms from expanding the self-energies here from the very start. Following the same steps as for the fermions, we first decompose the retarded and advanced components of the self-energy according to $\Pi^{R/A} = \Re \Pi^R \pm i\Im \Pi^R$. This allows us to write Eq.~\eqref{eq:Keldyshequationboson} as 
	\begin{equation}
		\left[D_0^{-1}-\Re\Pi^R\overset{\circ}, B\right] = -\Pi^K +i\{\Im \Pi^R \overset{\circ}, B\}.
	\end{equation}
	After a Wigner transformation and keeping the Moyal product to first non-trivial order, we obtain 
	\begin{eqnarray}\label{eq:gradientexpansionkeldyshequationboson}
		&&i\Big\{\partial_{\vec{x}}\left(D_0^{-1}-\Re\Pi^R\right) \cdot \partial_{\vec{p}} B - \partial_{\vec{p}}\left(D_0^{-1}-\Re\Pi^R\right) \cdot \partial_{\vec{x}} B\nonumber\\ &&-\partial_{t}\left(D_0^{-1}-\Re\Pi^R\right) \partial_{\omega} B + \partial_{\omega}\left(D_0^{-1}-\Re\Pi^R\right) \partial_{t} B\Big\} \nonumber\\&& \hspace{4.5cm}= -\Pi^K + 2i\Im\Pi^R  B.\nonumber\\
	\end{eqnarray}

Next, we use the parametrization
  \begin{equation}
  	B = 1+2b \;,
  	\label{eq:distributionfunctionB}
  \end{equation}
  where $b$ plays the role of the bosonic distribution function exactly in equilibrium reduces to the Bose-Einstein distribution. This leads to
 \begin{eqnarray}
 		&&\partial_{\epsilon}\left(D_{0}^{-1}(\vec{x},t,\vec{p},\epsilon)-\Re \Pi^R(\vec{x},t,\vec{p},\epsilon)\right)\partial_t b(\vec{x},t,\vec{p}) \nonumber\\&&+\;\partial_{\vec{x}}\left(D_{0}^{-1}(\vec{x},t,\vec{p},\epsilon)-\Re\Pi^R(\vec{x},t,\vec{p},\epsilon)\right) \cdot \partial_{\vec{p}} b(\vec{x},t,\vec{p}) \nonumber\\&&-\; \partial_{\vec{p}}\left(D_{0}^{-1}(\vec{x},t,\vec{p},\epsilon)-\Re\Pi^R(\vec{x},t,\vec{p},\epsilon)\right) \cdot \partial_{\vec{x}} b(\vec{x},t,\vec{p})\nonumber\\&& = \frac{i}{2}\Pi^K(\vec{x},t,\vec{p},\epsilon) + \Im\Pi^R(\vec{x},\vec{p},t,\epsilon)  (1+2b(\vec{x},t,\vec{p}))\;.\nonumber\\
 		\label{eq:Keldyshboson}
 \end{eqnarray}
The missing piece to transform Eqs.~\eqref{eq:Keldyshfermion} and~\eqref{eq:Keldyshboson} into Boltzmann equations is to integrate them over the respective spectral functions, as we will discuss later on.

\section{Part A: Electron-hole hydrodynamics in the weak-coupling limit}
\label{sec:HartreeBornapproximation}

In this section we review the equations of hydrodynamics in the weakly interacting limit. We first discuss the non-interacting limit to define the basic Green function and the projection into the appropriate quasi-particle basis of electrons and holes. Afterwards we discuss the Hartree-Fock approximation. We show that this is equivalent to the renormalized collisionless Boltzmann equation after a series of approximations. 

 \subsection{The non-interacting limit}\label{subsec:noninteracting}

We consider a model of interacting Dirac fermions according to Eq.~\eqref{eq:electrononlyaction} where we first neglect the interactions. By inserting Eq.~\eqref{eq:noninteractinginverseGreenfunction} into Eq.~\eqref{eq:generalinverseGreensunction}, we find the retarded Green function of the non-interacting Dirac theory according to
 \begin{eqnarray}
 G^{R/A}_{0}(\vec{x},\vec{p},\omega) = \left(\omega\pm i\delta - v_F \vec{\sigma}\cdot{\vec{p}} + \mu-V_{\rm{ex}}(\vec{x})\right)^{-1}.\nonumber\\
 \end{eqnarray}
  It is convenient to work in the quasi-particle basis. There, the Dirac Hamiltonian as well as the Green function are diagonal. This leads to
\begin{eqnarray}\label{eq:fermionGreenfucntioninquasiparticlebasis}
	g^{R/A}_{0,\lambda\lambda',ij}(\vec{x},\vec{p},\omega) &=& \left(\mathcal{U}^\dagger_{\vec{p}} G^{R/A}_{0,ij}(\vec{x},\vec{p},\omega) \mathcal{U}_{\vec{p}}\right)_{\lambda\lambda'}\nonumber\\
	&=& (\omega\pm i\delta-\lambda v_F p +\mu-V_{\rm{ex}}(\vec{x}))^{-1}\delta_{\lambda\lambda'}\delta_{ij}\;,\nonumber\\
\end{eqnarray}
with
\begin{eqnarray}
	\mathcal{U}_{\vec{p}} = \frac{1}{\sqrt{2}}\begin{pmatrix}
		-\exp(-i\theta_{\vec{p}})&\exp(-i\theta_{\vec{p}})\\1&1
	\end{pmatrix}\;\;.
	\label{eq:tranformationmatrix}
\end{eqnarray}
Here, $p=|\vec{p}|$ and $\tan(\theta_{\vec{p}})=p_y/p_x$.
The dispersion relation can be extracted from the poles of the Green function in Eq.~\eqref{eq:fermionGreenfucntioninquasiparticlebasis}: the non-interacting Dirac theory has two linear dispersing energy bands $\epsilon_\pm(\vec{x},\vec{p}) = \pm v_F p +V_{\rm{ex}}(\vec{x}) - \mu$ with the local electro-chemical potential given by $\mu-V_{\rm{ex}}(\vec{x})$.  The spectral function, defined by $\Im g^R_{0,\lambda\lambda'}(\vec{x},\vec{p},\omega) = -\pi  \delta(\omega-\epsilon_{\lambda}(\vec{x},\vec{p}))\delta_{\lambda\lambda'}$, exhibits resonances at $\omega = \epsilon_\pm(\vec{x},\vec{p})$.  The Wigner transform of the Keldysh Green function of Eq.~\eqref{eq:KeldyshGreenfunction} reads
\begin{eqnarray}
	&&g^K_{0,\lambda\lambda'}(\vec{x},t,\vec{p},\omega) =2i\Im g^R_{0,\lambda\lambda''}(\vec{x},\vec{p},\omega)F_{\lambda''\lambda'}(\vec{x},\vec{p},\omega)\nonumber\\ && \hspace{0.5cm}= -2\pi i \delta(\omega-\epsilon_\lambda(\vec{x},\vec{p}))\left(1-2f_{\lambda}(\vec{x},\vec{p},t)\right)\delta_{\lambda\lambda'}.\nonumber\\
\end{eqnarray} 
where $f(\vec{x},\vec{p},t)$ is the Fermi Dirac distribution in equilibrium. The Keldysh component thus contains the information about the occupation numbers, whereas the retarded and advanced  components only contain information about the resonances and the energy levels.

\subsection{Hartree-Fock approximation: The collisionless limit} \label{subsec:collisionless}
  In what follows, we will discuss the corrections of the energy spectrum due to the Coulomb interaction. To this end, we study the Hartree and Fock self-energies~\cite{Kotov2012} depicted in Fig.~\ref{fig:Hartreediagram} and Fig.~\ref{fig:Fockdiagram}. 
  
\subsubsection{The Hartree diagram}  
\begin{figure}[htb!]
	\centering
	\begin{subfigure}[b]{0.1\textwidth}
		\centering
		\includegraphics[width=\textwidth]{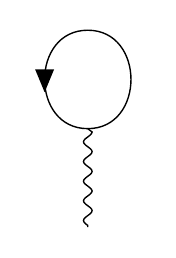}
		\subcaption{}
		\label{fig:Hartreediagram}
	\end{subfigure}
	\begin{subfigure}[b]{0.1\textwidth}
		\centering
		\includegraphics[width=\textwidth]{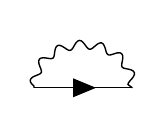}
		\subcaption{}
		\label{fig:Fockdiagram}
	\end{subfigure}
	\caption{The self-energies at first order in the interaction. The left diagram is known as the direct
		or Hartree contribution and the right diagram is known as the exchange or Fock contribution.}
	\label{fig:HartreeFock}
\end{figure}

 The Hartree diagram in Fig.~\ref{fig:Hartreediagram}, has the following algebraic expression:
  \begin{widetext}
\begin{eqnarray}	
-i\Sigma^H_{ab;\lambda\lambda'}(\vec{x},\vec{x}';t,t')= -\delta_{\lambda\lambda'} \delta(\vec{x}-\vec{x}')\delta(t-t')\int dt''d\vec{x}''\;\gamma^e_{ab}D_{0,ef}(\vec{x}',\vec{x}'',t',t'')\gamma^{f}_{cd}G_{0,dc;\lambda''\lambda'';ii}(\vec{x}'',\vec{x}'',t'',t'')\;.\nonumber\\ \label{eq:Hartreeselfenergy}
\end{eqnarray}
Since $D_{0,ef}(\vec{x},\vec{x}',t,t') = \sigma_{ef}^x \delta(t-t')\frac{V(\vec{x}-\vec{x}')}{2}$ is off-diagonal in Keldysh space, only two terms survive when we sum over $e$ and $f$. This gives
\begin{eqnarray}
-i\Sigma^H_{ab;\lambda\lambda'}(\vec{x},\vec{x}';t,t')
	&=& -\delta_{\lambda\lambda'}\delta(\vec{x}-\vec{x}')\delta(t-t') \int dt''d\vec{x}'' \frac{V(\vec{x}'-\vec{x}'')}{2}\gamma^1_{ab}\gamma^{2}_{cd}G_{0,dc;\lambda''\lambda'',ii}(\vec{x}'',\vec{x}'',t'',t'')\nonumber \\
	&-&\delta_{\lambda\lambda'}\delta(\vec{x}-\vec{x}')\delta(t-t') \int dt''d\vec{x}'' \frac{V(\vec{x}'-\vec{x}'')}{2}\gamma^2_{ab}\gamma^{1}_{cd}G_{0,dc;\lambda\lambda,ii}(\vec{x}'',\vec{x}'',t'',t'')\;.\nonumber\\
\end{eqnarray}
\end{widetext}
We find that the second term vanishes due to
\begin{widetext}
\begin{eqnarray} 				
\gamma^2_{ab}\gamma^1_{cd}G_{0,dc;\lambda''\lambda'',ii}(\vec{x}',\vec{x}',t',t')&=& G^R_{0;\lambda''\lambda'',ii}(\vec{x}',\vec{x}',t',t')+G^A_{0;\lambda''\lambda'',ii}(\vec{x}',\vec{x}',t',t')\nonumber \\ &=& \int \frac{d\omega}{2\pi} \frac{d\vec{p}}{(2\pi)^2} \Re G^R_{0;\lambda''\lambda'',ii}(\vec{x}',t',\vec{p},\omega)=0\;.
\end{eqnarray}
\end{widetext}
The last equality follows because $\Re G^R_{0;\sigma\sigma}(\vec{x}',t',\vec{p},\omega)$ is an analytic function in either the upper or the lower complex frequency half-plane. Consequently, we have
\begin{eqnarray}
	&&-i\Sigma^H_{ab;\lambda\lambda'}(\vec{x},\vec{x}',t,t')= -\delta_{\lambda\lambda'}\delta(\vec{x}-\vec{x}')\delta(t-t')\delta_{ab}\nonumber\\&&\int d\vec{x}'' \frac{V(\vec{x}'-\vec{x}'')}{2}G^K_{0;\lambda''\lambda'',ii}(\vec{x}'',\vec{x}'';t,t).\nonumber\\
\end{eqnarray}
Finally, after a Wigner transformation, we rotate the self-energy into the quasi-particle basis. This gives
\begin{eqnarray}
	&&\hspace{-0.5cm}-i \sigma^H_{ab;\lambda\lambda'}(\vec{x},t,\vec{p},\omega) = -i \left(\mathcal{U}^\dagger_{\vec{p}} \Sigma^{H}(\vec{x},t,\vec{p},\omega) \mathcal{U}_{\vec{p}}\right)_{\lambda\lambda'}\nonumber\\&=& -\delta_{\lambda\lambda'}\delta_{ab}\int d\vec{x}'' \frac{V(\vec{x}'-\vec{x}'')}{2}g^K_{0;\lambda''\lambda'',ii}(\vec{x}'' ,\vec{x}';t,t)\;. \nonumber \\
\end{eqnarray}
The Wigner transformation of the Keldysh Green function with the same time and spatial argument is associated with the electron density according to
\begin{equation}
g_{0;\lambda''\lambda'',ii}^K(\vec{x}'',\vec{x}'';t,t)= 2in(\vec{x}'',t),
\end{equation}
 where $n(\vec{x}'',t)=n_+(\vec{x}'',t)+n_-(\vec{x}'',t)$ defines the total charge density at position $\vec{x}''$ and time  $t$. Here $n_+(\vec{x}',t)= 
N\int\frac{d\vec{k}}{(2\pi)^2}f_+(\vec{x}',\vec{k},t)$ is the electron density and $n_-(\vec{x}',t) =  N\int\frac{d\vec{k}}{(2\pi)^2}\left(f_-(\vec{x}',\vec{k},t)-1\right)$ is the hole density. The electron charge $-e$ is henceforth set to 1. In total, the Hartree self-energy produces the classical Coulomb potential of all electrons in the system exerted on an electron located at a position $\vec{x}$ according to
\begin{eqnarray}
	\sigma^H_{ab;\lambda\lambda'}(\vec{x},t,\vec{p},\omega)  =\delta_{\lambda\lambda'}\delta_{ab} \int d\vec{x}' V(\vec{x}-\vec{x}') n(\vec{x}',t).\nonumber\\ 
	\label{eq:Hartreeselfenergyinthequasiparticlebasis}
\end{eqnarray}
The Hartree term is real-valued and independent of the momentum and frequency variables. Furthermore, it is diagonal in Keldysh space, meaning its Keldysh component is zero.  

\subsubsection{The Fock diagram}

\begin{figure}[t]
	\centering
	\centering
	\includegraphics[width=0.5\textwidth]{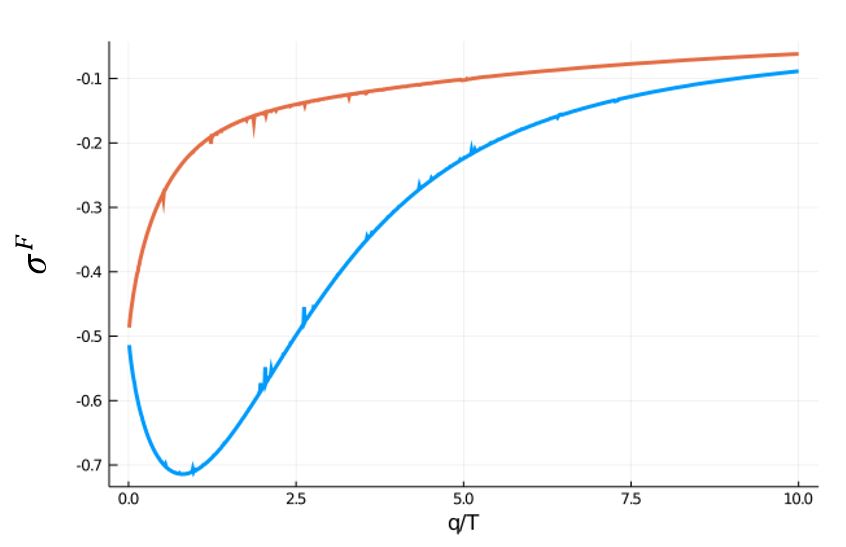}
	\caption{Exchange self-energies for Dirac fermions with the chemical potential $\mu/T = 1$ as functions of
		wave vector $q/T$. The fine structure constant $\alpha$ is chosen to be $1$. The blue curve represents $(\sigma^{F,2})^R_{++}$ showing the exchange energy correction of the electrons. The orange curve shows $(\sigma^{F,2})^R_{--}$\, the exchange energy of the holes.}
	\label{fig:exchangeenergy}
\end{figure}

\begin{figure}[t]
	\centering
	\includegraphics[width=0.5\textwidth]{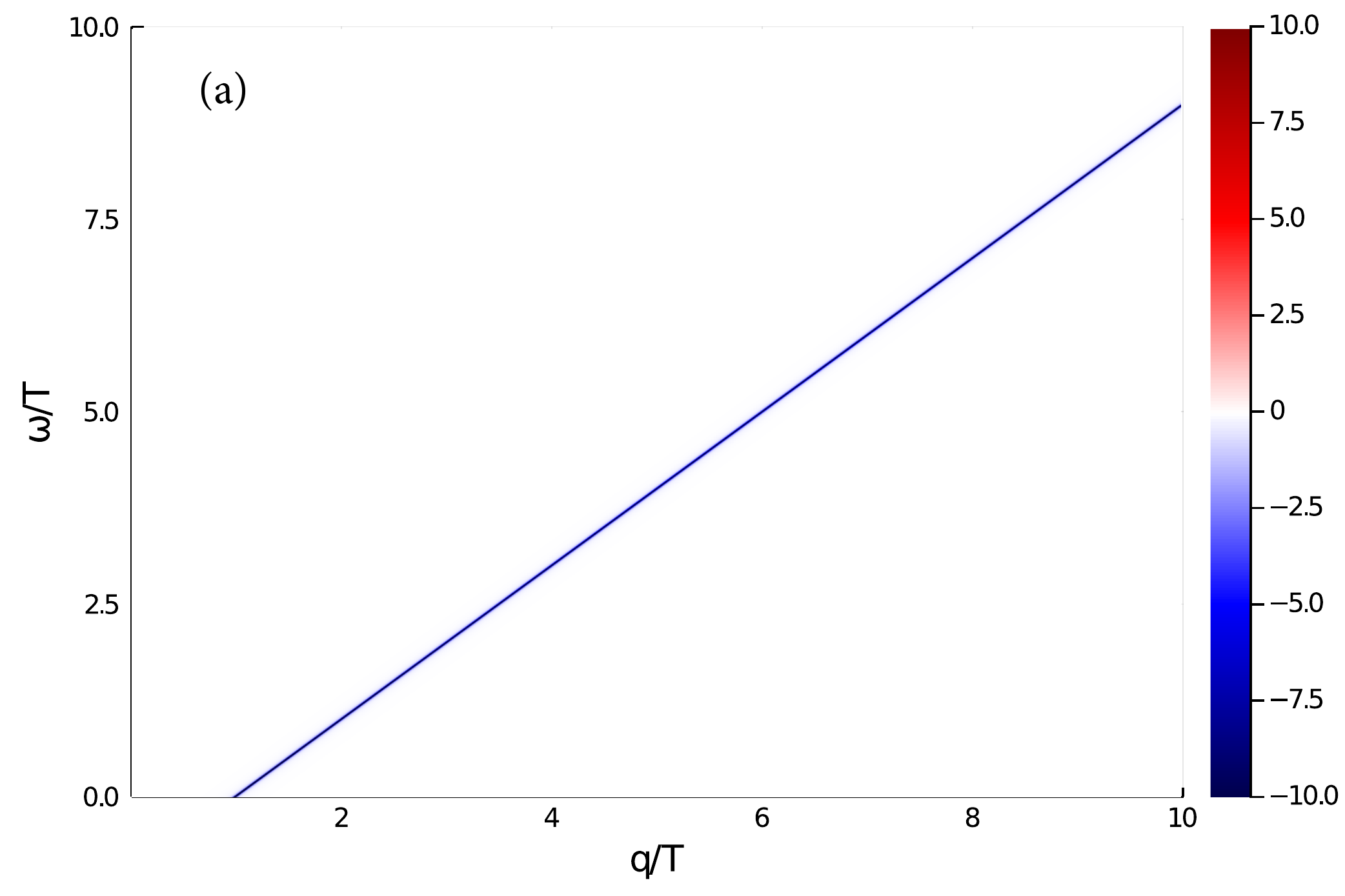}
	\includegraphics[width=0.5\textwidth]{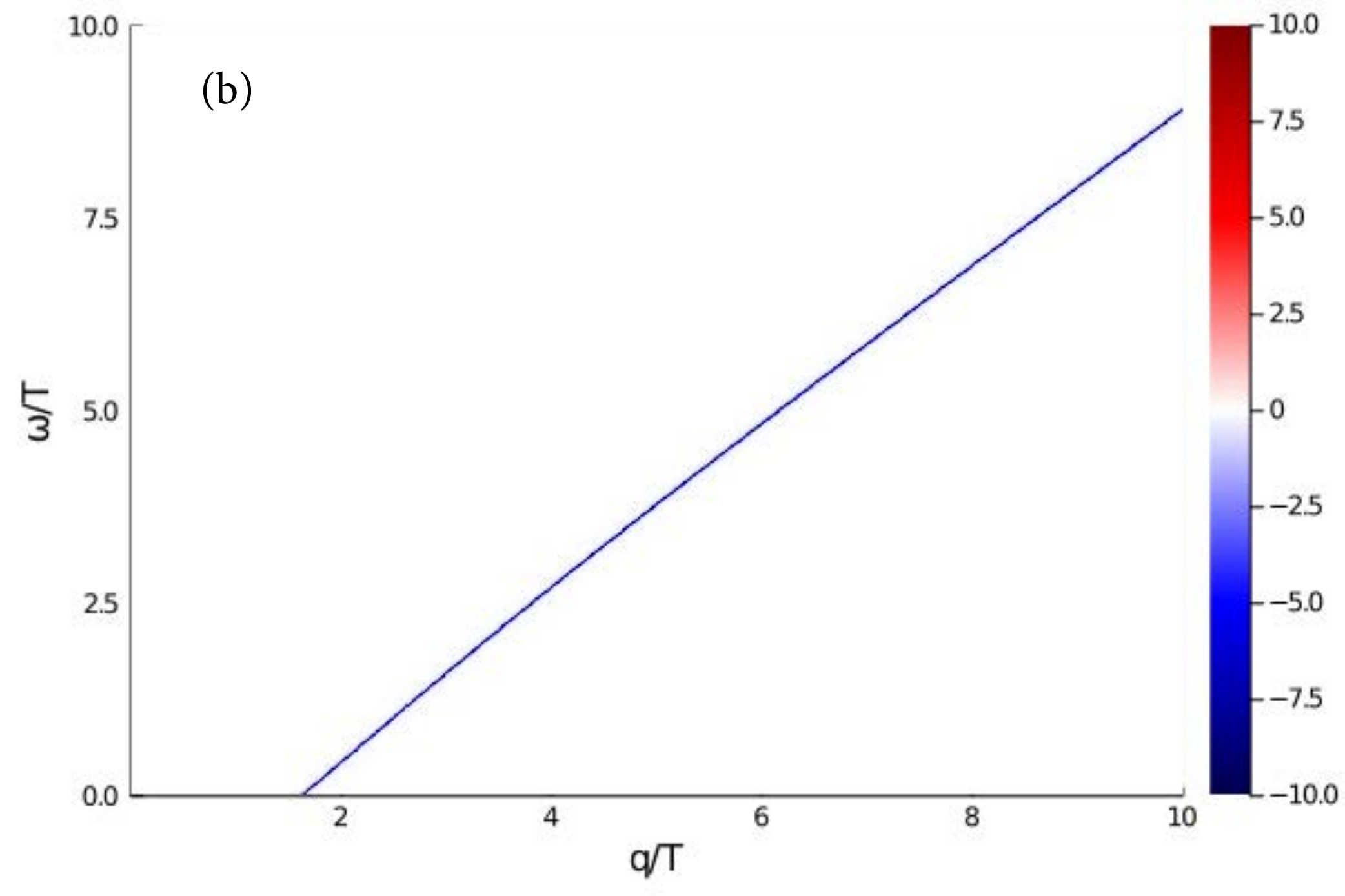}
	\caption{Spectral function at $\mu/T =1$ and $\alpha=1$. (a) The spectral function of the non-interacting electrons follows $\Im g_0^R(\vec{q},\omega) = -2\pi i \delta(\omega-v_F q+\mu)$. The blue line in the middle of the plot manifests the delta peak at $\omega = v_Fq-\mu$ ($v_F$ is set to 1 in the plot). (b) The spectral function with the inclusion of the exchange conventional energy in Eq.~\eqref{explicitfockpotential2}. Since the exchange energy is a real-valued function, the spectral function still has the delta-peak feature as shown by the blue line in the middle of the plot.}
	\label{fig:spectralfunction}
\end{figure}

	Next, we are going to sketch the calculation of the Fock self-energy diagram depicted in Fig.~\ref{fig:Fockdiagram}. It reads
	\begin{eqnarray}	
		&&-i\Sigma^F_{ab;\lambda\lambda'}(\vec{x},\vec{x}';t,t') \nonumber\\ &&\hspace{1.5cm}=\gamma^{\alpha}_{ac} D_{0,\alpha\beta}(\vec{x},\vec{x}';t,t')G_{0,cd;\lambda\lambda'}(\vec{x},\vec{x}';t,t')\gamma^{\beta}_{db}.\nonumber\\
	\end{eqnarray}
	After a Wigner transformation followed by a few steps of algebraic manipulations, we find that
	\begin{eqnarray}
		&&-i\Sigma^F_{ab;\lambda\lambda'}(\vec{x},\vec{p};t,\nu) 
		\nonumber\\&&= \delta_{ab}\int\frac{d\vec{p}_1}{(2\pi)^2}\frac{V(\vec{p}-\vec{p}_1)}{2}  \int\frac{d\nu_1}{2\pi}G^K_{0,\lambda\lambda'}(\vec{x},\vec{p}_1;t,\nu_1).\nonumber\\
	\end{eqnarray}
Subsequently, we transform it into the quasi-particle basis according to
	 $-i\sigma^F_{ab;\lambda\lambda'}(\vec{x},\vec{p};t,\nu)=-i\left(\mathcal{U}^\dagger_{\vec{p}}\Sigma^F_{ab}(\vec{x},\vec{p};t,\nu)\mathcal{U}_{\vec{p}}\right)_{\lambda\lambda'}$. 
We are interested in elements on the main diagonal of the spinor space. They are given by

\begin{equation}
		\sigma^F_{ab;\lambda\lambda}(\vec{x},\vec{p};t,\nu)= \sigma^{F,1}_{ab;\lambda\lambda}(\vec{x},\vec{p};t,\nu)+\sigma^{F,2}_{ab;\lambda\lambda}(\vec{x},\vec{p};t,\nu),
		\label{eq:Fockpotential}
	\end{equation}
	where
	\begin{equation}	
		\sigma^{F,1}_{ab;\lambda\lambda}(\vec{x},\vec{p};t,\nu)=\delta_{ab}\lambda\int \frac{d\vec{p}_1}{(2\pi)^2} \cos(\theta_{\vec{p}_1}-\theta_{\vec{p}}) \frac{V(\vec{p}-\vec{p}_1)}{2},
	\end{equation}
	and
	\begin{eqnarray}
		&&\sigma^{F,2}_{ab;\lambda\lambda}(\vec{x},\vec{p};t,\nu)\nonumber\\&&\hspace{1.5cm}=\delta_{ab} \int\frac{d\vec{p}_1}{(2\pi)^2}\frac{1+\cos(\theta_{\vec{p}_1}-\theta_{\vec{p}})}{2}\frac{V(\vec{p}-\vec{p}_1)}{2}\nonumber\\&&\hspace{4.5cm} \left((1-\lambda)-2f_{\lambda}(\vec{x},t,\vec{p}_1)\right), \nonumber\\
		\label{eq:Fockpotential2}
	\end{eqnarray}
	where $\theta_{\vec{p}}$ was introduced in Eq.~\eqref{eq:tranformationmatrix}.
The first term diverges logarithmically and is responsible for the renormalization of the Fermi velocity. To see this, we explicitly substitute the Coulomb potential in Eq.~\eqref{coulombpotential} followed by a transformation of the momentum variable from Cartesian to polar coordinates. This gives
\begin{eqnarray}
	\sigma^{F,1}_{ab;\lambda\lambda}(\vec{x},\vec{p};t,\nu)&=&\lambda\frac{\alpha v_F}{4\pi} \int p_1dp_1d\theta \frac{\cos\theta}{\sqrt{p^2+p_1^2-2pp_1\cos\theta}}  \nonumber\\&\approx& \lambda	\frac{\alpha v_F p}{4} \int_{p}^{\mathcal{K}} dp_1  \frac{1}{p_1} = \frac{\alpha}{4} \ln(\mathcal{K}/p) \lambda v_F p.\nonumber\\
	\label{eq:velocityrenormalization}
\end{eqnarray}
$\theta = \theta_{\vec{p}_1}-\theta_{\vec{p}}$ is the angle of the momentum $\vec{p}_1$ measured with respect to $\vec{p}$. In order to extract the logarithmic divergence of this integral, we expanded the square root to the first power in $p/p_1$, {\it i.e.}, $1/\sqrt{p^2+p_1^2-2pp_1\cos\theta} \approx 1/p_1 \left(1+p\cos\theta/p_1\right)$. After the angular integral, we find the result in the second line of Eq.~\eqref{eq:velocityrenormalization}.
The divergence of the integral is cut off at the inverse lattice spacing $\mathcal{K}$. The lower boundary of integration is
consistently set to $p$ (we require $p/p_1 \ll 1$). When substituting Eq.~\eqref{eq:velocityrenormalization} into the Dyson equation, we find 
\begin{equation}
	g^{-1}_0-\sigma^F = \omega - \lambda \left(1+\alpha\ln(\mathcal{K}/p)/4\right)v_F p.
\end{equation}
the Fermi velocity is  renormalized accordingly to $v^R_F = \left(1+\alpha\ln(\mathcal{K}/p)/4\right)v_F$.
Such logarithmic renormalization was first discussed within the renormalization group approach in Ref.~\cite{vozmediano} and measured in graphene in Ref.~\cite{Elias2011}.
The second contribution, Eq.~\eqref{eq:Fockpotential2}, describes the conventional exchange energy coming from both electrons and holes. It varies with the doping $\mu$ and the temperature $T$ of the system. Moreover, different from the exchange energy of a conventional two-dimensional electron gas, there is a factor $(1\pm \cos(\theta_{\vec{p}_1}-\theta_{\vec{p}}))/2$ arising from the wavefunction overlap, where $\theta_{\vec{p}_1}-\theta_{\vec{p}}$ is the angle between
$\vec{p}$ and $\vec{p}_1$~\cite{DasSarma2007}.
In total, it reads
\begin{eqnarray}
	&&(\sigma^{F,2})^R_{\pm\pm}(\vec{p})\nonumber\\&&\hspace{0.5cm}= -\int\frac{d\vec{p}_1}{(2\pi)^2}\frac{1\pm\cos(\theta_{\vec{p}_1}-\theta_{\vec{p}})}{2}V(\vec{p}-\vec{p}_1)f^0_+(\vec{p}_1)\nonumber\\&&\hspace{0.8cm}+\int\frac{d\vec{p}_1}{(2\pi)^2}\frac{1\mp\cos(\theta_{\vec{p}_1}-\theta_{\vec{p}})}{2}V(\vec{p}-\vec{p}_1)(1-f^0_-(\vec{p}_1)). \nonumber\\
	\label{explicitfockpotential2}
\end{eqnarray}
After a series of manipulations, we find
\begin{eqnarray}
	&&(\sigma^{F,2})^R_{\pm\pm}(\vec{p})\nonumber\\&&= -\frac{\alpha v_F}{4\pi}\int dp_1 \sqrt{\frac{p_1}{2p}} \left(g\left(\frac{p^2+p_1^2}{2pp_1}\right) \pm h\left(\frac{p^2+p_1^2}{2pp_1}\right)\right)\nonumber\\&& \hspace{6cm}  f^0_+(\vec{p}_1)\nonumber\\&&+\frac{\alpha v_F}{4\pi}\int dp_1 \sqrt{\frac{p_1}{2p}} \left(g\left(\frac{p^2+p_1^2}{2pp_1}\right) \mp h\left(\frac{p^2+p_1^2}{2pp_1}\right)\right)\nonumber\\&& \hspace{6cm} (1-f^0_-(\vec{p}_1)),\nonumber\\
	\label{eq:exchangeenergy}
\end{eqnarray}
with 
\begin{eqnarray}
g(x) &=& \int_0^{2\pi} \frac{d\theta}{\sqrt{x-\cos\theta}}\;, \nonumber \\ h(x) &=& \int_0^{2\pi} \frac{d\theta\; \cos\theta}{\sqrt{x-\cos\theta}}\;.
\end{eqnarray}
We proceed by evaluating the $p_1$-integral numerically. We show in Fig.~\ref{fig:exchangeenergy} the exchange energies for electrons,  $(\sigma^{F,2})^{R}_{++}$, and holes, $(\sigma^{F,2})^{R}_{--}$, at the chemical potential $\mu/T = 1$ for various values of momenta. We see that the self-energy becomes less important at high momenta. In Fig.~\ref{fig:spectralfunction} we show the spectral function of the electrons with and without the Fock correction. Since the exchange energy is a real-valued function, there remains the delta-peak feature manifested as the blue line in the middle of the plot. Compare to the spectral function of the non-interacting theory in Fig.~\ref{fig:spectralfunction} (a), the exchange energy plays two roles: (i) it shifts the chemical potential and (ii) it increases the Fermi velocity~\cite{DasSarma2007}. However, the effect of $(\sigma^{F,2})^R_{\pm\pm}(\vec{p})$ is relatively small compared to the Fermi velocity renormalization coming from the logarithmic divergence. Therefore, we keep only the latter effect in the following. In practice, this leads to replacing $v_F$ with $v_F^R$, the renormalized Fermi velocity, in all expressions. 

\subsubsection{Energy spectrum}
We are now ready to evaluate the energy spectrum of quasi-electrons and quasi-holes in equilibrium within the Hartree-Fock approximation. The dispersion relation of the excitations can be found from the roots of the inverse Green function. In the presence of the interaction, the inverse Green function is the solution of the Dyson equation in Eq.~\eqref{eq:Dysonequationfor retardedandadvancedcomponents}. We first transform the Dyson equation into the quasi-particle basis, which leads to
\begin{equation}
	(g^{-1})^R_{\lambda\lambda'} = (g^{-1}_0)^R_{\lambda\lambda'} - (\sigma^H)^R_{\lambda\lambda'}.
\end{equation} 
where the retarded component of the Hartree self-energy is given by the first row and first column element of Eq.~\eqref{eq:Hartreeselfenergyinthequasiparticlebasis} that is $(\sigma^H)_{\lambda\lambda'}^R = \sigma^H_{11,\lambda\lambda'}$. This gives 
\begin{eqnarray}
		(g^{-1})^R_{\lambda\lambda'}(\vec{x},t,\vec{p},\omega) &=& \Big(\omega+i\delta \mp  \lambda v_F^R p +\mu- V_{\rm{ex}}(\vec{x}) \nonumber\\&-&   \int d\vec{x}' V(\vec{x}-\vec{x}') n(\vec{x}',t)\Big)\delta_{\lambda\lambda'}\;,\nonumber\\
\end{eqnarray}
where $n(\vec{x},t)$ is the total electron density.
By inserting $V_{\rm{ex}}(\vec{x})$ from Eq.~\eqref{eq:jelliumpotential}, we find that 
\begin{equation}
	(g^{-1})^{R/A}_{\lambda\lambda'}(\vec{x},t,\vec{p},\omega) = \omega\pm i\delta -\lambda v_F^R p  + \mu - V^H(\vec{x},t),
	\label{eq:retardadvanceGreenfunction}
\end{equation}
where the Hartree potential $V^H(\vec{x},t)$ is defined as
\begin{eqnarray}\label{eq:Hartreepotential}
	V^H(\vec{x},t)=\int d\vec{x}' V(\vec{x}-\vec{x}')\left(n(\vec{x}',t)-n_0\right).
\end{eqnarray}
The Hartree term represents the potential energy of an electron at a position $\vec{x}$. This potential is produced by all  other electrons with the density $n(\vec{x}',t)$ at the other positions $\vec{x}'$ through the Coulomb interaction. This contribution is partially canceled by the potential $V_{\rm{ex}}(\vec{x})$ arising from the interaction between electrons and the underlying jellium background. This
fixed uniformly distributed positively charged background guarantees the over-all electrical neutrality of the system. Here $n_0$ is the ion charge density which is identical to the electron charge density in thermal equilibrium. Hence, in global thermal equilibrium, the Hartree potential vanishes and thus the energy spectrum of electrons is given by
\begin{equation}
	\epsilon_\lambda(\vec{p}) = \lambda v_F^R p \;,
	\label{eq:electronenergy}
\end{equation}
as expected.

\subsubsection{The kinetic equation for Dirac fermions}

Next, we derive the Bolzmann equation within the Hartree-Fock approximation. 
To this end, we consider Eq.~\eqref{eq:Keldyshfermion}. 
 Using the Hartree-Fock approximation in Eq.~\eqref{eq:retardadvanceGreenfunction}, we find
\begin{eqnarray} 
	\partial_\epsilon \left(g^{-1}_0 - (\Re\sigma^H)^R\right)&=&1,\nonumber\\
	\partial_{\vec{p}} \left(g^{-1}_0 - (\Re\sigma^H)^R\right)&=&-\lambda v_F^R \hat{p},\nonumber\\
	\partial_{\vec{x}} \left(g^{-1}_0 - (\Re\sigma^H)^R\right)&=&-\partial_{\vec{x}}V^H(\vec{x},t).
\end{eqnarray}
We first substitute these derivatives into Eq.~\eqref{eq:Keldyshfermion} followed by a multiplication of the resulting equation with the spectral function $\Im g^{-1}_{\lambda\lambda}(\vec{p},\epsilon) = -2\pi i \delta(\epsilon - \lambda v_F^R p)$. Subsequently, we integrate it over the frequency which amounts to the quasi-particle approximation. In the end, we obtain the mean-field collisionless Boltzmann equation, also known as the Vlasov equation~\cite{Vlasov1968},  for electrons ($\lambda=+$) and holes ($\lambda=-$) 
\begin{eqnarray}
		&&\partial_t f_\lambda(\vec{x},t,\vec{p}) +\lambda v_F^R \hat{p}\cdot \partial_{\vec{x}}f_\lambda(\vec{x},t,\vec{p}) \nonumber\\&&\hspace{2.5cm}-\;\partial_{\vec{x}}V^H(\vec{x},t)\cdot \partial_{\vec{p}} f_\lambda(\vec{x},t,\vec{p})
		= 0,
		\label{eq:Vlasovequation}
\end{eqnarray}
where $\hat{p}$ denotes the unit vector in the direction of the momentum $\vec{p}$. In the above equation, the equilibrium value of the distribution functions is given by $f_\lambda^0(\vec{p})=(1+\exp(\frac{\lambda v_F^R p - \mu}{T}))^{-1}$. 
The potential $V^H(\vec{x},t)=\int d\vec{x}'V(\vec{x}-\vec{x}') \delta n(\vec{x}',t)$ results from the Hartree self-energy, where $\delta n = n - n_0$ is the density fluctuation. This potential is also the solution of the classical Poisson equation for the internal electric field.

To summarize, we found that the Hartree-Fock diagrams lead to the Vlasov equation.

\subsection{Second-order perturbation theory: Born approximation}\label{subsec:born}

\begin{figure}[h]
	\centering
	\begin{subfigure}[]{0.1\textwidth}
		\centering
		\includegraphics[width=\textwidth]{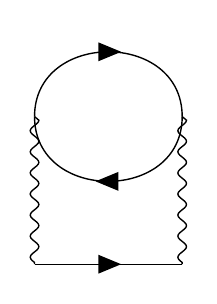}
		\subcaption{}
		\label{fig:sigma2a}
	\end{subfigure}
	\begin{subfigure}[]{0.2\textwidth}
		\centering
		\includegraphics[width=\textwidth]{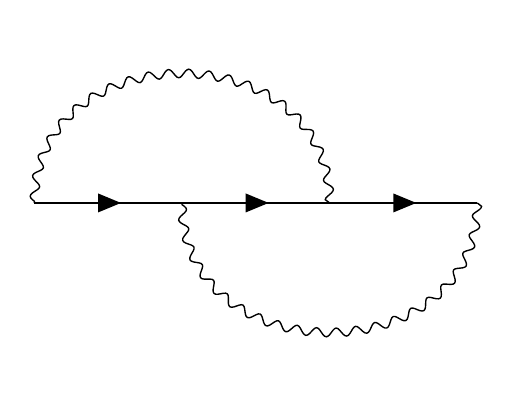}
		\subcaption{}
		\label{fig:sigma2b}
	\end{subfigure}
	\begin{subfigure}[]{0.2\textwidth}
		\centering
		\includegraphics[width=\textwidth]{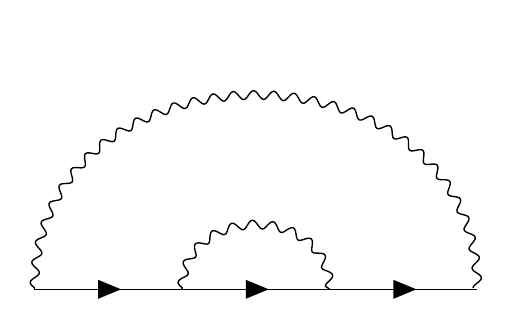}
		\subcaption{}
		\label{fig:sigma2c}
	\end{subfigure}
	\caption{Self-energy to second order in the interaction.}
	\label{fig:bornapproximation}
\end{figure}

An important role in hydrodynamic systems is played by the relaxation processes towards local equilibrium that conserve particle number, momentum, and energy. These collisions occur beyond first order in the interaction, Eq.~\eqref{eq:interaction}. The lowest non-vanishing order is second order and the contributions are pictorially shown in Fig.~\ref{fig:bornapproximation}. This is called the Born approximation for the cross section~\cite{Fritz2008a,Kashuba2008,Landau1965,KadanoffBaym1962}. In principle, these diagrams play two roles: (i) they describe the aforementioned relaxations due to collisions and (ii) they renormalize the quasi-particle properties~\cite{Kotov2012,Mishchenko}. 

The calculation of these diagrams is tedious but straightforward. Here we summarize our final results and present the full derivation in Appendix \ref{appendix:A}. The Boltzmann equation reads
\begin{widetext}
\begin{eqnarray}
	&&\partial_t f_\lambda(\vec{x},t,\vec{p}) +\lambda v_F^R \hat{p}\cdot \partial_{\vec{x}}f_\lambda(\vec{x},t,\vec{p})-\;\partial_{\vec{x}}V^H(\vec{x},t)\cdot \partial_{\vec{p}} f_\lambda(\vec{x},t,\vec{p})
	\nonumber\\&&	\hspace{3cm}= -\int \frac{d\vec{k}_1}{(2\pi)^2}\frac{d\vec{q}}{(2\pi)^2} 2\pi\delta(\lambda\epsilon_{\vec{k}}-\lambda_1\epsilon_{\vec{k}-\vec{q}}-\lambda_2\epsilon_{\vec{k}_1+\vec{q}}+\lambda_3\epsilon_{\vec{k}_1})R_{\lambda\lambda_1\lambda_3\lambda_2}(\vec{k},\vec{k}_1,\vec{q})\nonumber\\&&\hspace{3cm}\Big[f_{\lambda}(\vec{k})f_{\lambda_3}(\vec{k}_1)(1-f_{\lambda_1}(\vec{k}-\vec{q}))(1-f_{\lambda_2}(\vec{k}_1+\vec{q}))-(1-f_{\lambda}(\vec{k}))(1-f_{\lambda_3}(\vec{k}_1))f_{\lambda_1}(\vec{k}-\vec{q})f_{\lambda_2}(\vec{k}_1+\vec{q})\Big],\nonumber\\
	\label{eq:fermionboltzmannequation}
\end{eqnarray}
where we introduced the shorthand notation
\begin{eqnarray}
	R_{\lambda\lambda_1\lambda_3\lambda_2}(\vec{k},\vec{k}_1,\vec{q})=2\Big[|T_{\lambda\lambda_1\lambda_3\lambda_2}-T_{\lambda\lambda_2\lambda_1\lambda_3}|^2+(N-1)\left(|T_{\lambda\lambda_1\lambda_3\lambda_2}|^2+|T_{\lambda\lambda_2\lambda_1\lambda_3}|^2\right) \Big].
\end{eqnarray}
\end{widetext}
In this expression, we have used
\begin{eqnarray}
	T_{\lambda \lambda_1\lambda_2\lambda_3}(\vec{k},\vec{k}_1,\vec{q}) = \frac{V(\vec{q})}{2} M^{\lambda\lambda_1}_{\vec{k},\vec{k}-\vec{q}}M^{\lambda_2\lambda_3}_{\vec{k}_1,\vec{k}_1+\vec{q}}\;,
\end{eqnarray}
where the coherence factor $M$ comes from the overlap of the single-particle wavefunctions. It is defined according to
\begin{eqnarray}
	M^{\lambda\lambda_1}_{\vec{k},\vec{k}_1}=\left(\mathcal{U}^\dagger_{\vec{k}}\mathcal{U}^{\phantom{\dagger}}_{\vec{k}_1}\right)_{\lambda\lambda_1}\;.
\end{eqnarray}
For brevity, we suppress the space and time variables of the distribution functions in the collision terms and have in mind that they all depend on the same set of variables that is $(\vec{x},t)$. The collision does not shift the center-of-mass and time coordinates. This effect indeed exists, but it will show up at higher order in the gradient expansion~\cite{pesin}. We can understand this collision integral in the following way: an electron in band $\lambda$ with momentum $\vec{k}$  is scattered into band $\lambda_1$ and momentum $\vec{k}+\vec{q}$ by a collision with another electron in band $\lambda_3$ and state $\vec{k}_1$ which is itself scattered into the energy band $\lambda_2$ and state $\vec{k}_1-\vec{q}$. For this event to take place, the initial states $\vec{k}$ and $\vec{k}_1$ have to be filled and the final states $\vec{k}-\vec{q}$ and $\vec{k}_1+\vec{q}$ must be empty. Thus, the factors $f_{\lambda}(\vec{k})$ and $f_{\lambda_3}(\vec{k}_1)$ are the occupation numbers of these state and $1-f_{\lambda_1}(\vec{k}-\vec{q})$ and $1-f_{\lambda_2}(\vec{k}_1+\vec{q})$ are the probabilities for the final states to be unoccupied. The conservation of energy is taken into account by the delta function. The transition probability of this event is
$R_{\lambda\lambda_1\lambda_3\lambda_2}$.
In total, we find that Coulomb interaction enters the Boltzmann transport equation in two ways: (i) as the Hartree potential produced by all the other particles in the system and (ii) inelastic and momentum-conserving electron-electron scatterings leading to local equilibration. Let us note that within this approximation, the contribution from the real part of the second-order diagrams to the left-hand side of Eq.~\eqref{eq:fermionboltzmannequation} is neglected.

Finally, let us note that the Born approximation is valid only when the kinetic energy of the electrons is large compared to the Coulomb interaction potential~\cite{Ziman2000}.  For the Dirac system, the ratio of the potential energy to the kinetic energy is characterized by the fine structure constant $\alpha=e^2/4\pi\epsilon v_F$. In condensed-matter systems, this constant is not necessarily small (for graphene $\alpha \approx 0.3-2.2$). In such a strong interaction limit, a perturbative series expansion in $\alpha$ may break down. Instead one can employ an alternative perturbative expansion in the other parameters. In the subsequent section, we will employ the random-phase approximation (RPA) and show that it gives a different, more complicated picture than the Hartree-Fock-Born result presented in this section.
\subsection{Conservation laws}\label{subsec:conservationlawsborn}
The collision integral on the right-hand side of the Boltzmann equation in Eq.~\eqref{eq:fermionboltzmannequation}, henceforth denoted $C_{\lambda}[f](\vec{k})$, has conservation laws encoded in it. These
conservation laws are important for two reasons: (i) They allow for an identification or derivation of physical quantities such as charge and current densities; (ii) it provides the basis of the derivation of conservation laws and even transport phenomena. 
When a system is driven away from equilibrium, the first thing that
happens is that collisions drive the system to local equilibrium. Afterwards, there is a much slower return to global equilibrium. The latter describes transport processes and is governed by the conservation laws. The conserved quantities in the system under consideration are particle number, momentum, and energy. The conservation laws of electric charge, momentum, and energy are obtained by multiplying the Boltzmann equation in Eq.~\eqref{eq:fermionboltzmannequation}, by $1$, $\vec{p}$ and $\epsilon_\lambda(\vec{x},\vec{p}) = \lambda v_F^R p +V^H(\vec{x})$ and then integrating the resulting equations over all momentum $\vec{p}$ as well as summing over energy bands $\pm$ and flavors. This leads to the following collisional invariants 
\begin{equation}
	N\sum_{\lambda=\pm}\int\frac{d\vec{p}}{(2\pi)^2} \; C_\lambda[f](\vec{p}) = 0,
	\label{chargeconservationofcollisionintegral}
\end{equation}
\begin{equation}
	N\sum_{\lambda=\pm}\int\frac{d\vec{p}}{(2\pi)^2} \; \vec{p} C_\lambda[f](\vec{p}) = 0,\label{momentumconservationofcollisionintegral}
\end{equation}
and 
\begin{equation}
	N\sum_{\lambda=\pm}\int\frac{d\vec{p}}{(2\pi)^2} \; \epsilon_\lambda(\vec{x},\vec{p}) C_\lambda[f](\vec{p}) = 0.\label{energyconservationofcollisionintegral}
\end{equation}
(i) The continuity equation of charge can be obtained by integrating the Boltzmann equation in Eq.~\eqref{eq:fermionboltzmannequation}, over momentum $\vec{p}$ followed by a summation over the band index and flavor index, see Eq.~\eqref{chargeconservationofcollisionintegral}. In contrast to the case of a one band system, there is a subtle point here relating to the infinite number of particles in the filled band which is unbounded from below. This infinite constant vanishes upon differentiation and does not contribute to the continuity equation. Therefore, we can subtract the infinite contribution coming from the Dirac sea and instead consider the population of holes defined as $f(\vec{x},t,\vec{k})-1$. First, we consider the time-derivative term. Integrating this term over all states gives
\begin{eqnarray}
	&&N\sum_{\lambda=\pm} \int \frac{d\vec{p}}{(2\pi)^2} \partial_t f_\lambda(\vec{x},t,\vec{p})\nonumber\\&& \hspace{0.2cm}= \partial_t \; N\int\frac{d\vec{p}}{(2\pi)^2} \Big[ f_+(\vec{x},t,\vec{p}) + (f_-(\vec{x},t,\vec{p})-1)\Big]. 
\end{eqnarray}
We denote the charge density by
\begin{equation}
	n(\vec{x},t) = N\int \frac{d\vec{p}}{(2\pi)^2} \; \left[ f_+(\vec{x},t,\vec{p}) + (f_-(\vec{x},t,\vec{p})-1)\right].
	\label{eq:electrondensity}
\end{equation}
The second term can be integrated in a similar fashion and gives
\begin{eqnarray}
	&&N\sum_{\lambda=\pm} \int \frac{d\vec{p}}{(2\pi)^2} \lambda v_F^R\hat{p}\cdot\partial_{\vec{x}} f_\lambda(\vec{x},t,\vec{p})\nonumber\\&& = \partial_{\vec{x}} \cdot \; N\int\frac{d\vec{p}}{(2\pi)^2} v_F^R\hat{p}\Big[ f_+(\vec{x},t,\vec{p}) +  (f_-(\vec{x},t,\vec{p})-1)\Big].\nonumber\\ 
\end{eqnarray}
This term can be identified with the charge current density
\begin{eqnarray}
	\vec{j}(\vec{x},t) &=& N\int \frac{d\vec{p}}{(2\pi)^2} \; v_F^R \hat{p}\Big[ f_+(\vec{x},t,\vec{p})- (f_-(\vec{x},t,\vec{p})-1)\Big].\nonumber\\,
	\label{eq:chargecurrent}
\end{eqnarray}
The momentum-derivative term vanishes upon integration
\begin{eqnarray}
	&&-N\sum_{\lambda=\pm} \int \frac{d\vec{p}}{(2\pi)^2} \partial_{\vec{x}}V^H(\vec{x},t)\cdot\partial_{\vec{p}} f_\lambda(\vec{x},t,\vec{p})\nonumber\\&& \hspace{0.2cm}=- N\sum_{\lambda=\pm} \int \frac{d\vec{p}}{(2\pi)^2} \partial_{\vec{p}} \cdot \Big[ \partial_{\vec{x}}V^H(\vec{x},t) f_\lambda(\vec{x},t,\vec{p})\Big]=0\;,\nonumber\\ 
\end{eqnarray}
since it is a total derivative.
Combining the above equations together with Eq.~\eqref{chargeconservationofcollisionintegral}, we find the continuity equation for the electric charge according to
\begin{equation}
	\partial_t n(\vec{x},t) + \partial_{\vec{x}} \cdot \vec{j}(\vec{x},t) = 0.
	\label{eq:continuityequation}
\end{equation}
It is worthwhile noting here that the particle density is conserved locally.
\par

(ii) Momentum conservation can be obtained by multiplying the Boltzmann equation in Eq.~\eqref{eq:fermionboltzmannequation} by momentum before integrating the resulting equation over all states, see Eq.~\eqref{momentumconservationofcollisionintegral}. The time-derivative term yields 
\begin{eqnarray}
	&&N\sum_{\lambda=\pm} \int \frac{d\vec{p}}{(2\pi)^2} \vec{p} \partial_t f_\lambda(\vec{x},t,\vec{p})\nonumber\\&& \hspace{0.1cm}= \partial_t \; N\int\frac{d\vec{p}}{(2\pi)^2} \vec{p}\Big[ f_+(\vec{x},t,\vec{p}) +(f_-(\vec{x},t,\vec{p})-1)\Big].\nonumber \\ 
\end{eqnarray}
 This allows to define the momentum density,
 \begin{equation}
 	\vec{n}^{\vec{p}}(\vec{x},t) = N\int \frac{d\vec{p}}{(2\pi)^2} \vec{p}\Big[f_+(\vec{x},t,\vec{p}) + (f_-(\vec{x},t,\vec{p})-1)\Big].
 	\label{eq:momentumdensity}
 \end{equation}
The space-derivative term can be similarly integrated and gives
\begin{eqnarray}
	&&N\sum_{\lambda=\pm} \int \frac{d\vec{p}}{(2\pi)^2} \vec{p} \; \lambda v_F^R\hat{p}\cdot\partial_{\vec{x}} f_\lambda(\vec{x},t,\vec{p})\nonumber\\&& = \partial_{\vec{x}} \cdot \; N\int\frac{d\vec{p}}{(2\pi)^2} v_F^R\vec{p}\hat{p}\Big[ f_+(\vec{x},t,\vec{p}) -  (f_-(\vec{x},t,\vec{p})-1)\Big].\nonumber\\ 
\end{eqnarray}
From this result, we define the momentum-flux tensor, 
\begin{equation}
	\vec{\vec{\Pi}}(\vec{x},t)=N\int \frac{d\vec{p}}{(2\pi)^2} \;v_F^R\vec{p} \hat{p} \Big[f_+(\vec{x},t,\vec{p}) - (f_-(\vec{x},t,\vec{p})-1)\Big].
\end{equation}
The momentum-derivative term yields
\begin{eqnarray}
	&&-N\sum_{\lambda=\pm} \int \frac{d\vec{p}}{(2\pi)^2} \vec{p} \partial_{\vec{x}}V^H(\vec{x},t)\cdot\partial_{\vec{p}} f_\lambda(\vec{x},t,\vec{p})\nonumber\\&& \hspace{0.2cm}= \partial_{\vec{x}}V^H(\vec{x},t)  N\int \frac{d\vec{p}}{(2\pi)^2} \; \left[ f_+(\vec{x},t,\vec{p}) +(f_-(\vec{x},t,\vec{p})-1)\right]\nonumber\\&&\hspace{0.2cm}=\partial_{\vec{x}}V^H(\vec{x},t)n(\vec{x},t).\nonumber\\ 
\end{eqnarray}
which defines a force term.
Finally, combining the above equations, we find the momentum equation according to
\begin{equation}
	\partial_{t} \vec{n}^{\vec{p}}(\vec{x},t) + \partial_{\vec{x}}\cdot\vec{\vec{\Pi}}(\vec{x},t) = - \partial_{\vec{x}}V^H(\vec{x},t) n(\vec{x},t).
	\label{eq:momentumequation}
\end{equation}
This has a straightforward interpretation. The momentum of the electron fluid in any given volume element can be changed in two ways: (i) by means of momentum flow through the volume boundary accounting for the space-gradient term and (ii) by the internal electric field. Locally, the momentum density is not conserved and changed by the internal electric force $-\partial_{\vec{x}}V^H(\vec{x},t)$. This internal force, however, does not affect the total momentum of the entire system. We still expect that the total momentum of the system is conserved. This can be shown explicitly by integrating the momentum equation over all space. The integration of the momentum-flux gradient results in a surface term which vanishes. The Hartree force term also vanishes. To see this, we consider the property of the Coulomb potential
\begin{equation}
	\partial_{\vec{x}} \frac{1}{|\vec{x}-\vec{x}'|} = -\frac{\vec{x}-\vec{x}'}{|\vec{x}-\vec{x}'|^3} = -\partial_{\vec{x}'}\frac{1}{|\vec{x}-\vec{x}'|}.
\end{equation}
Based on this, one can show explicitly that
\begin{eqnarray}
	&&\int d\vec{x}\partial_{\vec{x}}V^H(\vec{x},t) n(\vec{x},t) \nonumber\\&&=	\int d\vec{x}d\vec{x}'\partial_{\vec{x}}V(\vec{x}-\vec{x}')(n(\vec{x}',t)-n_0) n(\vec{x},t) =0.\nonumber \\
\end{eqnarray}
Combining all the above equations, we obtain 
\begin{equation}
	\partial_t \vec{P} = 0,
\end{equation}
where the total momemtum of the entire system is given by
\begin{equation}
	\vec{P}(t) = \int d\vec{x}\; \vec{n}^{\vec{p}}(\vec{x},t).
\end{equation}
\par
(iii) Similarly, energy conservation is obtained by multiplying the Boltzmann equation, Eq.~\eqref{eq:fermionboltzmannequation} by energy $\epsilon(\vec{x},\vec{p})$ followed by integrating and summing the equation over all states, see Eq.~\eqref{energyconservationofcollisionintegral}. The time-derivative term yields 
\begin{eqnarray}
	&&N\sum_{\lambda=\pm} \int \frac{d\vec{p}}{(2\pi)^2} \epsilon_{\lambda}(\vec{x},\vec{p}) \partial_t f_\lambda(\vec{x},t,\vec{p})\nonumber\\&& \hspace{1cm} = \partial_t \; N\int\frac{d\vec{p}}{(2\pi)^2} \Big[\epsilon_{+}(\vec{x},\vec{p}) f_+(\vec{x},t,\vec{p})\nonumber\\&&\hspace{1cm} + \epsilon_{-}(\vec{x},\vec{p}) (f_-(\vec{x},t,\vec{p})-1)\Big]. 
\end{eqnarray}
This defines the energy density according to
\begin{eqnarray}
	n^\epsilon(\vec{x},t)&=&N\int\frac{d\vec{p}}{(2\pi)^2}\;  \Big[\epsilon_+(\vec{x},\vec{p})f_+(\vec{x},t,\vec{p})\nonumber\\&&\hspace{1cm}+\;\epsilon_-(\vec{x},\vec{p})(f_-(\vec{x},t,\vec{p})-1)\Big].
\end{eqnarray}
The spatial derivative leads to 
\begin{eqnarray}
	&&N\sum_{\lambda=\pm} \int \frac{d\vec{p}}{(2\pi)^2} \epsilon_\lambda(\vec{x},\vec{p}) \; \lambda v_F^R\hat{p}\cdot\partial_{\vec{x}} f_\lambda(\vec{x},t,\vec{p})\nonumber\\&& = \partial_{\vec{x}} \cdot \; N\int\frac{d\vec{p}}{(2\pi)^2} v_F^R\hat{p}\Big[\epsilon_+(\vec{x},\vec{p}) f_+(\vec{x},t,\vec{p}) \nonumber\\&&\hspace{0.5cm}-\;  \epsilon_-(\vec{x},\vec{p}) (f_-(\vec{x},t,\vec{p})-1)\Big] -\partial_{\vec{x}}V^H(\vec{x},t) \cdot \vec{j}.
	\label{spacederivativetermforderivationofenergyconservation}
\end{eqnarray}
We find that it consists of two terms. The first term describes a divergence of an energy current density defined as 
\begin{eqnarray}
	\vec{j}^\epsilon(\vec{x},t)&=& N\int\frac{d\vec{p}}{(2\pi)^2}\; v_F^R\hat{p} \Big[\epsilon_+(\vec{x},\vec{p})f_+(\vec{x},t,\vec{p})\nonumber\\&&-\;\epsilon_-(\vec{x},\vec{p})(f_-(\vec{x},t,\vec{p})-1)\Big],
\end{eqnarray}
whereas the second term describes Joule heating due to the internal force. 
Next, we consider the momentum-derivative term and find that it can also be rewritten as Joule heating due to the internal Coulomb force.
\begin{eqnarray}
	&&-N\sum_{\lambda=\pm} \int \frac{d\vec{p}}{(2\pi)^2} \epsilon_\lambda(\vec{x},\vec{p}) \; \partial_{\vec{x}}V^H(\vec{x},t)\cdot\partial_{\vec{p}} f_\lambda(\vec{x},t,\vec{p})\nonumber\\&&\hspace{4cm}=\partial_{\vec{x}}V^H(\vec{x},t) \cdot \vec{j}\;.
\end{eqnarray}
This term will thus be canceled by the second term on the right-hand side of Eq.~\eqref{spacederivativetermforderivationofenergyconservation}. Combining these equations, we find the continuity equation of energy:
\begin{equation}
	\partial_t n^\epsilon(\vec{x},t)+\partial_{\vec{x}}\cdot\vec{j}^\epsilon(\vec{x},t) =   0.
	\label{eq:energyequation}
\end{equation}
Eqs.~\eqref{eq:continuityequation},~\eqref{eq:momentumequation}, and~\eqref{eq:energyequation} constitute the main equations of electron hydrodynamics and can be used to derive the Navier-Stokes equations. Let us emphasize once again that, in contrast to the hydrodynamic equations of usual fluids, electrons interact among themselves via long-range Coulomb interactions and this effect shows up on the right-side of the momentum equation in Eq.~\eqref{eq:momentumequation}.

\subsection{Collective modes}\label{subsec:collectivemodes}
Now that we have the fundamental equations of hydrodynamics of weakly interacting charged Dirac electrons, let us study some of its properties. As will be presented in Appendix \ref{Appendix:hydrovariables}, there are three independent hydrodynamic variables. Here we choose the set of independent variables consisting of the charge density ($n$), the energy density  ($n^\epsilon$) and the hydrodynamic velocity $\vec{u}$ in terms of  which the other quantities can be written. To linear order in $\vec{u}$, we find that the charge current density is given by $\vec{j} = n\vec{u}$. The momentum flux is associated with the pressure by means of $\Pi_{ij}=P\delta_{ij}$ where the pressure is in turn proportional to the energy density according to $P = n^\epsilon/2$. One of the consequences of the linear spectrum of Dirac electrons is that the momentum density is decoupled from the charge current. Instead, it is proportional to the energy currents according to $n^{\vec{p}}=\vec{j}^\epsilon/v_F^2$  where the energy current is given by $j^{\epsilon} = (P+n^\epsilon)\vec{u}$ . We now consider an electron fluid at rest with constant $n = n_0$, $n^\epsilon=n_0^\epsilon$, and $\vec{u} = 0$. We are interested in  small fluctuations around the constant and homogeneous solution, and put $n = n_0 + \delta n$ $n^\epsilon = n_0^\epsilon + \delta n^\epsilon$ and assume small $\vec{u}$. We insert this solution into the hydrodynamic equations in Eqs. (\ref{eq:continuityequation}, (\ref{eq:momentumequation}), and (\ref{eq:energyequation}). By keeping terms up to linear order in the fluctuations,  we obtain 
\begin{eqnarray}
	\partial_t \delta n(\vec{x},t) + n_0 \partial_{\vec{x}} \cdot \vec{u}(\vec{x},t) &=& 0,
	\nonumber\\	(P_0+n^\epsilon_0)\partial_{t} \vec{u}(\vec{x},t)/v_F^2 + \partial_{\vec{x}}\delta P(\vec{x},t) &=& - \partial_{\vec{x}}V^H(\vec{x},t) n_0,
	\nonumber\\	\partial_t \delta n^\epsilon(\vec{x},t)+(P_0+n_0^\epsilon)\partial_{\vec{x}}\cdot\vec{u}(\vec{x},t) &=&   0.
	\label{eq:linearizedequation}
\end{eqnarray}
The solutions to these equations are propagating waves. As an ansatz, we insert
\begin{eqnarray}
	\begin{pmatrix}
		\delta n (\vec{x},t)\\ \vec{u}(\vec{x},t)\\ \delta n^\epsilon(\vec{x},t)
	\end{pmatrix} = 	\begin{pmatrix}
	\delta n (\vec{p},\omega)\\ \vec{u}(\vec{p},\omega)\\ \delta n^\epsilon(\vec{p},\omega)
\end{pmatrix} e^{i\vec{p}\cdot\vec{x}-i\omega t}
\end{eqnarray}
into the linearized hydrodynamic equations in Eq. (\ref{eq:linearizedequation}). This leads to
\begin{eqnarray}
	-i\omega \delta n(\vec{p},\omega) + i n_0 p  u_\parallel(\vec{p},\omega) &=& 0\;,\nonumber\\	-i\omega \frac{P_0+n^\epsilon_0}{v_F^2} \vec{u}(\vec{p},\omega) + i \vec{p}\;\delta P(\vec{p},\omega) &=& \nonumber\\ &&\hspace{-1cm}-   \frac{i2\pi\alpha v_F \vec{p}}{p} \delta n(\vec{p},\omega) n_0\;,
	\nonumber\\
	-i\omega \delta n^\epsilon(\vec{p},\omega)+(P_0+n_0^\epsilon)ipu_\parallel(\vec{p},\omega) &=&   0\;.
\end{eqnarray}
Here, we define $u_\parallel = \vec{u} \cdot \vec{p}/p$ which gives the component of the hydrodynamic velocity in the direction of the momentum $\vec{p}$. We are interested in the longitudinal propagating modes, so we project the momentum equation on the momentum direction $\vec{p}/p$. This gives
\begin{equation}
	-i\omega\frac{P_0+n^\epsilon_0}{v_F^2} u_\parallel(\vec{p},\omega) + i p\;\delta P(\vec{p},\omega) = - i 2\pi\alpha v_F \delta n(\vec{p},\omega) n_0.
\end{equation}
Together with the other two equations, we find that
\begin{equation}
		\begin{pmatrix}
	-i\omega&in_0 p &0\\i2\pi\alpha v_F n_0&-i\omega\frac{P_0+n^\epsilon_0}{v_F^2} &ip/2 \\0&ip(P_0+n_0^\epsilon)  &-i\omega
	\end{pmatrix}
	\begin{pmatrix}
		\delta n(\vec{p},\omega)\\ u_\parallel(\vec{p},\omega) \\ \delta n^\epsilon(\vec{p},\omega)
	\end{pmatrix} = \begin{pmatrix}
	0\\0 \\0
\end{pmatrix}.
\end{equation}
These equations have three solutions when the frequencies of the fluctuations satisfie the dispersion relations
\begin{eqnarray}
\omega(\vec{p})&=&0 \nonumber\\  \omega_{\pm}(\vec{p})&=&\pm\sqrt{2\pi\alpha v^3_F n_0^2p/(P_0+n^\epsilon_0)+v_F^2p^2/2}.
\end{eqnarray}
We observe, that in the long wavelength limit, there is a square-root dispersion which represents the hydrodynamic plasmon. At charge neutrality, $n_0 = 0$, the dispersion becomes linear representing sound waves with a velocity given by $v_F/\sqrt{2}$. 

This result, as we will see later, is in disagreement with the calculation within the RPA. The RPA calculation predicts the existence of thermal plasmons at charge neutrality and at non-zero temperature. The key step to reconcile these results relies on the observation that, for Dirac electrons, the momentum density and charge current are decoupled. This is in stark contrast to one-band systems with a parabolic dispersion where the momentum density is proportional to the charge current. To this end, let us additionally consider an equation of motion for the charge current, which is obtained by multiplying the Boltzmann equation, Eq.~\eqref{eq:fermionboltzmannequation}, by the corresponding group velocity $\partial_{\vec{p}}\epsilon_{\lambda}(\vec{x},\vec{p}) = \lambda v_F^R\hat{p}$ and integrating the resulting equation over all states. The time-derivative term yields 
\begin{equation}
	N\sum_{\lambda=\pm} \int \frac{d\vec{p}}{(2\pi)^2} \lambda v_F^R \hat{p} \partial_t f_\lambda(\vec{x},t,\vec{p})= \partial_t \vec{j}(\vec{x},t). 
\end{equation}
The space-derivative term can be similarly integrated and gives
\begin{eqnarray}
	&&N\sum_{\lambda=\pm} \int \frac{d\vec{p}}{(2\pi)^2}  \left( v^R_F\right)^2\hat{p} \hat{p}\cdot\partial_{\vec{x}} f_\lambda(\vec{x},t,\vec{p}) = \partial_{\vec{x}} \cdot  \vec{\vec{\Xi}}(\vec{x},t)\;,\nonumber \\   
\end{eqnarray}
where we define a second-rank tensor according to
\begin{eqnarray}
	&&\vec{\vec{\Xi}}(\vec{x},t)=\nonumber\\&&N\sum_{\lambda=\pm}\int\frac{d\vec{p}}{(2\pi)^2} v^2_F\hat{p}\hat{p}\Big[ f_+(\vec{x},t,\vec{p}) +  (f_-(\vec{x},t,\vec{p})-1)\Big].\nonumber\\
\end{eqnarray}
The momentum-derivative term yields
\begin{eqnarray}
	&&\hspace{-0.5cm}-N\sum_{\lambda=\pm} \int \frac{d\vec{p}}{(2\pi)^2} \lambda v_F^R\hat{p} \partial_{\vec{x}}V^H(\vec{x},t)\cdot\partial_{\vec{p}} f_\lambda(\vec{x},t,\vec{p})\nonumber= \nonumber\\&&\hspace{5cm} \partial_{\vec{x}}V^H(\vec{x},t) \cdot \vec{\vec{\Lambda}}(\vec{x},t)\;. \nonumber\\
\end{eqnarray}
where another second-rank tensor has components given by
\begin{eqnarray}
	&&\Lambda_{ij}(\vec{x},t)=\nonumber\\&&N\int \frac{d\vec{p}}{(2\pi)^2} \; \left(\frac{\delta_{ij}}{p}-\frac{p_ip_j}{p^3}\right)\left[ f_+(\vec{x},t,\vec{p}) + (f_-(\vec{x},t,\vec{p})-1)\right].\nonumber\\
\end{eqnarray}
In contrast to the conserved quantities discussed in the previous section, the group velocity is not a collisional invariant. Therefore, the current density is not conserved by electron-electron interactions. This is particularly true at charge neutrality. The integration of the collision term is, in general, very complicated, especially since the distribution function is unknown. For the purpose of this discussion, we resort to the relaxation-time approximation and assume that
\begin{equation}
	\sum_{\lambda=\pm}\int\frac{d\vec{p}}{(2\pi)^2} \;\lambda v_F^R p C_\lambda[f](\vec{p}) \approx -  \frac{\vec{j}}{\tau}\;.
\end{equation}
The value of the relaxation time $\tau$ may be approximated by the corresponding element of the collision operator in the Boltzmann equation (see for example in Ref.~\cite{Fritz2008a}).
Finally, combining the above calculations, we find the  equation of motion for the charge current as
\begin{eqnarray}
	\partial_{t} \vec{j}(\vec{x},t) + \partial_{\vec{x}}\vec{\vec{\Xi}}(\vec{x},t) + \partial_{\vec{x}}V^H(\vec{x},t) \vec{\vec{\Lambda}}(\vec{x},t)=- \frac{ \vec{j}(\vec{x},t)}{\tau}\;.\nonumber\\\label{eq:currentequation}
\end{eqnarray}
To linear order in $\vec{u}$, we find that $\Xi_{ij} =   nv_F^2/2 \;\delta_{ij}$ which is proportional to the charge density and $\Lambda_{ij} = \frac{N}{4\pi}\mathcal{N}\delta_{ij}$ where $\mathcal{N}=\int dp \left[ f_+^0(\vec{p})-(f^0_-(\vec{p})-1)\right] = T\log(2+2\cosh\mu/T)$. Here $f_\lambda^0(\vec{p})=(1+\exp(\frac{\lambda v_F p - \mu}{T}))^{-1}$ is the Fermi-Dirac distribution function. 

We are again interested in small density fluctuations around the constant and homogeneous value. Consequently, we assume  $n = n_0 + \delta n$ and small $\vec{u}$. We insert this solution into the current Eq.~\eqref{eq:currentequation} and keep the terms to linear order in the fluctuations. This gives
\begin{eqnarray}
	n_0\partial_{t} \vec{u}(\vec{x},t) + \frac{v_F^2}{2}\partial_{\vec{x}}\delta n(\vec{x},t) + \partial_{\vec{x}}V^H(\vec{x},t) \frac{N}{4\pi}\mathcal{N}\nonumber\\&&\hspace{-2.5cm}=- \frac{n_0 \vec{u}(\vec{x},t)}{\tau}.
	\label{eq:linearizedcurrentequation}
\end{eqnarray}
The solution to this equation are propagating waves of the form
\begin{equation}
	\begin{pmatrix}
		\delta n (\vec{x},t)\\ \vec{u}(\vec{x},t)
	\end{pmatrix} = 	\begin{pmatrix}
		\delta n (\vec{p},\omega)\\ \vec{u}(\vec{p},\omega)
	\end{pmatrix} e^{i\vec{p}\cdot\vec{x}-i\omega t}.
\end{equation}
We insert this solution into the linearized current Eq.~\eqref{eq:linearizedcurrentequation} and obtain
\begin{eqnarray}
	-i\omega n_0 \vec{u}(\vec{p},\omega) + i  \frac{v_F^2}{2} \vec{p}\delta n(\vec{p},\omega) + i\frac{2\pi\alpha v_F \vec{p}}{p} \delta n(\vec{p},\omega) \frac{N}{4\pi}\mathcal{N}&&\nonumber\\&&\hspace{-3cm}=- \frac{n_0 \vec{u}(\vec{p},\omega)}{\tau}.\nonumber\\
\end{eqnarray}
Next, we  project the  equation on the momentum direction $\vec{p}/p$, This gives
\begin{eqnarray}
	-i\omega n_0 u_\parallel(\vec{p},\omega) + i  \frac{v_F^2}{2} p\delta n(\vec{p},\omega) + i2\pi\alpha v_F \delta n(\vec{p},\omega) \frac{N}{4\pi}\mathcal{N}&&\nonumber\\&&\hspace{-3cm}=- \frac{n_0 u_\parallel(\vec{p},\omega)}{\tau}.\nonumber\\
\end{eqnarray}
Together with the continuity equation, we find that
\begin{equation}
	\begin{pmatrix}
		-i\omega&in_0 p \\i\alpha v_F \frac{N}{2}\mathcal{N}+i\frac{v_F^2}{2}p&-i\omega n_0+n_0/\tau 
	\end{pmatrix}
	\begin{pmatrix}
		\delta n(\vec{p},\omega)\\ u_\parallel(\vec{p},\omega) 
	\end{pmatrix} = \begin{pmatrix}
		0\\0 
	\end{pmatrix}.
\end{equation}
For this equation to be valid, the frequency of the density fluctuations has to
satisfy the dispersion relation
\begin{eqnarray}\label{eq:plasmondispersionhydro}
	&&\omega_{\pm}(\vec{p})=-\frac{i}{2\tau}\nonumber\\&&\pm\sqrt{\frac{N}{2}\alpha v_F T p \log(2+2\cosh\mu/T)+\frac{v_F^2p^2}{2}-\frac{1}{4\tau^2}} .\nonumber\\
\end{eqnarray}
Note that, in the long-wavelength limit, we recover the previous hydrodynamic result.  

\begin{figure}[t]
	\centering
	\includegraphics[width=0.4\textwidth]{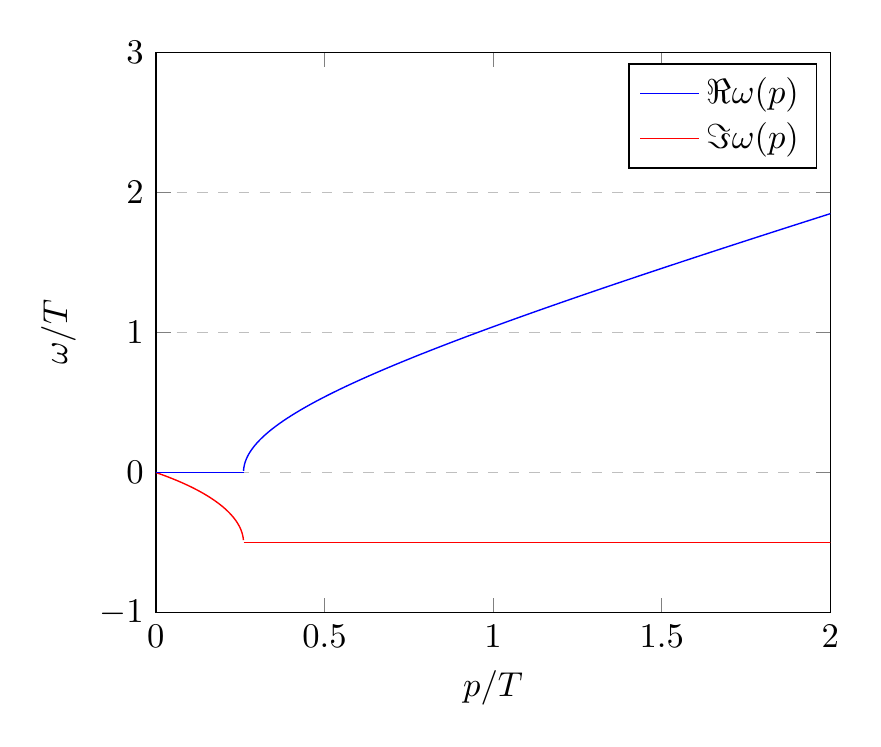}
	\caption{The figure shows the real and imaginary part of the frequency of the density fluctuations as a function of momentum at the charge neutrality point. In the plot, the momentum and frequency are in the units of $T$. The values the parameters are chosen for illustrative purpose: $\alpha=0.3$, $1/\tau =T$, and $N=4$. In the case of graphene $N=4$ counting spin and valley degrees of freedom.}
\end{figure}

\section{Part B: Electron-hole-plasmon hydrodynamics in the strong coupling limit}\label{sec:strongcoupling}

 In the previous section we focused on the weak-interaction limit of the action in Eq.~\eqref{eq:electrononlyaction}. We analyzed the theory by a straightforward pertubative expansion in the coupling constant. 
 We now consider a system of fermions interacting strongly via long-range Coulomb interactions. It is well known, that the interactions between electrons can generate plasma oscillations. Under certain circumstances, these plasma oscillations act as proper quasi-particles, as we will show below. Describing these oscillations starting from Eq.~\eqref{eq:electrononlyaction} requires to go beyond pure perturbation theory and to resort to a resummation scheme, such as the random-phase approximation (RPA)~\cite{Pines&Bohm1952}. Formally, this can be achieved by a Hubbard-Stratonovich transformation~\cite{Kleinert1978}, which is the formulation we choose here. The results in the following sections are the most important new results of this paper.

\subsection{The effective field theory: The random-phase approximation}\label{subsec:rpa}
We introduce a real scalar boson field $\phi_a(\vec{r},t)$ to decouple the quartic Coulomb interaction using the Hubbard-Stratonovich identity
\begin{eqnarray}\label{eq:HStransformation}
	&&\exp\left(-\frac{i}{2}\int dtdt'd\vec{x}d\vec{x}' 	 \rho^a(\vec{x},t)D_{0,ab}(\vec{x},\vec{x}',t,t')\rho^b(\vec{x}',t')\right)\nonumber\\ &&\hspace{0.5cm} = \int \mathcal{D}\phi \exp\Bigg(\frac{i}{2}\int dtdt'd\vec{x}d\vec{x}'  \phi_{a}(\vec{x}, t)D_{0,ab}^{-1}\phi_{b}(\vec{x}',t')\nonumber\\&&\hspace{1.5cm}-i\int dt d\vec{x} \phi_{a}(\vec{x},t)\rho^a(\vec{x},t)\Bigg)\;.
\end{eqnarray}
All the manipulations in this section are performed on the Schwinger-Keldysh closed time contour, meaning the indices $a$ and $b$ are the previously defined Keldysh indices.
In writing down Eq.~\eqref{eq:HStransformation}, we absorb an irrelevant normalization constant into the functional integration measure. The real scalar field is conjugate to the electron density and therefore directly captures the dynamics of the plasmons. Therefore, we henceforth refer to this boson field as the plasmon field. Inserting the identity of Eq.~\eqref{eq:HStransformation} into the partition function of Eq.~\eqref{eq:partitionfuntionelectrononlytheory} leads to
\begin{equation}
	Z = \int \mathcal{D}\psi^\dagger\mathcal{D}\psi\mathcal{D}\phi \exp(i S[\psi^\dagger,\psi,\phi])\;,
	\label{eq:partitionfunction}
\end{equation}  
where the action reads
\begin{widetext}
	\begin{eqnarray}\label{electronplasmonaction}
		S[\psi^\dagger,\psi,\phi] &=& \int_{-\infty}^{\infty} dtdt'  \Big[ \int d\vec{x}d\vec{x}' \psi_{a,\lambda}^\dagger(\vec{x},t)\Big( G^{-1}_{0,ab;\lambda \lambda'}-\gamma^c_{ab}\delta_{\lambda\lambda'}\phi_c(\vec{x},t)\delta(\vec{x}-\vec{x}')\delta(t-t')\Big)\psi_{b,\lambda'}(\vec{x}',t') \nonumber\\&& +\frac{1}{2}\int d\vec{x}d\vec{x}'  \phi_a(\vec{x},t)D^{-1 }_{0,ab}(\vec{x},\vec{x}',t,t')\phi_{b}(\vec{x}',t')\Big].\nonumber\\
	\end{eqnarray}	
\end{widetext}
Philosophically, we have traded a theory of electrons interacting amongst themselves for a field theory where electrons interact with the electric potential and the plasmon field. 
	
\subsubsection{Green functions}	
	The bare inverse Green function of the boson field reads $D_{0,ab}^{-1}(\vec{x},\vec{x}',t,t') =  4\epsilon\sigma^x_{ab} \delta(\vec{x}-\vec{x}')\delta(t-t') \sqrt{-\nabla^2}/e^2$. The square root of the Laplacian, $\sqrt{-\nabla^2}$, can be understood in the following way: it is the inverse Fourier transform of the absolute value of the momentum, $p \equiv |\vec{p}|$. The Fourier transform of $D^{-1}_{0,ab}(\vec{x},\vec{x}',t,t')$ consequently is given by 
	\begin{equation} 
		D^{-1}_{0,ab}(\vec{p},\omega)= \sigma^x_{ab} 2V^{-1}(\vec{p}) = \sigma^x_{ab} 2p/2\pi\alpha v_F\;.
		\label{eq:barecoulombpropagator}
	\end{equation} 
	There are two things worthwhile noting here. First, this zeroth-order Green function has no dynamics. The dynamics will only be generated upon integrating out fermions or, equivalently, in perturbation theory. Second, it comes with a factor of $2$ due to our choice of convention that $D_{0,ab} \propto V(\vec{x}-\vec{x}')/2$, see Eq.~\eqref{eq:coulombGreenfunction}.
There also is the fermionic propagator, that we have to evaluate in the effective field theory of the plasmons. To obtain this field theory, we have to integrate out the fermions. This suggests that the fermionic propogator is given by the non-interacting one and all the renormalization effects are in the plasmon sector. This, however, is not true, and the generated dynamics feeds back into the fermion sector. To see this, it proves advantageous to introduce source terms in the action in Eq.~\eqref{electronplasmonaction}, according to
\begin{equation}
	S_J[\psi^\dagger,\psi]=\int dt d\vec{x} \left(\psi_a^\dagger(\vec{x},t)J_a(\vec{x},t)+J_a^\dagger(\vec{x},t)\psi_a(\vec{x},t)\right).
\end{equation}
This allows to recover the fermionic Green function for any level of approximation of the plasmon field, even once the electrons are integrated out. This is very important in the section about the coupled quantum kinetic equations. 
The generating function for the fermionic Green function reads
\begin{equation}
	Z[J,J^\dagger] = \int \mathcal{D}\psi^\dagger\mathcal{D}\psi \mathcal{D}\phi \exp(i S[\psi^\dagger,\psi,\phi]+iS_J[\psi^\dagger,\psi])\;.
\end{equation}
From this, we can determine the fermion Green function by means of a functional derivative with respect to the source field, according to
\begin{equation}
	iG(\vec{r},\vec{r}\;',t,t') = \frac{-1}{Z[J,J^\dagger]}\frac{\delta^2\;Z[J,J^\dagger]}{\delta  J(\vec{x}',t')\delta J^\dagger(\vec{x},t)}\Big|_{J=J^\dagger=0}\;.
	\label{eq:Greenfunctionformula}
\end{equation}
We continue to integrate out the fermion fields, which gives an effective theory of the boson field associated with the density fluctuations. We find
\begin{equation}
	Z[J,J^\dagger] = \int \mathcal{D}\phi \exp(i S_{\rm{eff}}[\phi,J,J^\dagger]),
	\label{eq:generatingfunction}
\end{equation}
with the effective action given by
\begin{eqnarray}
	&&S_{\rm{eff}}[\phi,J,J^\dagger] = -i\text{Tr}\left[\ln\left(-iG^{-1}\right)\right] \nonumber\\&&+\frac{1}{2}\int dt dt' d\vec{x}d\vec{x}'  \phi_a(\vec{x},t)D^{-1 }_{0,ab}(\vec{x},\vec{x}'t,t')\phi_{b}(\vec{x}',t').\nonumber\\&&-\int dtdt'd\vec{x}d\vec{x}' J^\dagger(\vec{x},t)G(\vec{x},\vec{x}',t,t';\phi)J(\vec{x}',t').
	\label{eq:effectiveaction}
\end{eqnarray}
where the Green function $G_{ab;\lambda\lambda'}(\vec{x},\vec{x}',t,t';\phi)$ is a functional of the plasmon field~\cite{Stoof2009} as
\begin{widetext}
\begin{eqnarray}
	G^{-1}_{ab;\lambda\lambda'}(\vec{x},\vec{x}',t,t';\phi) = G^{-1}_{0,ab;\lambda\lambda'}(\vec{x},\vec{x}',t,t') -\gamma_{ab}^{c}\phi_{c}(\vec{x},t)\delta_{\lambda\lambda'} \delta(\vec{x}-\vec{x}')\delta(t-t')\;.
\end{eqnarray}
While this is formally exact, the presence of the dynamical field requires an approximation scheme to evaluate it. 
Using Eq.~\eqref{eq:Greenfunctionformula} and the generating function introduced in Eq.~\eqref{eq:generatingfunction},
we find that the fermion Green function can be calculated according to 
\begin{eqnarray}
	G_{ab;\lambda\lambda'}(\vec{x},\vec{x}\;',t,t') = \int \mathcal{D}\phi \;G_{ab;\lambda\lambda'}(\vec{x},\vec{x}\;',t,t';\phi) \exp(i S_{\rm{eff}}[\phi,0,0])\;.
	\nonumber\\
	\label{eq:fermionGreenfunctionfromtheeffectiveaction}
\end{eqnarray}
\end{widetext}
This equation consequently shows in a very explicit manner that the generated plasmon dynamics feeds back into the fermion dynamics through the field $\phi$ and its associated dynamics encoded in $S_{\rm{eff}}$. Consequently, the next step is to determine $S_{\rm{eff}}$.

\subsubsection{The saddle-point equation}

\begin{figure}[t]
		\centering
		\includegraphics[width=0.5\textwidth]{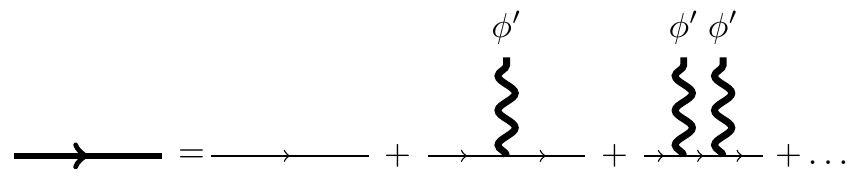}
		\caption{Diagram representing the solution of the Dyson equation (\ref{eq:Dysonequationforfermionwithkappap}) in terms of a series expansion of the quantum fluctuation $\phi'$. }
		\label{fig:solutiontodysonequation}
\end{figure}

\begin{figure}[t]
	\centering
	\begin{subfigure}[]{0.2\textwidth}
		\centering
		\includegraphics[width=\textwidth]{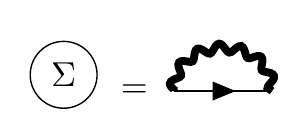}
		\subcaption{electron self-energy}
		\label{fig:fermionselfenergy}
	\end{subfigure}
	\begin{subfigure}[]{0.2\textwidth}
		\centering
		\includegraphics[width=\textwidth]{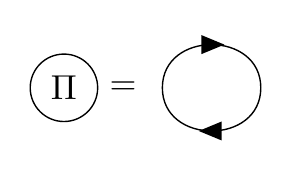}
		\subcaption{Boson self-energy}
		\label{fig:plasmonselfenergy}
	\end{subfigure}
	\caption{Self-energy for electron and plasmon fields within the RPA approximation. The set of the GW-diagram and polarization function constitutes a conserving approximation.}
	\label{fig:}
\end{figure}

The effective action $S_{\rm{eff}}$, introduced formally in Eq.~\eqref{eq:effectiveaction}, can be obtained after integrating out the fermions. It is exact but also very complicated. The problem is that the trace cannot be evaluated in an easy manner due to the presence of the plasmon field in the Green function of the fermions. Consequently, we require an approximation scheme. 
The saddle-point contribution to the partition function is given by the configuration that minimizes the action $S_{\rm{eff}}[\phi,0,0]$ (formally this manipulation is equivalent to Hartree or mean-field approximation). This can be obtained from the condition
\begin{eqnarray}
\frac{\delta S_{\rm{eff}}[\phi,0,0]}{\delta \phi}\Big|_{\langle \phi \rangle} = 0
\end{eqnarray}
This directly leads to
\begin{eqnarray}\label{eq:saddlepoint}
	&&\langle\phi_{c}(\vec{x},t)\rangle = \nonumber\\&&-i\int dt'd\vec{x}' D_{0,cd}(\vec{x},\vec{x}';t,t')G_{ab;\lambda\lambda}(\vec{x}';\vec{x}',t',t';\langle\phi\rangle) \gamma^{d}_{ba}.\nonumber\\
\end{eqnarray} 
This is a self-consistency equation for the local charge density. Philosophically, Eq.~\eqref{eq:saddlepoint} corresponds to a self-consistent version of the Hartree diagram previously discussed in Sec.~\ref{subsec:collisionless}. 
The fermion propagator at this level of approximation reads
\begin{widetext}
\begin{eqnarray}
	G^{-1}_{ab;\lambda\lambda'}(\vec{x},\vec{x}',t,t';\langle\phi\rangle) = G^{-1}_{0,ab;\lambda\lambda'}(\vec{x},\vec{x}',t,t') -\gamma_{ab}^{c}\langle \phi_{c}(\vec{x},t) \rangle \delta_{\lambda\lambda'} \delta(\vec{x}-\vec{x}')\delta(t-t')\;.
\end{eqnarray}
\end{widetext}
The expectation value of the plasmon field can now be identified with the self-energy within the Hartree approximation, already given in Eq.~\eqref{eq:Hartreeselfenergy}. It thus recovers the dispersion of Eq.~\eqref{eq:electronenergy}. 
 We proceed to expand the plasmon field in deviations from the mean-field value, {\it i.e.}, $\phi_a(\vec{x},t) = \langle\phi_a(\vec{x},t) \rangle+\phi_a'(\vec{x},t) $ where $\langle \phi_a(\vec{x},t)  \rangle$ is the saddle-point configuration and $\phi_a'$ is associated with fluctuations around the saddle point. As a result, we have 
 \begin{widetext}
\begin{eqnarray}
	G^{-1}_{ab;\lambda\lambda'}(\vec{x},\vec{x}\,',t,t';\phi)= G^{-1}_{ab;\lambda\lambda'}(\vec{x},\vec{x}',t,t';\langle\phi\rangle)  -  \gamma_{ab}^{\alpha} \phi'_{\alpha}(\vec{x},t)  \delta_{\lambda\lambda'} \delta(\vec{x}-\vec{x}')\delta(t-t')\;.
	\label{eq:Dysonequationforfermionwithkappap}
\end{eqnarray}
\end{widetext}
We proceed to expand the effective action, Eq.~\eqref{eq:effectiveaction}, to second order in the fluctuations $\phi'$. 
 Substituting Eq.~\eqref{eq:Dysonequationforfermionwithkappap} into Eq.(\ref{eq:effectiveaction}) and using the series expansion of the logarithm schematically, we suppress the Keldysh indices here, we find to second order in fluctuations that
 \begin{widetext}
\begin{eqnarray}\label{eq:logexpansion}
S_{\rm{eff}}[\phi,0,0] &=& -i\text{Tr}\left[\ln \left(-iG^{-1} (\langle \phi \rangle+\phi')\right) \right] +\frac{1}{2}\int dt dt' d\vec{x}d\vec{x}'  \left( \langle \phi_a(\vec{x},t)\rangle +\phi'_a(\vec{x},t) \right)D^{-1 }_{0,ab}(\vec{x},\vec{x}'t,t')\left( \langle \phi_{b}(\vec{x}',t')\rangle +\phi'_{b}(\vec{x}',t') \right) \nonumber \\ &\approx& -i\text{Tr}[\ln(-iG^{-1}(\langle{\phi}\rangle))] +i\text{Tr}[G(\langle{\phi}\rangle)\phi'] + \int dt dt' d\vec{x}d\vec{x}' \langle\phi_a(\vec{x},t)\rangle D^{-1 }_{0,ab}(\vec{x},\vec{x}';t,t')\phi'_{b}(\vec{x}',t')\nonumber \\ &&+i\text{Tr}[\frac{1}{2}G(\langle \phi \rangle)\phi'G(\langle{\phi}\rangle)\phi'].\nonumber\\
\end{eqnarray}
\end{widetext}
Here, we use for brevity the shorthand notation $G^{-1}(\langle{\phi}\rangle) \equiv G^{-1}_{ab;\lambda\lambda'}(\vec{x},\vec{x}',t,t';\langle\phi\rangle)$. The first term gives an irrelevant constant which will be absorbed in the functional integration measure.
The linear terms in the fluctuations sum to zero at the saddle point. Their cancelation is equivalent to the saddle-point condition in Eq.~\eqref{eq:saddlepoint}.
The remaining term, consequently, is the last term in Eq.~\eqref{eq:logexpansion}. It is the term that accounts for quadratic fluctuations around the saddle point. At the same time, however, it determines the effective plasmon propagator with the RPA approximation according to
\begin{eqnarray}
	&&S_{\rm{eff}}[\phi'] =
	\nonumber\\&&\frac{1}{2}\sum_{a,b=1,2}\int dt dt_1 d\vec{x}d\vec{x}'  \phi'_a(\vec{x},t)D^{-1 }_{ab}(\vec{x},\vec{x}';t,t')\phi'_b(\vec{x}',t')\;,\nonumber\\
\end{eqnarray}
where the inverse plasmon Green function satisfies
\begin{equation}
	D^{-1 }_{ab}(\vec{x},\vec{x}',t,t') = D^{-1 }_{0;ab}(\vec{x},\vec{x}';t,t') - \Pi_{ab}(\vec{x},\vec{x}',t,t').
	\label{eq:bosondysonequation}
\end{equation}
The self-energy $\Pi$ has the diagrammatic representation shown in Fig.~\ref{fig:plasmonselfenergy}. It is commonly referred to as the polarization diagram. The corresponding algebraic expression reads 
\begin{equation}
	\label{eq:bosonselfenergyfreefermionRPA}
	\Pi_{ab} = -i N \text{Tr}[\gamma^a G(\vec{x},\vec{x}',t,t',\langle \phi \rangle)\gamma^b G(\vec{x}',\vec{x},t',t,\langle \phi \rangle)].
\end{equation}
The retarded component of Eq.~\eqref{eq:bosondysonequation} gives dynamics to the plasmon: it allows to determine the dispersion and decay rate of the plasmons in the next section. Its Keldysh components has the form of  Eq.~\eqref{eq:Keldyshequationboson} which will be the starting point for the derivation of the Boltzmann equation for the plasmons. Now that we have the above effective action of the plasmon and its dynamics, it is time to return to the fermion Green function $G_{ab;\lambda\lambda'}(\vec{x},\vec{x}',t,t')$ in Eq.~\eqref{eq:fermionGreenfunctionfromtheeffectiveaction}. We can graphically represent the fermion Green function expansion in terms of $\phi'$ according to Fig.~\ref{fig:solutiontodysonequation}. We now have to `average' this fermion propagator over the Gaussian action of the bosons and resum it. It turns out that we have to choose the series corresponding to the Fock-like diagram, Fig.~\ref{fig:fermionselfenergy}, to obtain a conserving approximation~\cite{KadanoffBaym1961}. This is also known as the GW approximation. The resulting Green function is the solution of the Dyson equation
\begin{eqnarray}
&&	G^{-1 }_{ab;\lambda\lambda'}(\vec{x},\vec{x}',t,t') = \nonumber\\&&\hspace{0.8cm}G^{-1 }_{ab;\lambda\lambda'}(\vec{x},\vec{x}';t,t';\langle\phi\rangle) - \Sigma_{ab;\lambda\lambda'}(\vec{x},\vec{x}',t,t')\;,\nonumber\\
	\label{eq:fermiondysonequation}
\end{eqnarray}
where
\begin{eqnarray}
	&&\Sigma_{ab;\lambda\lambda'}(\vec{x},\vec{x}',t,t') = \nonumber\\&& i \left(\gamma^\alpha G(\vec{x},\vec{x}',t,t';\langle\phi\rangle)\gamma^\beta\right)_{ab;\lambda\lambda'} D_{\alpha\beta}(\vec{x},\vec{x}',t,t')\;.
	\label{eq:fermionselfenergyfreefermionRPA}
\end{eqnarray}

The Dyson equations in  Eqs.~\eqref{eq:bosondysonequation} and \eqref{eq:fermiondysonequation}, together with the self-energies in Eqs.~\eqref{eq:bosonselfenergyfreefermionRPA} and \eqref{eq:fermionselfenergyfreefermionRPA}, are the minimal set of equations that describes the interplay between the collective modes and the single-particle components of the interacting Dirac electron. This approximation can be derived from a single free energy diagram. It was shown in Ref.~\cite{KadanoffBaym1961} that this implies that it constitutes a conserving approximation. As such, it respects the conservation laws of total energy and momentum in the combined system of electrons and plasmons, as we show explicitly in Sec.~\ref{subsec:conservationlawsrpa}. 

\subsection{The plasmons}\label{sec:plasmons}

In this section we discuss the plasmon dynamics at non-zero temperature and non-zero chemical potential. To that end, we analyze the polarization function numerically. We then proceed to find an approximate analytical description that we use to determine the plasmon spectrum and the quasi-particle life-times. 
\subsubsection{Non-zero temperature polarization function}\label{subsec:polarization}
Here, we consider the retarded component of the polarization function, Eq.~\eqref{eq:bosonselfenergyfreefermionRPA}, at non-zero temperature. After a Wigner transformation, we obtain
\begin{eqnarray}
		&&\Pi^{R}(\vec{p},\omega) \nonumber\\&&=-i N \int\frac{d\vec{q}}{(2\pi)^2} \frac{d\nu}{2\pi}  \Big[G^{R}_{\lambda\lambda'}(\vec{p}+\vec{q},\omega+\nu,\langle\phi\rangle)G^K_{\lambda'\lambda}(\vec{q},\nu,\langle\phi\rangle)\nonumber\\&&\hspace{2.5cm}+ \; G^K_{\lambda\lambda'}(\vec{p}+\vec{q},\omega+\nu,\langle\phi\rangle)G^A_{\lambda'\lambda}(\vec{q},\nu,\langle\phi\rangle) \Big].\nonumber\\
\end{eqnarray}
Here, for brevity, we suppress the space and time variables of the functions involved. Next, we use the transformation matrix in Eq.~\eqref{eq:tranformationmatrix} to transform the objects within the polarization function into the quasi-particle basis. After that, we integrate over the frequency variable $\nu$ by making use of the Dirac delta function coming from $g_0^K$. Finally, by a straightforward algebraic manipulation, we find the polarization function expressed in the form of the Lindhard formula
\begin{eqnarray}	
	&&\Pi^{R}(\vec{p},\omega) =\nonumber\\&&2N\sum_{\lambda\lambda'=\pm1} \int\frac{d\vec{q}}{(2\pi)^2} 
	\frac{ \mathcal{F}_{\lambda\lambda'}(\vec{p},\vec{q}) \left(f_\lambda(\vec{q})-f_{\lambda'}(\vec{p}+\vec{q})\right)}{\omega+i0^++\lambda v_F^R q-\lambda' v_F^R|\vec{p}+\vec{q}|}\;.\nonumber\\
	\label{retpol}
\end{eqnarray}
The coherence factor is defined according to
\begin{equation} \mathcal{F}_{\lambda\lambda'}(\vec{p},\vec{q})= \frac{1}{2}(1+\lambda\lambda'\cos(\theta_{\vec{p}+\vec{q}}-\theta_{\vec{q}}))\;.
\end{equation}
Let us note that, strictly speaking, compared to the conventional Lindhard formula, there is an extra factor $2$ in our result. This is consistent within our convention that the inverse bare boson Green function comes with the same factor, {\it i.e.}, $D_0 \propto V(\vec{x}-\vec{x}')/2$.

\begin{figure}
		\centering
		\includegraphics[width=0.4\textwidth]{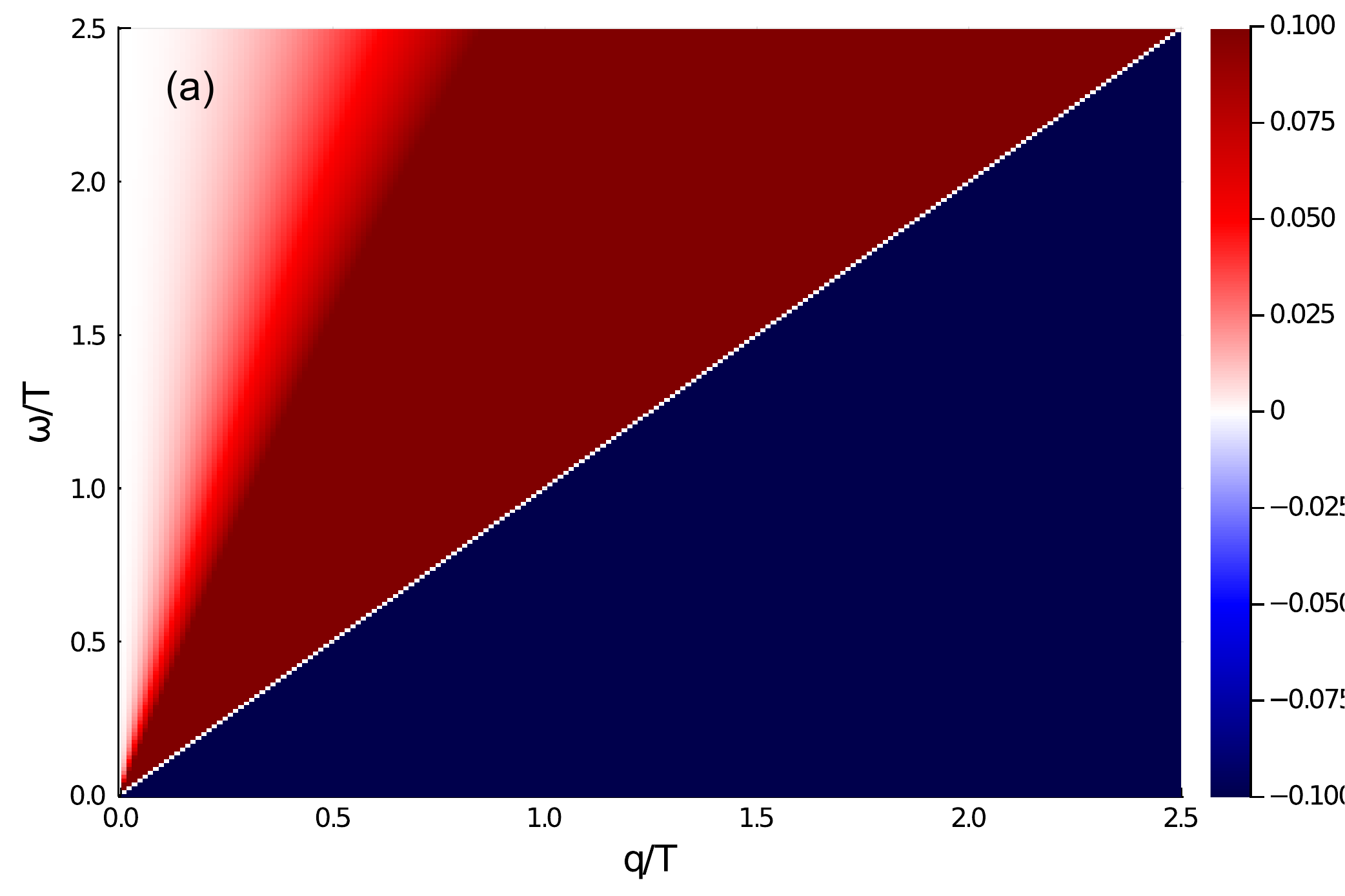}
		\includegraphics[width=0.4\textwidth]{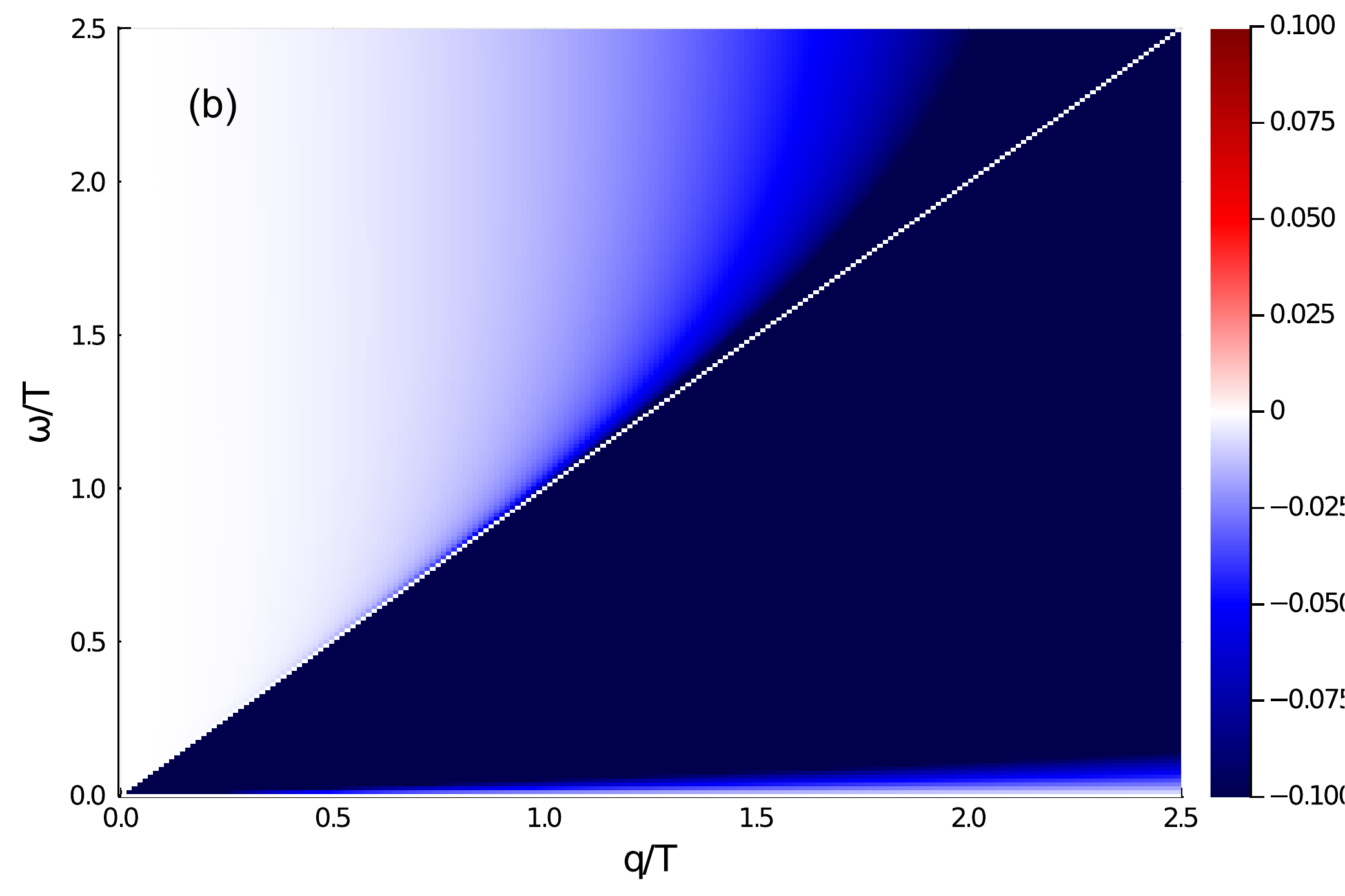}
	\caption{Polarization function, (a) real part and (b) imaginary part, at a non-zero temperature and chemical potential. We show the polarization function when the chemical potential $\mu/T$ = 1.}
	\label{fig:pol}
\end{figure}

The polarization function at zero temperature and arbitrary chemical potential ($\Pi_{\mu,0}$) can be calculated exactly. Here, we focus on non-zero temperatures. There is no analytical expression for the non-zero temperature polarization. However, there exists a relation between zero temperature and non-zero temperature polarization functions~\cite{Sarma2013},
\begin{equation}
	\Pi^R_{\mu,T}(\vec{p},\omega) = \int_0^\infty \; d\mu' \sum_{\lambda=\pm1} \frac{\Pi^R_{\mu',0}(\vec{p},\omega)}{4T\cosh^2\left(\frac{\mu'+\lambda\mu}{2T}\right)}.
\end{equation}
Using this, we numerically solve the polarization function at arbitrary temperatures and subsequently replace it in the Dyson equation in Eq.~\eqref{eq:bosondysonequation}. This allows us to determine the energy dispersion ($\omega_p$) and the decay rate ($\gamma_p$) of the plasmon mode. The plasmon frequency ($\omega=\omega_p-i\gamma_p$) is obtained by equating the inverse Green function to zero.
\begin{equation}\label{eq:plasmonenergyequation}
	(D^{-1})^R(\vec{p},\omega)=(D_0^{-1} )^{R}(\vec{p},\omega)-\Pi(\vec{p},\omega) = 0.
\end{equation}
Defined this way, the decay rate $\gamma_p$ is positive. If the damping is sufficiently weak ($\gamma_p \ll \omega_p$), one can expand the polarization function to leading order in $\gamma_p$ 
\begin{eqnarray}
	&&\Pi(\vec{p},\omega_p-i \gamma_p) \nonumber\\&&\approx \Re\Pi(\vec{p},\omega_p)-i\gamma_p \partial_\omega \Re\Pi(\vec{p},\omega)\Big|_{\omega=\omega_p} + i \Im\Pi(\vec{p},\omega_p).\nonumber\\ 
\end{eqnarray}
The energy of the plasmon can be determined from the real part of Eq.~\eqref{eq:plasmonenergyequation}
\begin{equation}\label{eq:realpartofplasmonenergyequation}
	(D_0^{-1} )^{R}(\vec{p},\omega_p)-\Re\Pi(\vec{p},\omega_p) = 0\;,
\end{equation}
whereas the decay rate is a solution of the imaginary part 
\begin{equation}\label{eq:plasmonlifetime}
	\gamma_p = \frac{\Im \Pi(\vec{p},\omega)}{\partial_\omega \Re\Pi(\vec{p},\omega)}\Bigg|_{\omega=\omega_p}\;.
\end{equation}
 In general, the non-interacting Green function, $D^{-1}_0$ can be a function of both momentum $\vec{p}$ and frequency $\omega$. However, in our case $D^{-1}_0$ describes the bare Coulomb potential which is non-dynamical and hence does not depend on the frequency variable. Fig.~\ref{fig:pol} (a) shows the real part of the polarization function at non-zero chemical potential in the momentum-frequency plane. Solutions to Eq.~\eqref{eq:realpartofplasmonenergyequation} exist only when $\Re \Pi > 0$. This is the case in the upper triangle of the plot in Fig.~\ref{fig:pol} (a).  As discussed before, a stable plasmon requires $\Im \Pi \approx 0$. In Fig.~\ref{fig:pol}(b) we plot the imaginary part of the polarization function at the same value of parameters as in Fig.~\ref{fig:pol} (a). Although it is not zero, it is still negligibly small in the low-momentum limit. As a result, one may expect a long-wavelength underdamped plasmon mode with for all pratical purposes almost infinitely long life time. This implies that plasmons behave like quasi-particles for practical matters.

 In the next section, we find an approximate description of the dispersion of the plasmon and its decay rate in the low-momentum limit. We furthermore determine the value of the momentum cutoff beyond which the plasmons are overdamped.  
\begin{figure}
	\centering
		\centering
		\includegraphics[width=0.4\textwidth]{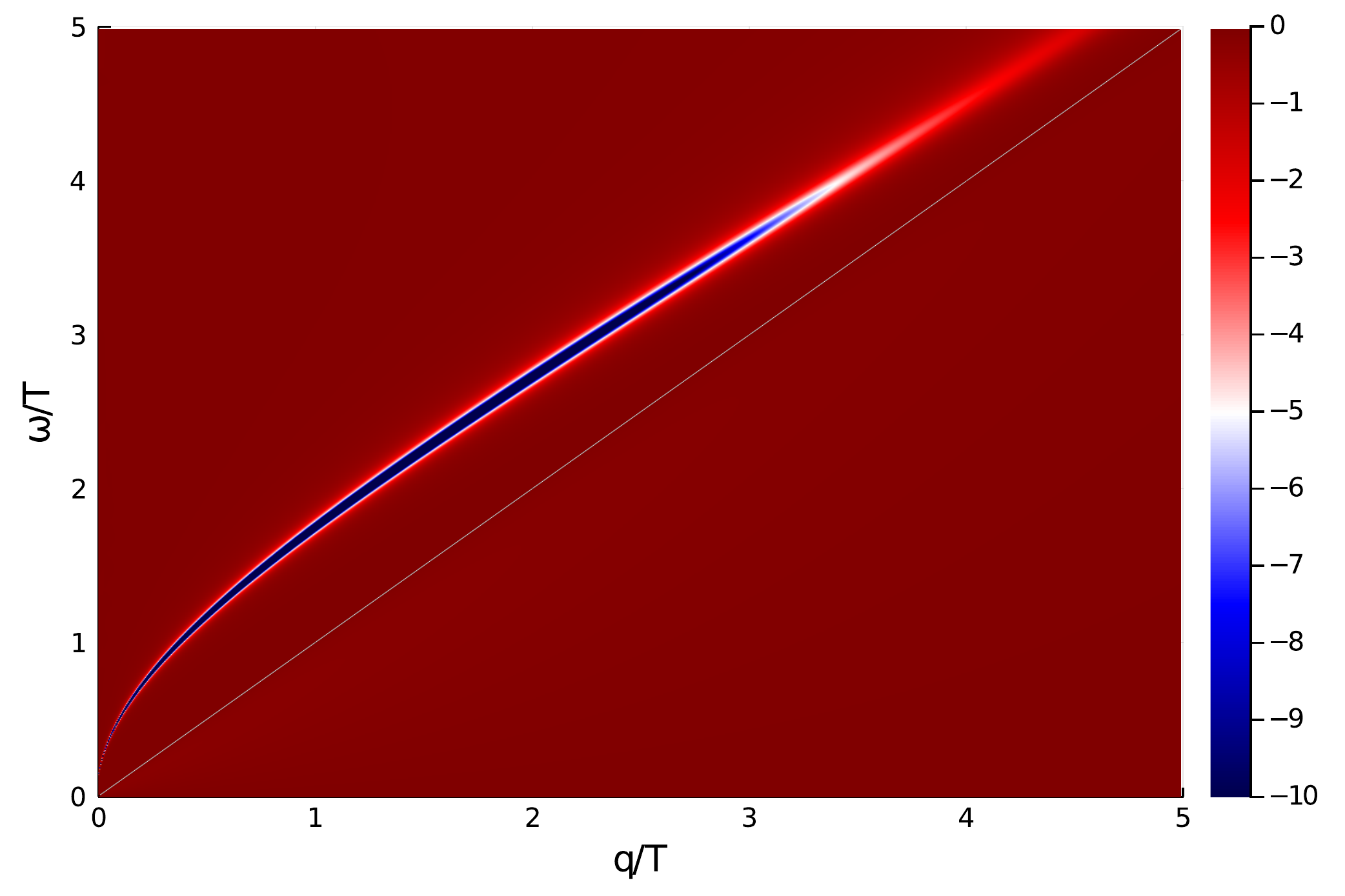}
	\caption{Spectral function of the plasmon at a non-zero temperature and chemical potential $\mu/T=4$. The value of fine structure constant is chosen to mimic graphene device sandwiched in hBN ($\alpha = 0.3$).}
	\label{fig:spectralfunctionomegamomentumplane}.
\end{figure}

\begin{figure}[t]
	\includegraphics[width=0.5\textwidth]{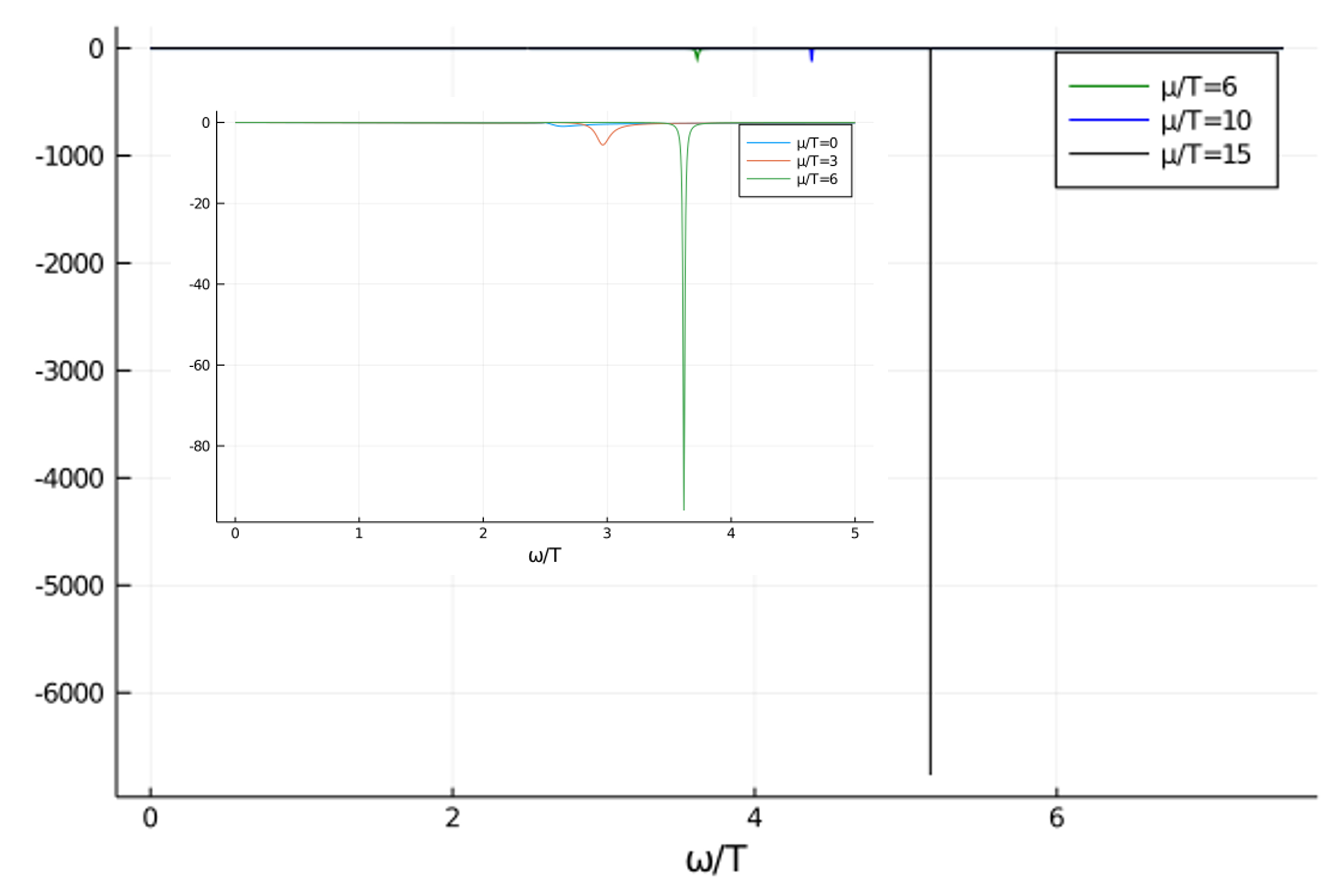}
	\caption{Spectral function of the plasmons at various electron chemical potential at a small value of momentum $q/T=2.5$}. Here the fine structure constant $\alpha=0.3$.
	\label{fig:spectralfunctionatp2.5alpha03}
\end{figure}

 \subsubsection{Analytical approximation} \label{subsec:analyticalplasmon}
 In this section, we will evaluate the energy spectrum and decay rate of the plasmon in the long-wavelength limit. To this end, we first decompose the polarization function into a sum of two terms accordingly to
$\Pi^R(\vec{p},\omega) = \Pi^R_+(\vec{p},\omega)+\Pi^R_-(\vec{p},\omega)$. The first term, $\Pi_+(\vec{p},\omega)$, describes the contribution from intraband particle-hole pairs ($\lambda'=\lambda$), whereas $\Pi_-(\vec{p},\omega)$ comes from the interband transitions ($\lambda' =-\lambda $). 
 The imaginary part of the polarization function accordingly reads
 \begin{equation}
 	\Im \Pi^R(\vec{p},\omega) = \Im \Pi^R_+(\vec{r},t) + \Im \Pi_-(\vec{p},\omega)
	\label{eq:imagpi}
\end{equation}
 and it amounts to a decay rate of the plasmon mode with
 \begin{eqnarray}	
	&&\Im\Pi^{R}_\pm(\vec{p},\omega) =-2N \pi\sum_{\lambda=\pm1} \int\frac{d\vec{q}}{(2\pi)^2} \mathcal{F}_{\lambda \;\pm\lambda}(\vec{p}+\vec{q},\vec{q}) \nonumber\\&&\Big[ 
	f_\lambda(\vec{q})-f_{\pm\lambda}(\vec{p}+\vec{q}) \Big]  \delta(\omega+\epsilon_\lambda(\vec{q})-\epsilon_{\pm\lambda}(\vec{p}+\vec{q})).
		\label{eq:imagpipm}
\end{eqnarray}
The conservation of energy enters through the delta function with the argument $\omega+\lambda v_F^Rq-\lambda' v_F^R|\vec{p}+\vec{q}|$. This function forms either an ellipse or a hyperbola in the $q_xq_y$-plane. It may be easy to see this by means of a transformation of the momentum variables into the elliptic coordinate system, $(\theta,\mu)$, where $\theta \in [0,2\pi)$ and $\mu \in [0,\infty)$. When $\lambda=-$ and $\lambda'=+$, this equation forms an ellipse with the value of $\mu$ determined by $\cosh \mu = \omega/ p$. The size of the ellipse depends on the value of $\cosh\mu \le 1$. We find that when $\omega$ is slightly bigger than $p$, the available phase space is restricted since the size of the ellipse is small whereas when $\omega$ is much bigger than $p$, the available phase space in turn grow bigger and allows the decay process more likely to occur. However, when $\lambda' = \lambda$, this equation becomes an equation of a hyperbola  centered at $(-p_x/2,-p_y/2)$. The width of the hyperbola is determined from the ratio of the frequency to the momentum given by $\cos\theta = \lambda \omega/ p$. Therefore, the available phase space for a plasmon with the energy $\omega$ to decay into two fermions with the energies $\lambda p $ and $\lambda |\vec{p}+\vec{q}|$ is extended. This process is thus the main mechanism for plasmon decay. 
We  expand all the quantities appearing in the polarization function up to first order in $\vec{p}$. This gives
\begin{eqnarray}
	\mathcal{F}_{\pm\pm}(\vec{p},\vec{q})&=&\frac{1}{2}\left(1+\cos\left(\theta_{\vec{p}+\vec{q}}-\theta_{\vec{q}}\right)\right) \approx 1 ,
	\nonumber\\		
	\mathcal{F}_{\pm\mp}(\vec{p},\vec{q})&=&\frac{1}{2}\left(1-\cos\left(\theta_{\vec{p}+\vec{q}}-\theta_{\vec{q}}\right)\right) \approx \frac{1}{4}\left(\vec{p}\cdot\vec{\nabla}_{\vec{q}}\theta_\mathbf{q}\right)^2,
	\nonumber\\		
	f_\lambda(\vec{p}+\vec{q})&\approx& f_\lambda(\vec{q})+\vec{p}\cdot\vec{\nabla}_{\vec{q}}f_\lambda(\vec{q}),
	\nonumber\\		
	\epsilon_\lambda(\vec{p}+\vec{q})&\approx& \epsilon_\lambda(\vec{q})+\vec{p}\cdot\vec{\nabla}_{\vec{q}}\epsilon_\lambda(\vec{q}).
	\label{eq:longwavelengthexpansion}
\end{eqnarray}
By substituting Eq.~\eqref{eq:longwavelengthexpansion} into the imaginary part of the polarization followed by a straightforward calculation, we find
\begin{equation}
	\Im\Pi^{R}(\vec{p},\omega)\approx -\frac{2N}{16}\frac{p^2}{\omega} \Big(f_+(|\omega/2|)-f_{-}(|\omega/2|)\Big),
\end{equation} 
which provides the main contribution to the decay rate whereas $\Im\Pi^{R}_+(\vec{p},\omega)$ gives an unimportant correction. In thermal equilibrium, the distribution function becomes the Fermi-Dirac distribution function at a temperature $T$ and  chemical potential $\mu$, {\it i.e.}, $f_{\lambda}(\omega) = \frac{1}{\exp((\omega-\mu)/T)+1)}$. We present more details of the calculation in Appendix~\ref{Appendix:longwavelengthapproximationplasmon}.

At zero temperature, this becomes
\begin{equation}
	\Im\Pi_-^{R}(\vec{p},\omega) \approx -\frac{2N}{16}\frac{p^2}{\omega} \Theta(|\omega|-2|\mu|).
\end{equation}
It vanishes when $|\omega|<2|\mu|$, meaning in that region a long-lived plasmon mode exists.
The real part of the polarization function is given by
\begin{eqnarray}	
	&&\Re\Pi_\pm^{R}(\vec{p},\omega)\nonumber\\&& =2N\sum_{\lambda=\pm1} \int\frac{d\vec{q}}{(2\pi)^2} \mathcal{F}_{\lambda\;\pm\lambda}(\vec{p},\vec{q})
	\frac{f_\lambda(\vec{q})-f_{\pm \lambda}(\vec{p}+\vec{q})}{\omega+\lambda v_F^R q\mp \lambda v_F^R|\vec{p}+\vec{q}|}.\nonumber\\
\end{eqnarray} 
We substitute these expressions into Eq.~\eqref{eq:longwavelengthexpansion} followed by expanding its denominator to first order in $p/\omega$. Based on numerics, we expect to find a stable plasmon mode in this limit. We can approximate the expression as
\begin{eqnarray}
		&&\Re\Pi_+^{R}(\vec{p},\omega)\nonumber\\&& \approx 2N\sum_{\lambda=\pm1} \int\frac{d\vec{q}}{(2\pi)^2} 
		\frac{-\vec{p}\cdot\vec{\nabla}_{\vec{q}}f_{\lambda}(\vec{q})}{\omega}\Big[1+\frac{\vec{p}\cdot\vec{\nabla}_{\vec{q}}\epsilon_\lambda(\vec{q})}{\omega}\Big],\nonumber\\
\end{eqnarray}
which results in
\begin{equation}
	\Re\Pi^{R}_+(\vec{p},\omega)\approx \frac{2Np^2}{4\pi \omega^2}\mathcal{N},
	\label{realpartpolarizationfunction}
\end{equation} 
where $\mathcal{N} = T\log\left(2+2\cosh{\mu/T}\right)$.
When the Dirac system is exposed to external perturbations, the polarization function varies in position and time via the distribution function. In that case $
	\mathcal{N}=\int dq \Big[ 
	f_+(\vec{q})-\left(f_-(\vec{q})-1\right)\Big].
$ In contrast, the interband term gives a logarithmic correction which will be neglected from evaluating the plasmon energy as
\begin{equation}
	\Re\Pi^-_{R}(\vec{p},\omega) \approx \frac{Np^2}{2\pi} \int dq \Big[f_+(\vec{q})-f_{-}(\vec{q})\Big] \Big[\frac{1}{4\left(v_F^R\right)^2q^2-\omega^2}\Big].
\end{equation}
At non-zero doping and zero temperature, we find that
$\Re \Pi^R_-(\vec{p},\omega)=\frac{Np^2}{16\pi\omega} \log \left(\Big|\frac{\omega-2\mu}{\omega+2\mu}\Big|\right)$.
If we substitute Eqs.~\eqref{realpartpolarizationfunction} and \eqref{eq:barecoulombpropagator} into Eq.~\eqref{eq:realpartofplasmonenergyequation}, we obtain the dispersion relation of the plasmon at non-zero $T$,
\begin{eqnarray}\label{eq:plasmondispersionrpa}
	\omega_{p}(\vec{p})=	\sqrt{\frac{N}{2} \alpha T p \log\left(2+2\cosh{\mu/T}\right)  }\;,
\end{eqnarray}
with the decay rate
\begin{equation}
	\gamma_{p} = \frac{\pi\omega_p(\vec{p})^2}{16\log(2+2\cosh\mu/T)}\Big(f_+(\omega_p(\vec{p})/2)-f_{-}(\omega_p(\vec{p})/2)\Big).
\end{equation}
\begin{figure}
	\centering
	\begin{subfigure}[]{0.45\textwidth}
		\centering
		\includegraphics[width=1\textwidth]{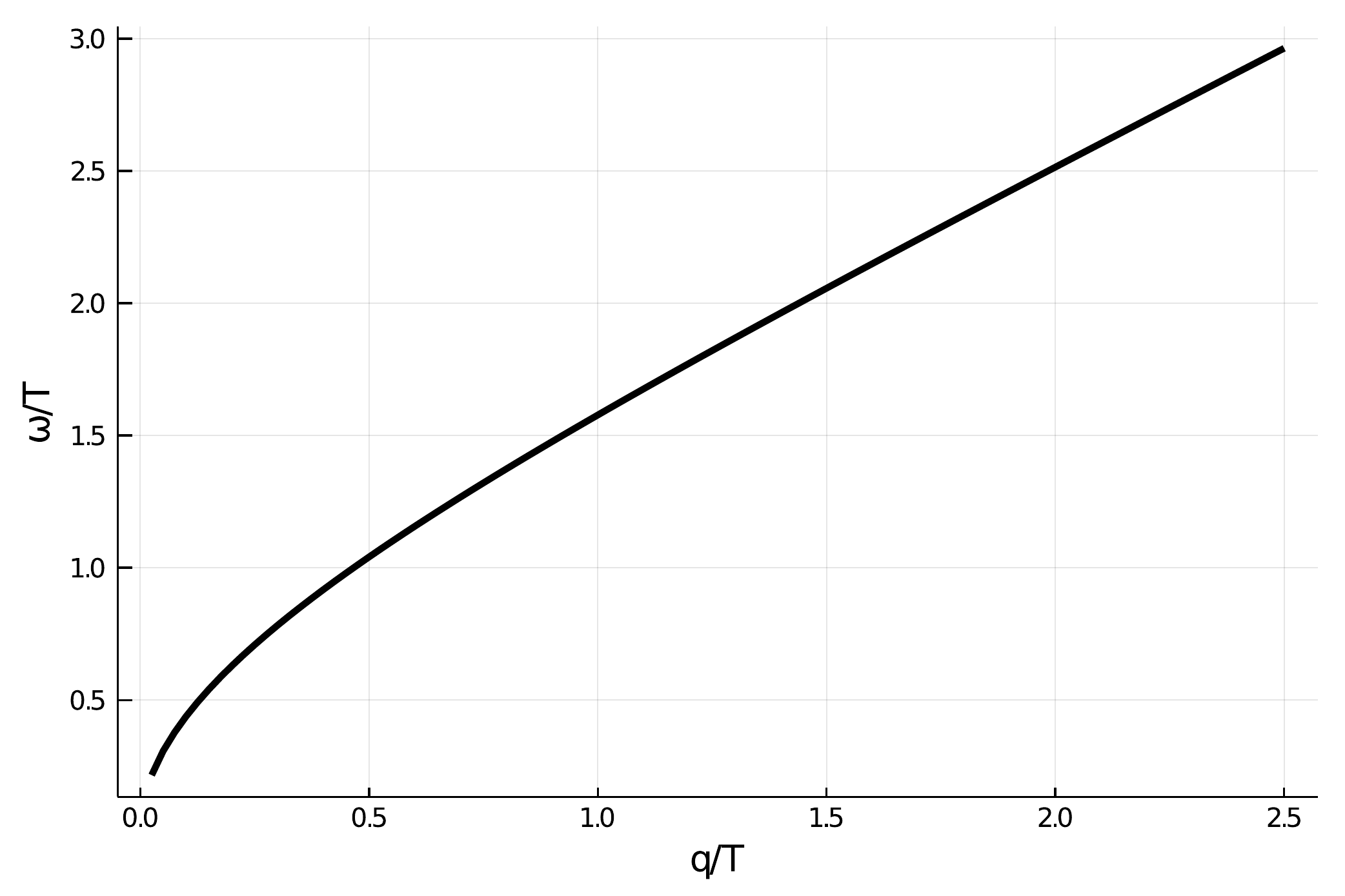}
		\subcaption{Energy dispersion}
		\label{fig:plasmondispersionmu3alpha03}
	\end{subfigure}
	\begin{subfigure}[]{0.45\textwidth}
		\centering
		\includegraphics[width=1\textwidth]{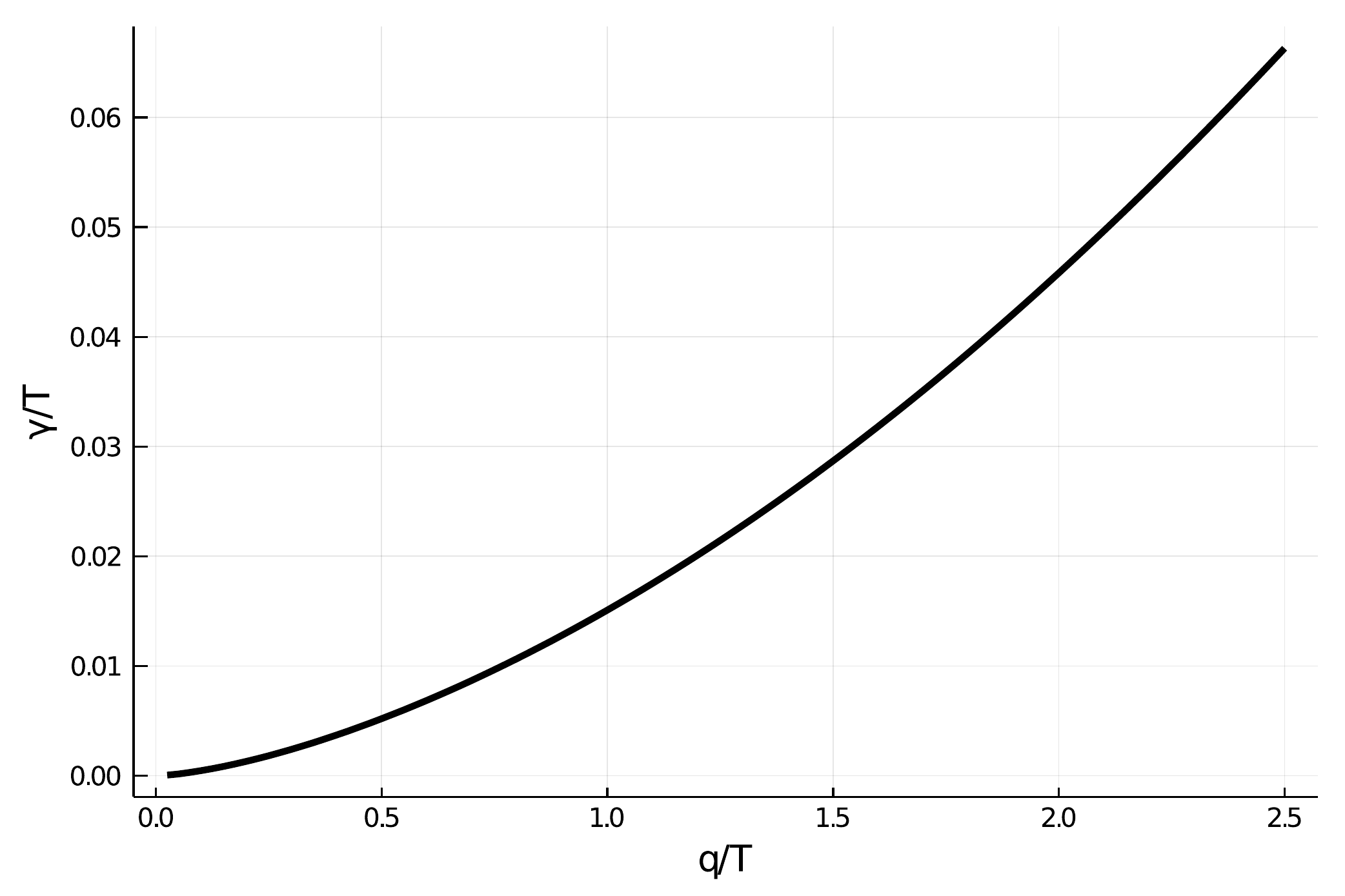}
		\subcaption{decay rate}
		\label{fig:decayrate}
	\end{subfigure}
	\caption{Numerical solution of the Dyson equation (\ref{eq:plasmonenergyequation}) at $\mu/T=3$  Here we choose a relatively small value of the fine structure constant $\alpha = 0.3$.}
	\label{}
\end{figure}
\noindent It is worthwhile pointing out that Eq.~\eqref{eq:plasmondispersionrpa} agrees with the plasmon dispersion from the beyond hydrodynamic treatment in Eq.~\eqref{eq:plasmondispersionhydro} in the collisionless regime.
 In Fig.~\ref{fig:plasmondispersionmu3alpha03} and  Fig.~\ref{fig:decayrate}, we show the dispersion relation and the decay rate of the plasmon for a  relatively small fine structure constant of $\alpha=0.3$, solved numerically from the Dyson equation in Eq.~\eqref{eq:plasmonenergyequation}. The dispersion follows a square-root relation in the small momentum limit, as in the approximate solution. In that limit, the plasmon becomes one of the relevant quasi-particles for the interacting Dirac electron since its decay rate is parametrically small. Fig.~\ref{fig:decayrate} shows that the decay rate as obtained from Eq.~\eqref{eq:plasmonlifetime} is much smaller than the plasmon energy, see Fig.~\ref{fig:plasmondispersionmu3alpha03}. Additionally, we plot the spectral functions of the plasmons at non-zero chemical potential in the $q\omega$-plane in Fig.~\ref{fig:spectralfunctionomegamomentumplane}. It shows that the spectral function is pronounced in the small momentum region. Therefore, it is possible to treat the plasmon as a proper quasi-particle emerging from the interacting Dirac electron gas. 
 This also allows to define a cutoff. It follows from the condition that at the momentum cutoff $p_c$, we have $\frac{p_c}{\omega_p(\vec{p})} = 1$ which invalidates the quasi-particle picture. As a result, we find $p_c= \frac{N}{2}\alpha T \log(2+2\cosh\mu/T)$.
 It is interesting to note that the combination of the linear dispersion relation for electrons and the plasmon dispersion kinematically allows a plasmon to decay into two electrons and hence contribute to its lifetime. Moreover, we observe that the plasmon decay rate decreases significantly and therefore our quasi-particle assumption is more accurate as the electron density increases. This can be clearly seen in Fig.(\ref{fig:spectralfunctionatp2.5alpha03}). We show the spectral function of the plasmons at various electron chemical potential ($\mu/T$). We find that the spectral function at a high doping away from the Dirac point manifests a narrow spike shape resemble the Dirac delta function.

\subsection{Coupled kinetic equations}\label{subsec:kineticplasmons}

	The starting point of our discussion is the gradient expanded version of the plasmon Keldysh equation, Eq.~\eqref{eq:Keldyshboson}. We have estimated the real part of the polarization in the long-wavelength limit given by Eq.~\eqref{realpartpolarizationfunction}. Using it here leads to
\begin{equation}
	D^{-1}_0-\Re\Pi^R
	= \frac{p}{\pi\alpha}- \frac{2Np^2}{4\pi\omega^2}\mathcal{N} = 	\frac{p}{\pi\alpha\omega^2}\left(\omega^2-\omega^2_p(\vec{p})\right).
	\label{eq:plasmoninverseGreenfunction}
\end{equation}
As a result, the derivatives appearing on the left-hand side of the above Keldysh equation are obtained as follows
\begin{eqnarray} 
	\partial_\omega \left(D^{-1}_0-\Re\Pi^R\right)&=&\frac{Np^2}{\pi \omega^3}\mathcal{N},\nonumber\\
	\partial_{\vec{p}} \left(D^{-1}_0-\Re\Pi^R\right)&=&\frac{\hat{p}}{\pi\alpha}-\frac{Np \hat{p}}{\pi\omega^2}\mathcal{N},\nonumber\\
	\partial_{\vec{x}} \left(D^{-1}_0-\Re\Pi^R\right)&=&-\frac{Np^2}{2\pi\omega^2}\partial_{\vec{x}}\mathcal{N}.
\end{eqnarray}
	We substitute these expressions into Eq.~\eqref{eq:Keldyshboson}. Next, we divide the resulting equation by the spectral weight that is $\partial_\omega(D_0^{-1}-\Re\Pi^R)$ and evaluate the resulting equation on-shell at the frequency $\omega=\omega_p(\vec{p})$. We assume that the excitations are long-lived and therefore the spectral function is sharply peaked at the energy dispersion $\omega_p(\vec{p})$. Finally,  we obtain the Boltzmann equation for the plasmons according to
	\begin{widetext}
		\begin{eqnarray}\label{eq:boltzmannequationboson}
			&& \partial_{t} b(\vec{x},\vec{p},t) + \frac{\omega_p(\vec{p})}{2p}\hat{p}\cdot \partial_{\vec{x}} b(\vec{x},\vec{p},t) -\frac{\omega_p(\vec{p})}{2\mathcal{N}}\partial_{\vec{x}}\mathcal{N} \cdot \partial_{\vec{p}} b(\vec{x},\vec{p},t)  \nonumber\\&&\hspace{4cm}= -\frac{2\alpha \pi^2 N \omega_p(\vec{p})}{p} \int \frac{d\vec{q}}{(2\pi)^2}  \mathcal{F}_{\lambda\lambda'}(\vec{p}+\vec{q},\vec{q})\delta(\omega_p(\vec{p})+\epsilon_\lambda(\vec{q})-\epsilon_{\lambda'}(\vec{p}+\vec{q})) \nonumber\\&&\hspace{4.5cm} \Big[f_\lambda(\vec{q})\Big(1-f_{\lambda'}(\vec{p}+\vec{q})\Big)b(\vec{p})  - \Big(1-f_{\lambda}(\vec{q})\Big)f_{\lambda'}(\vec{p}+\vec{q})\Big(1+b(\vec{p})\Big) \Big].\nonumber\\
		\end{eqnarray}
	\end{widetext}
	The left-hand side of the equation describes the changes of the distribution function by the streaming of the plasmon distribution with the group velocity $\vec{v}_p = \omega_p(\vec{p})\hat{p}/2p $. This is consistently  identical to calculating the group velocity from taking a derivative of the plasmon energy with respect to its momentum, {\it i.e.}, $\vec{v}_p= \partial_{\vec{p}}\omega_p(\vec{p})$. The fluctuations of the underlying electron density have an effect on the plasmon dispersion and enter as a force given by $\vec{F}=-\frac{\omega_{\vec{p}}\partial_{\vec{x}}\mathcal{N}}{2\mathcal{N}}$. 
	Next, let us examine how the plasmon and the electrons coexist in the system. To this end, we consider the Keldysh equation of the fermions given by Eq.~\eqref{eq:fermiondysonequation}. We proceed in exactly the same steps as in the previous section to arrive at the kinetic equation for the fermions in Eq.~\eqref{eq:Keldyshfermion}. Below, we only show the final result and give its full derivation in Appendix~\ref{Appendix:Derivationof fermionequationRPA}. We find that 
\begin{widetext}
	\begin{eqnarray}
	&&\partial_t f_\lambda(\vec{x},t,\vec{p}) +\left(\lambda v_F \hat{p}+\lambda\partial_{\vec{p}}\Re\sigma^R(\vec{x},t,\vec{p})\right)\cdot \partial_{\vec{x}}f_\lambda(\vec{x},t,\vec{p})-\;\partial_{\vec{x}}\left(V^H(\vec{x},t)+\lambda\Re\sigma^R(\vec{x},t,\vec{p})\right)\cdot \partial_{\vec{p}} f_\lambda(\vec{x},t,\vec{p})
	\nonumber\\&&	\hspace{3cm}= -2\alpha \pi^2 \int \frac{d\vec{q}}{(2\pi)^2} \frac{  \omega_p(\vec{q})}{q}  \mathcal{F}_{\lambda\lambda'}(\vec{p}+\vec{q},\vec{q})\delta(\omega_p(\vec{q})+\epsilon_\lambda(\vec{p})-\epsilon_{\lambda'}(\vec{p}+\vec{q})) \nonumber\\&&\hspace{4.5cm} \Big[f_\lambda(\vec{p})\Big(1-f_{\lambda'}(\vec{p}+\vec{q})\Big)b(\vec{q})  - \Big(1-f_{\lambda}(\vec{p})\Big)f_{\lambda'}(\vec{p}+\vec{q})\Big(1+b(\vec{q})\Big) \Big]\nonumber\\ && \hspace{3.3cm}-2\alpha \pi^2 \int \frac{d\vec{q}}{(2\pi)^2} \frac{  \omega_p(\vec{q})}{q}\mathcal{F}_{\lambda\lambda'}(\vec{p}-\vec{q},\vec{q})\delta(-\omega_p(\vec{q})+\epsilon_\lambda(\vec{p})-\epsilon_{\lambda'}(\vec{p}-\vec{q})) \nonumber\\&&\hspace{4.5cm} \Big[f_\lambda(\vec{p})\Big(1-f_{\lambda'}(\vec{p}-\vec{q})\Big)\Big(1+b(\vec{q})\Big)  - \Big(1-f_{\lambda}(\vec{p})\Big)f_{\lambda'}(\vec{p}-\vec{q})b(\vec{q}) \Big].
	\label{eq:fermionboltzmannequationRPA}
\end{eqnarray}
\end{widetext}
 In this expression, $V^H(\vec{x},t)$ is the Hartree potential as defined in Eq.~\eqref{eq:Hartreepotential} and  
\begin{eqnarray}
\Re\sigma^R\approx\frac{\pi\alpha}{2p}\int\frac{d\vec{q}}{(2\pi)^2}\frac{q}{\omega_p(\vec{q})}\left(1+2b(\vec{q})\right) \sin^2\theta
\end{eqnarray}
is the correction to the fermion energy resulting from the electron-plasmon interactions in the GW approximation. We evaluate this self-energy using the long-wavelength approximation of the plasmon dispersion, Eq.~\eqref{eq:plasmondispersionrpa}. We define the angle $\theta$ between $\vec{p}$ and $\vec{q}$. Details of the calculation are presented in Appendix (\ref{Appendix:Derivationof fermionequationRPA}). The collision term comes from the Fock-like diagram shown in Fig.~\ref{fig:fermionselfenergy}.   
It is a sum of two terms. The first term describes a scattering process of an electron from the momentum state $\vec{p}$ into another momentum state $\vec{p}+\vec{q}\;$ by absorbing a plasmon of momentum $\vec{q}$. The second term describes an emission of a plasmon of momentum $\vec{q}$ from an electron of momentum $\vec{p}$ and as a result the electron scatters into the momentum state $\vec{p}-\vec{q}$. We need to write these two terms of the collision integral separately because from the perspective of the electron in $\vec{k}$, the two events, the emission and absorption of a plasmon are essentially different. The coupled system of Boltzmann equations in Eq.~\eqref{eq:boltzmannequationboson} and Eq.\eqref{eq:fermionboltzmannequationRPA} constitutes one of the central results of this paper. 

\subsection{Conservation laws}	\label{subsec:conservationlawsrpa}

In this section we check whether our level of approximation indeed respects all the conservation laws. Compared to the weak-coupling consideration based on Eq.~\eqref{eq:fermionboltzmannequation} we now have two coupled Boltzmann equations, Eq.~\eqref{eq:boltzmannequationboson} and Eq.~\eqref{eq:fermionboltzmannequationRPA}. Let us denote the collision integrals, {\it i.e.}, the right-hand side of the Boltzmann equations of Eq.~\eqref{eq:boltzmannequationboson} and Eq.~\eqref{eq:fermionboltzmannequationRPA} by  $C^b[f,b](\vec{p})$ and 
$C^f_{\lambda}[f,b](\vec{p})$, respectively. The collision integrals again have three collisional invariants that correspond to the conservation of electric charge, momentum, and energy.  In that order, they read
\begin{equation}
	\sum_{\lambda=\pm}\int\frac{d\vec{p}}{(2\pi)^2} \; C^f_\lambda[f,b](\vec{p}) = 0\;,
\end{equation}
\begin{equation}
	\sum_{\lambda=\pm}\int\frac{d\vec{p}}{(2\pi)^2} \; \vec{p} \;C^f_\lambda[f,b](\vec{p})+\int\frac{d\vec{p}}{(2\pi)^2} \; \vec{p}\; C^b[f,b](\vec{p}) = 0\;,
\end{equation}
and 
\begin{eqnarray}
	&&\sum_{\lambda=\pm}\int\frac{d\vec{p}}{(2\pi)^2} \; \epsilon_\lambda(\vec{p}) C^f_\lambda[f,b](\vec{p})\nonumber\\&&\hspace{1cm}+\int\frac{d\vec{p}}{(2\pi)^2} \; \omega_p(\vec{p}) C^b_\lambda[f,b](\vec{p})  = 0\;,
\end{eqnarray}
and these statements can be checked in a straightforward manner. 
By integrating Eq.~\eqref{eq:fermionboltzmannequationRPA} over all momenta $\vec{p}$ and then summing over the energy bands $\pm$ we obtain the continuity equation of charge
\begin{equation}
	\partial_t n(\vec{x},t) + \partial_{\vec{x}} \cdot \vec{j}(\vec{x},t) = 0,
\end{equation}
where the total charge density is given by
\begin{equation}
	n(\vec{x},t) = N\int \frac{d\vec{p}}{(2\pi)^2} \; \left[ f_+(\vec{x},t,\vec{p}) + (f_-(\vec{x},t,\vec{p})-1)\right],
\end{equation}
whereas the total charge current density reads
\begin{eqnarray}
	\vec{j}(\vec{x},t) &=& N\int \frac{d\vec{p}}{(2\pi)^2} \; (v_F \hat{p}+\partial_{\vec{p}}\Re\sigma^R_{+})\nonumber\\&&\hspace{1cm}\left( f_+(\vec{x},t,\vec{p}) - (f_-(\vec{x},t,\vec{p})-1)\right).\nonumber\\
\end{eqnarray}
In writing the above expression, we again subtract the infinite contribution from the Dirac sea and define the distribution function of holes as $f_-(\vec{x},t)-1$. This is allowed since subtracting this infinite constant does not affect the conservation law.

In addition, we multiply Eq.~\eqref{eq:boltzmannequationboson} and Eq.~\eqref{eq:fermionboltzmannequationRPA} by momentum $\vec{p}$ and then integrate the resulting equations over all momentum $\vec{p}$. We add them together and find the law of momentum conservation as
\begin{eqnarray}
&&	\partial_{t} \vec{n}^{\vec{p}}(\vec{x},t) + \partial_{\vec{x}}\cdot\vec{\vec{\Pi}}(\vec{x},t) =\nonumber\\&& - \partial_{\vec{x}}V^H(\vec{x},t) n(\vec{x},t)-\int\frac{d\vec{p}}{(2\pi)^2} \frac{\omega_p(\vec{p})}{2}\frac{\partial_{\vec{x}}\mathcal{N}}{\mathcal{N}}b(\vec{x},t,\vec{p})\nonumber\\&&-N\int\frac{d\vec{p}}{(2\pi)^2} \partial_{\vec{x}} \Re\sigma^R(\vec{x},t,\vec{p})\left( f_+(\vec{x},t,\vec{p}) + (1-f_-(\vec{x},t,\vec{p}))\right)
. \nonumber\\
\end{eqnarray}
The right-hand side of the equation has three terms. The first term is an internal force due to the Hartree potential from the other electrons. The second term describes a force on a plasmon due to electron inhomogeneity. It is proportional to a gradient of the electron distribution function via  $\partial_{\vec{x}}\mathcal{N}$ where $\mathcal{N} = \int dq\Big[f_+(\vec{x},t,\vec{q})+(1-f_-(\vec{x},t,\vec{q}))\Big]$. The third term  is a reaction on the force in the second term. 
The total momentum density is
\begin{eqnarray}
	\vec{n}^{\vec{p}}(\vec{x},t) &=& N\int \frac{d\vec{p}}{(2\pi)^2} \vec{p}\Big[f_+(\vec{x},t,\vec{p}) - (1-f_-(\vec{x},t,\vec{p}))\Big]\nonumber\\&&+\int\frac{d\vec{p}}{(2\pi)^2}\;\vec{p} \; b(\vec{x},t,\vec{p}),
\end{eqnarray}
and the total momentum flux is
\begin{eqnarray}
	\vec{\vec{\Pi}}(\vec{x},t)&=&N\int \frac{d\vec{p}}{(2\pi)^2} \; (v_F \hat{p}+\partial_{\vec{p}}\Re\sigma) \vec{p}\nonumber\\&&\hspace{1cm}\left( f_+(\vec{x},t,\vec{p}) + (1-f_-(\vec{x},t,\vec{p}))\right)\nonumber\\&&+\int\frac{d\vec{p}}{(2\pi)^2}\vec{v}_p \; \vec{p}\; b(\vec{x},t,\vec{p}).
\end{eqnarray}
Here $\vec{v}_p = \omega_p(\vec{p})\hat{p}/2p $ defines the group velocity of a plasmon. We find that the total momentum density is not locally conserved but changed by the internal electric forces on the right-hand side of the equation. However, integrating over all space ${\vec{x}}$, the force terms vanish and the Hartree potential cancels by virtue of being a total derivative, 
\begin{equation}
	\int d\vec{x} \partial_{\vec{x}}V^H(\vec{x},t) n(\vec{x},t) = 0.
\end{equation}
In contrast, the other two forces do not vanish individually, but instead cancel each other as
\begin{eqnarray}
	&&-\int d\vec{x}\;\int\frac{d\vec{p}}{(2\pi)^2} \frac{\omega_p(\vec{p})}{2}\frac{\partial_{\vec{x}}\mathcal{N}}{\mathcal{N}}b(\vec{x},t,\vec{p}) \nonumber\\&&-N\int d\vec{x}\int\frac{d\vec{p}}{(2\pi)^2} \partial_{\vec{x}} \Re\sigma^R(\vec{x},t,\vec{p})\Big( f_+(\vec{x},t,\vec{p}) \nonumber\\&&\hspace{3cm}+ (1-f_-(\vec{x},t,\vec{p}))\Big) = 0.
\end{eqnarray}
On general grounds, this is a result of the Kadanoff-Baym conditions for approximate Green function to maintain the macroscopic conservation laws~\cite{KadanoffBaym1961}. As we discussed earlier, a self-energy included in the approximate Green function must be generated from a diagram of a free energy functional and the contributions from those self-energies generated from the same free-energy diagram will cancel out to ensure the conservation laws.
 We find the conservation of total momentum of the electrons
\begin{equation}
	\partial_t \vec{P} = 0\;,
\end{equation}
where the total momemtum of the whole system
\begin{equation}
	\vec{P} = \int d\vec{x}\; \vec{n}^{\vec{p}}(\vec{x},t).
\end{equation}	
\noindent Next, we multiply the electron Boltzmann equation of Eq.~\eqref{eq:boltzmannequationboson} by the energy  $\epsilon_{\lambda}(\vec{x},t,\vec{p}) = \lambda p + V^H(\vec{x},t) + \lambda\Re\sigma^R(\vec{x},t,\vec{p})$, integrate the resulting equation over all momentum $\vec{p}$ and then sum over the energy bands $\pm$. Similarly, we multiply the plasmon Boltzmann equation of Eq.~\eqref{eq:boltzmannequationboson} by its energy dispersion given by $\omega(\vec{x},\vec{p},t) = \sqrt{\frac{N}{2}\alpha p \mathcal{N}(\vec{x},t)}$ and integrate over the momentum. We add the resulting equations together, and find the conservation law of energy for the total system.
\begin{equation}
	\partial_t n^\epsilon(\vec{x},t)+\partial_{\vec{x}}\cdot\vec{j}^\epsilon(\vec{x},t) = 0,
\end{equation}
where the total energy density is
\begin{eqnarray}
	&&n^\epsilon(\vec{x},t)=N\int\frac{d\vec{p}}{(2\pi)^2}\;  \Big[\epsilon_+(\vec{x},t,\vec{p})f_+(\vec{x},t,\vec{p})\nonumber\\&&-\epsilon_-(\vec{x},t,\vec{p})(1-f_-(\vec{x},t,\vec{p}))\Big]+\int\frac{d\vec{p}}{(2\pi)^2}\omega(\vec{x},t,\vec{p})b(\vec{x},t,\vec{p}),\nonumber\\
\end{eqnarray}
and the total energy current density is
\begin{eqnarray}
	\vec{j}^\epsilon(\vec{x},t)&=& N\int\frac{d\vec{p}}{(2\pi)^2}\;  \Big[\vec{v}_+\epsilon_+(\vec{x},t,\vec{p})f_+(\vec{x},t,\vec{p})\nonumber\\&&-\vec{v}_-(\vec{x},t,\vec{p})\epsilon_-(\vec{x},t,\vec{p})(1-f_-(\vec{x},t,\vec{p}))\Big] \nonumber\\&&+ \int \frac{d\vec{p}}{(2\pi)^2} \vec{v}_p \omega_p b(\vec{x},t,\vec{p}).
\end{eqnarray}

\section{Conclusion and Outlook}\label{sec:conclusionandoutlook} 

In this work we have studied the basic equations of hydrodynamics in ultraclean interacting two-dimensional Dirac systems. Our approach was based on non-equilibrium quantum field theory. We first derived the hydrodynamic equations in weakly interacting systems, based on low-order perturbation theory. This allows to recover mostly known literature expressions. In the second part we go beyond a weak coupling analysis. We use the random-phase approximation which naturally leads to the notion of a coupled field theory of electrons, holes, and plasmons. Contrary to in three dimensional metals, the emerging plasmons constitute proper low-energy degrees of freedom without an excitation gap. Furthermore, these plasmons are stable and do not decay easily. Based on this, we study a set of coupled Boltzmann equations. We explicitly establish in that framework, that the approach provides a consistent conserving approximation which respects the conservation of electrical charge, momentum, and energy. Our main findings are that, compared to weak-coupling theories, there are direct low-energy contributions of the plasmons to the heat current and the energy-momentum tensor that have to be treated on equal footing with electronic excitations. In a companion paper we show that this implies that they should be measurable in transport experiments in encapsulated graphene devices that achieve the hydrodynamic regime. 
While we do not expect a similarl effect in three-dimensional metals, we expect an enhancement close to the Dirac point of three dimensional Dirac and Weyl systems or in bilayer graphene systems. 
A study of the thermo-electric transport properties of this theory has appeared recently~\cite{Kitinan2020}. There also is part I of this paper that studies the most salient features of the coupled theory on a phenomenological level.


\section{Acknowledgments}
 We acknowledge former collaborations and discussions with R.A. Duine, M. Fremling, P.M. Gunnink, J.S. Harms, E.I. Kiselev, A. Lucas, J. Lux, A. Mitchell, M. M\"uller, A. Rosch, S. Sachdev, J. Schmalian, M. Vojta, G. Wagner, and J. Waissman. KP thanks the Institute for the Promotion of Teaching Science and Technology (IPST) of Thailand for a Ph.D. fellowship. TL thanks the Dutch Research Council (NWO) for (partly) financing this work as a part of the research programme Fluid Spintronics with project number 182.069. This work is part of the D-ITP consortium, a program of the Netherlands Organisation for Scientific Research (NWO) that is funded by the Dutch Ministry of Education, Culture and Science
	(OCW).


\pagebreak
\clearpage
\newpage
\widetext
\appendix

\section{Derivation of the collision integral of Eq.(\ref{eq:fermionboltzmannequation})}
\label{appendix:A}
The algebraic expression for the diagram in Fig.~\ref{fig:sigma2a} is given by 
\begin{eqnarray}
	-i\Sigma^{(2a)}_{ah}(\vec{x},\vec{x}';t,t')&=& -iN \int dt_1 dt_2d\vec{x}_1d\vec{x}_2 \; \gamma^{a'}_{ab} \gamma^{b'}_{cd} \gamma^{c'}_{ef} \gamma^{d'}_{gh}   D_{0,a'b'}(\vec{x},\vec{x}_1;t,t_1)   D_{0,\gamma\delta}(\vec{x}_2,\vec{x}';t_2,t') \nonumber\\&& G_{0,bg}(\vec{x},\vec{x}';t,t')  G_{0,de}(\vec{x}_1,\vec{x}_2;t_1,t_2)  G_{0,fc}(\vec{x}_2,\vec{x}_1;t_2,t_1). 
\end{eqnarray} After a Wigner transformation, its Keldysh component in the quasi-particle basis reads
\begin{eqnarray}\label{eq:sigmaK2a}
	(\sigma_{\lambda\lambda}^{(2a)})^{K}(\vec{x},\vec{k},t,\omega)&=&iN\int \frac{d\vec{k}_1}{(2\pi)^2}\frac{d\vec{q}}{(2\pi)^2} 2\pi\delta(\omega-\lambda_1\epsilon_{\vec{k}-\vec{q}}-\lambda_2\epsilon_{\vec{k}_1+\vec{q}}+\lambda_3\epsilon_{\vec{k}_1})|T_{\lambda \lambda_1\lambda_2\lambda_3}(\vec{k},\vec{k}_1,\vec{q})|^2\nonumber\\&& \Big(F_{\lambda_3}(\vec{x},t,\vec{k}_1)-F_{\lambda_2}(\vec{x},t,\vec{k}_1+\vec{q})+F_{\lambda_1}(\vec{x},t,\vec{k}-\vec{q})\Big[F_{\lambda_2}(\vec{x},t,\vec{k}_1+\vec{q})F_{\lambda_3}(\vec{x},t,\vec{k}_1)-1\Big]\Big).\nonumber\\
\end{eqnarray}
Here we assume that the distribution function within the quasi-particle approximation has no off-diagonal elements in the spinor space. Furthermore, we introduce the shorthand notation for the Coulomb interaction transition probability amplitude 
\begin{equation}
	T_{\lambda \lambda_1\lambda_2\lambda_3}(\vec{k},\vec{k}_1,\vec{q}) = \frac{V(\vec{q})}{2} M^{\lambda\lambda_1}_{\vec{k},\vec{k}-\vec{q}}M^{\lambda_2\lambda_3}_{\vec{k}_1,\vec{k}_1+\vec{q}}\;,
\end{equation}
where the coherence factor coming from the overlap of the wavefunction is defined according to
\begin{equation}
	M^{\lambda\lambda_1}_{\vec{k},\vec{k}_1}=\left(\mathcal{U}^\dagger_{\vec{k}}\mathcal{U}_{\vec{k}_1}\right)_{\lambda\lambda_1}\;.
\end{equation}
The retarded component of the self-energy is given by
\begin{eqnarray}\label{eq:sigmaR2a}
	2i\Im(\sigma_{\lambda}^{(2a)})^{R}(\vec{k},\omega)&=&iN\int \frac{d\vec{k}_1}{(2\pi)^2}\frac{d\vec{q}}{(2\pi)^2} 2\pi\delta(\omega-\lambda_1\epsilon_{\vec{k}-\vec{q}}-\lambda_2\epsilon_{\vec{k}_1+\vec{q}}+\lambda_3\epsilon_{\vec{k}_1})|T_{\lambda \lambda_1\lambda_2\lambda_3}(\vec{k},\vec{k}_1,\vec{q})|^2 \nonumber\\&& \Big(F_{\lambda_2}(\vec{x},t,\vec{k}_1+\vec{q})F_{\lambda_3}(\vec{x},t,\vec{k}_1)-1+F_{\lambda_1}(\vec{x},t,\vec{k}-\vec{q})\Big[F_{\lambda_3}(\vec{x},t,\vec{k}_1)-F_{\lambda_2}(\vec{x},t,\vec{k}_1+\vec{q})\Big]\Big).\nonumber\\
\end{eqnarray}

The algebraic expression for the diagram in Fig.~\ref{fig:sigma2b} is given by 
\begin{eqnarray}
	-i\Sigma^{(b)}_{ah}(\vec{x},\vec{x}';t,t') &=& i\int dt_1dt_2d\vec{x}_1d\vec{x}_2 \gamma^{a'}_{ab}\;\gamma^{b'}_{cd}\gamma^{c'}_{ef} \gamma^{d'}_{gh}  D_{0,a'c'}(\vec{x},\vec{x}_2;t,t_2)D_{0,b'd'}(\vec{x}_1,\vec{x}';t_1,t')\nonumber\\&& G_{0,bc}(\vec{x},\vec{x}_1;t,t_1)  G_{0,de}(\vec{x}_1,\vec{x}_2;t_1,t_2) G_{0,fg}(\vec{x}_2,\vec{x}';t_2,t').
\end{eqnarray}
After a Wigner transformation, the Keldysh component of the self-energy in the quasi-particle basis reads
\begin{eqnarray}\label{eq:sigmaK2b}
	(\sigma_{\lambda}^{(2b)})^{K}(\vec{x},\vec{k},t,\omega)&=&-i\int \frac{d\vec{k}_1}{(2\pi)^2}\frac{d\vec{q}}{(2\pi)^2} 2\pi\delta(\omega-\lambda_1\epsilon_{\vec{k}-\vec{q}}-\lambda_2\epsilon_{\vec{k}_1+\vec{q}}+\lambda_3\epsilon_{\vec{k}_1})T_{\lambda \lambda_1\lambda_3\lambda_2}(\vec{k},\vec{k}_1,\vec{q}) T^*_{\lambda \lambda_2\lambda_1\lambda_3}(\vec{k},\vec{k}_1,\vec{k}-\vec{q}-\vec{k}_1)\nonumber\\&&\Big(F_{\lambda_3}(\vec{x},t,\vec{k}_1)-F_{\lambda_2}(\vec{x},t,\vec{k}_1+\vec{q})+F_{\lambda_1}(\vec{x},t,\vec{k}-\vec{q})\Big[F_{\lambda_2}(\vec{x},t,\vec{k}_1+\vec{q})F_{\lambda_3}(\vec{x},t,\vec{k}_1)-1\Big]\Big),\nonumber\\
\end{eqnarray}
whereas the retarded component of the self-energy is given by
\begin{eqnarray}\label{eq:sigmaR2b}
	2i\Im(\sigma_{\lambda}^{(2b)})^{R}(\vec{k},\omega)&=&-i\int \frac{d\vec{k}_1}{(2\pi)^2}\frac{d\vec{q}}{(2\pi)^2} 2\pi\delta(\omega-\lambda_1\epsilon_{\vec{k}-\vec{q}}-\lambda_2\epsilon_{\vec{k}_1+\vec{q}}+\lambda_3\epsilon_{\vec{k}_1})T_{\lambda \lambda_1\lambda_3\lambda_2}(\vec{k},\vec{k}_1,\vec{q}) T^*_{\lambda \lambda_2\lambda_1\lambda_3}(\vec{k},\vec{k}_1,\vec{k}-\vec{q}-\vec{k}_1)\nonumber\\&&\Big(F_{\lambda_2}(\vec{x},t,\vec{k}_1+\vec{q})F_{\lambda_3}(\vec{x},t,\vec{k}_1)-1+F_{\lambda_1}(\vec{x},t,\vec{k}-\vec{q})\Big[F_{\lambda_3}(\vec{x},t,\vec{k}_1)-F_{\lambda_2}(\vec{x},t,\vec{k}_1+\vec{q})\Big]\Big).
\end{eqnarray}
The algebraic expression for the diagram Fig.~\ref{fig:sigma2c} is given by 
\begin{eqnarray}
	-i\Sigma^{(c)}_{ah}(\vec{x},\vec{x}';t,t') &=& i\int dt_1dt_2d\vec{x}_1d\vec{x}_2\; \gamma^{a'}_{ab} \gamma^{b'}_{cd} \gamma^{c'}_{ef} \gamma^{d'}_{gh} D_{0,a'd'}(\vec{x},\vec{x}',t,t')  D_{0,b'c'}(\vec{x}_1,\vec{x}_2;t_1,t_2) \nonumber\\ &&G_{0,bc}(\vec{x},\vec{x}_1;t,t_1) G_{0,de}(\vec{x}_1,\vec{x}_2;t_1,t_2) G_{0,fg}(\vec{x}_1,\vec{x}';t_2,t').
\end{eqnarray}
This contribution to the collision intergral is omitted because $2i\Im(\sigma_{\lambda}^{(2c)})^{R}$ and $(\sigma_{\lambda}^{(2c)})^{K}$ are both zero.
After combining these contributions from the diagram in Figs.~$2a$-$2c$ and substituting the distribution function to zero order in the Berry connection given by Eq.~\eqref{eq:distributionfunctionzeroorderinA}, we find the collision integral for electron-electron scattering within the Born approximation. Hereafter we for brevity suppress space and time variables in the distribution function since the collision integral is local in spacetime.
\begin{eqnarray}
	&&\sigma^K_{\lambda}(\vec{p}) - 2i\Im \sigma_{\lambda}^R(\vec{p})(1-2f_\lambda(\vec{p}))\\&&\hspace{1cm}= -8i\int \frac{d\vec{k}_1}{(2\pi)^2}\frac{d\vec{q}}{(2\pi)^2} 2\pi\delta(\omega-\lambda_1\epsilon_{\vec{k}-\vec{q}}-\lambda_2\epsilon_{\vec{k}_1+\vec{q}}+\lambda_3\epsilon_{\vec{k}_1})\nonumber\\&&\hspace{1cm}\Big[N|T_{\lambda \lambda_1\lambda_3\lambda_2}(\vec{k},\vec{k}_1,\vec{q})|^2-T_{\lambda \lambda_1\lambda_3\lambda_2}(\vec{k},\vec{k}_1,\vec{q}) T^*_{\lambda \lambda_2\lambda_1\lambda_3}(\vec{k},\vec{k}_1,\vec{k}-\vec{q}-\vec{k}_1)\Big]\nonumber\\&&\hspace{1cm}\Big[f_{\lambda}(\vec{k})f_{\lambda_3}(\vec{k}_1)(1-f_{\lambda_1}(\vec{k}-\vec{q}))(1-f_{\lambda_2}(\vec{k}_1+\vec{q}))-(1-f_{\lambda}(\vec{k}))(1-f_{\lambda_3}(\vec{k}_1))f_{\lambda_1}(\vec{k}-\vec{q})f_{\lambda_2}(\vec{k}_1+\vec{q})\Big].\nonumber\\
\end{eqnarray}
 We make connection to the Golden rule result by shifting the variables appropriately, This gives
\begin{eqnarray}
	&&\sigma^K_{\lambda}(\vec{p}) - 2i\Im \sigma_{\lambda}^R(\vec{p})(1-2f_\lambda(\vec{p}))= -4i\int \frac{d\vec{k}_1}{(2\pi)^2}\frac{d\vec{q}}{(2\pi)^2} 2\pi\delta(\omega-\lambda_1\epsilon_{\vec{k}-\vec{q}}-\lambda_2\epsilon_{\vec{k}_1+\vec{q}}+\lambda_3\epsilon_{\vec{k}_1})\nonumber\\&&\Big[|T_{\lambda\lambda_1\lambda_3\lambda_2}(\vec{k},\vec{k}_1,\vec{q})-T_{\lambda\lambda_2\lambda_1\lambda_3}(\vec{k},\vec{k}_1,\vec{k}-\vec{q}-\vec{k}_1)|^2+(N-1)\Big(|T_{\lambda\lambda_1\lambda_3\lambda_2}(\vec{k},\vec{k}_1,\vec{q})|^2+|T_{\lambda\lambda_2\lambda_1\lambda_3}(\vec{k},\vec{k}_1,\vec{k}-\vec{q}-\vec{k}_1)|^2\Big) \Big]\nonumber\\&&\Big[f_{\lambda}(\vec{k})f_{\lambda_3}(\vec{k}_1)(1-f_{\lambda_1}(\vec{k}-\vec{q}))(1-f_{\lambda_2}(\vec{k}_1+\vec{q}))-(1-f_{\lambda}(\vec{k}))(1-f_{\lambda_3}(\vec{k}_1))f_{\lambda_1}(\vec{k}-\vec{q})f_{\lambda_2}(\vec{k}_1+\vec{q})\Big].\nonumber\\
	\label{eq:fermioncollisionintegral}
\end{eqnarray}
We then multiply Eq.(\ref{eq:fermioncollisionintegral}) by the spectral function followed by an integration over the frequency. In the end, we obtain a coupled system of Boltzmann equations for electrons ($\lambda=+$) and holes ($\lambda=-$).
\section{Hydrodynamic variables}
\label{Appendix:hydrovariables}
The underlying assumption for the electron hydrodynamics is that inelastic electron-electron collisions occur much faster than momentum-relaxing scatterings of electrons against impurities and/or phonons. As a result, electrons establish the local equilibrium  and the corresponding distribution function can be written as
\begin{equation}
	f_\lambda(\vec{p}) = \frac{1}{\exp(\frac{\lambda v_Fp - \mu -\vec{u}\cdot\vec{p}}{T})+1}.
\end{equation}
We insert this distribution function into the charge density defined in Eq.~\eqref{eq:electrondensity}
\begin{eqnarray}
	n(\vec{x},t) &=& N\int \frac{d\vec{p}}{(2\pi)^2} \; \left[ f_+(\vec{x},t,\vec{p}) - (1-f_-(\vec{x},t,\vec{p}))\right] \nonumber\\&=&N\int \frac{d\vec{p}}{(2\pi)^2} \; \left[  \frac{1}{\exp(\frac{ v_Fp - \mu -\vec{u}\cdot\vec{p}}{T})+1} - \frac{1}{\exp(\frac{ v_Fp + \mu +\vec{u}\cdot\vec{p}}{T})+1}\right] \nonumber\\ &=& N\int \frac{d\vec{p}}{(2\pi)^2} \; \left[  \frac{1}{\exp(\frac{ v_Fp - \mu }{T})+1} - \frac{1}{\exp(\frac{ v_Fp + \mu }{T})+1}\right] + \mathcal{O}(u) \nonumber\\
	 &=& \frac{N}{2\pi}\int pdp \; \left[  \frac{1}{\exp(\frac{ v_Fp - \mu }{T})+1} - \frac{1}{\exp(\frac{ v_Fp + \mu }{T})+1}\right] + \mathcal{O}(u^2) \nonumber\\ &=& \frac{N T^2}{2\pi v_F^2}\left(-\text{Li}_2(-e^{\mu/T})+\text{Li}_2(-e^{-\mu/T})\right).
\end{eqnarray}
To arrive at the third line, we expand the distribution functions to linear order in $u$.
\begin{equation}
	 \frac{1}{\exp(\frac{ v_Fp - \mu -\vec{u}\cdot\vec{p}}{T})+1} \approx \frac{1}{\exp(\frac{ v_Fp - \mu}{T})+1} + \frac{\exp(\frac{v_F p -\mu}{T})}{\left(\exp(\frac{ v_Fp - \mu}{T})+1\right)^2} \frac{\vec{u}\cdot\vec{p}}{T} +\mathcal{O}(u^2).
\end{equation}
After performing the angular integral, the terms linear in $\vec{u}$ vanish. The remaining integrals can be written in terms of the polylogarithmic function by means of
\begin{equation}
	\int_0^{\infty} dx \frac{x^{n-1}}{e^{x-\mu}+1} = -\Gamma(n)\text{Li}_n(-e^{\mu}).
\end{equation}

Next, let us evaluate the charge current density defined in Eq.(\ref{eq:chargecurrent})
\begin{eqnarray}
	\vec{j}(\vec{x},t) &=&  N\int \frac{d\vec{p}}{(2\pi)^2} \; v_F \hat{p}\Big[ f_+(\vec{x},t,\vec{p})+ (1-f_-(\vec{x},t,\vec{p}))\Big]\nonumber\\
	&=&  N\int \frac{d\vec{p}}{(2\pi)^2} \; v_F \hat{p}\Big[ \frac{1}{\exp(\frac{ v_Fp - \mu}{T})+1} + \frac{\exp(\frac{v_F p -\mu}{T})}{\left(\exp(\frac{ v_Fp + \mu}{T})+1\right)^2} \frac{\vec{u}\cdot\vec{p}}{T} + \frac{1}{\exp(\frac{ v_Fp + \mu}{T})+1} - \frac{\exp(\frac{v_F p +\mu}{T})}{\left(\exp(\frac{ v_Fp + \mu}{T})+1\right)^2} \frac{\vec{u}\cdot\vec{p}}{T} \Big].\nonumber
\end{eqnarray}
The terms of zeroth order in $\vec{u}$ vanish after the angular integration. We observe that the charge current is parallel to the hydrodynamic velocity $\vec{u}$. Its component is given by
\begin{eqnarray}
	\vec{j}(\vec{x},t)=\vec{j}(\vec{x},t) \cdot \hat{u} \hat{u} &=& N\int \frac{d\vec{p}}{(2\pi)^2} \; v_F \hat{p}\cdot \hat{u}\hat{u}\Big[  \frac{\exp(\frac{v_F p -\mu}{T})}{\left(\exp(\frac{ v_Fp - \mu}{T})+1\right)^2} \frac{\vec{u}\cdot\vec{p}}{T}  - \frac{\exp(\frac{v_F p +\mu}{T})}{\left(\exp(\frac{ v_Fp + \mu}{T})+1\right)^2} \frac{\vec{u}\cdot\vec{p}}{T} \Big]\nonumber\\&=&
	\frac{Nv_Fu}{4\pi^2T}\int p^2dpd\theta \; \cos^2\theta \Big[  \frac{\exp(\frac{v_F p -\mu}{T})}{\left(\exp(\frac{ v_Fp - \mu}{T})+1\right)^2}   - \frac{\exp(\frac{v_F p +\mu}{T})}{\left(\exp(\frac{ v_Fp + \mu}{T})+1\right)^2}  \Big]\hat{u}\nonumber\\
	&=&
	\frac{Nv_F\vec{u}}{4\pi T}\int p^2dp \Big[  \frac{\exp(\frac{v_F p -\mu}{T})}{\left(\exp(\frac{ v_Fp - \mu}{T})+1\right)^2}   - \frac{\exp(\frac{v_F p +\mu}{T})}{\left(\exp(\frac{ v_Fp + \mu}{T})+1\right)^2}  \Big]\nonumber\\
	&=&
	\frac{NT^2\vec{u}}{4\pi v_F^2}\int x^2dx \Big[  \frac{\exp(x-\mu/T)}{\left(\exp(x-\mu/T)+1\right)^2}   - \frac{\exp(x+\mu/T)}{\left(\exp(x+\mu/T)+1\right)^2}  \Big]
	\nonumber\\
	&=&
	-\frac{NT^2\vec{u}}{4\pi v_F^2}\int x^2dx \frac{d}{dx}\Big[  \frac{1}{\exp(x-\mu/T)+1}   - \frac{1}{\exp(x+\mu/T)+1}  \Big]
	\nonumber\\
	&=&
	\frac{NT^2\vec{u}}{2\pi v_F^2}\int xdx \Big[  \frac{1}{\exp(x-\mu/T)+1}   - \frac{1}{\exp(x+\mu/T)+1}  \Big]
	\nonumber\\
	&=&
	\frac{NT^2\vec{u}}{2\pi v_F^2}\Big[  -\text{Li}_2(-e^{\mu/T})   +\text{Li}_2(-e^{-\mu/T})\Big]
		\nonumber\\
	&=&
	n(\vec{x},t)\vec{u}.
\end{eqnarray}  

The other quantities can be calculated in a similar way. We find that to linear order in $\vec{u}$ the energy density is given by
\begin{equation}
	n^{\epsilon}(\vec{x},t) = -\frac{NT^3}{2\pi v_F^2} \Gamma(3) \left[\text{Li}_3(-e^{\mu/T})+\text{Li}_3(-e^{-\mu/T})\right].
\end{equation}
The momentum flux defines the pressure by means of
\begin{equation}
	\Pi_{ij}(\vec{x},t) = -\frac{NT^3}{4\pi v_F^2}\Gamma(3)\left[\text{Li}_3(-e^{\mu/T})+\text{Li}_3(-e^{-\mu/T})\right]\delta_{ij}=\frac{n^\epsilon(\vec{x},t)}{2}\delta_{ij} \equiv P(\vec{x},t) \delta_{ij}. 
\end{equation}
The momentum density and the energy current read
\begin{equation}
	n^{\vec{p}}(\vec{x},t) = -\frac{NT^3\vec{u}}{4\pi v_F^4}\Gamma(4)\left[\text{Li}_3(-e^{\mu/T}+\text{Li}_3(-e^{-\mu/T}))\right] = \frac{n^\epsilon(\vec{x},t)+P(\vec{x},t)}{v_F^2}\vec{u},
\end{equation}
\begin{equation}
	\vec{j}^{\epsilon}(\vec{x},t) = -\frac{NT^3\vec{u}}{4\pi v_F^2}\Gamma(4)\left[\text{Li}_3(-e^{\mu/T})+\text{Li}_3(-e^{-\mu/T})\right] =v_F^2n^{\vec{p}}(\vec{x},t).
\end{equation} 
\section{Properties of real boson Green functions}
According to Eq.(\ref{eq:bosongrennfunctionstructure}), the Keldysh component of the boson Green function is in the first row and the first column. By making use of the reality of the fields, one conclude from the definition that
\begin{eqnarray}
	D^K(\vec{x}_1,\vec{x}_2,t_1,t_1) &=& \int \mathcal{D}\phi\; \phi_1(\vec{x}_1,t_1)\phi_1(\vec{x}_2,t_2)\exp(iS[\phi])\nonumber\\ &=& \int \mathcal{D}\phi\; \phi_1(\vec{x}_2,t_2)\phi_1(\vec{x}_1,t_1)\exp(iS[\phi]) \nonumber\\ &=& D^K(\vec{x}_2,\vec{x}_1,t_2,t_1).
\end{eqnarray}
Consider their Wigner transformation
\begin{equation}
	D^K(\vec{x}_1,\vec{x}_2,t_1,t_2) = \int \frac{d\vec{p}}{(2\pi)^2}\frac{d\omega}{2\pi} D^K(\vec{x},t,\vec{p},\omega)e^{i\vec{p}\cdot(\vec{x}_1-\vec{x}_2)-i\omega(t_1-t_2)},
\end{equation}
\begin{equation}
	D^K(\vec{x}_2,\vec{x}_1,t_2,t_1) = \int \frac{d\vec{p}}{(2\pi)^2}\frac{d\omega}{2\pi} D^K(\vec{x},t,\vec{p},\omega)e^{-i\vec{p}\cdot(\vec{x}_1-\vec{x}_2)+i\omega(t_1-t_2)},
\end{equation}
which implies that
\begin{equation}
	D^K(\vec{x},t,\vec{p},\omega) = 	D^K(\vec{x},t,-\vec{p},-\omega).
\end{equation}

We find that, in the long-wavelength limit, the plasmon retarded Green function read

\begin{equation}
	D^R(\vec{q},\nu)=\frac{\pi\alpha \nu^2}{q} \frac{1}{(\nu+i\delta)^2-\omega_p^2(\vec{q})}= \frac{\pi\alpha\nu}{2q}\left(\frac{1}{\nu+i\delta+\omega_p(\vec{q})}+\frac{1}{\nu+i\delta-\omega_p(\vec{q})}\right).
\end{equation}
As a result, the imaginary part is given by
\begin{equation}
	\Im D^R(\vec{q},\nu) = - \frac{\pi^2\alpha \nu}{2q}\left(\delta(\nu+\omega_p(\vec{q}))+\delta(\nu-\omega_p(\vec{q}))\right).
\end{equation}
which is an odd function of the frequency and momentum variables 
\begin{equation}
	\Im D^R(-\vec{q},-\nu)=-\Im D^R(\vec{q},\nu).
\end{equation}
It follows that
\begin{equation}
	B(-\vec{q},-\nu)=-B(\vec{q},\nu).
\end{equation}

\section{Derivation of the polarization function in the long-wavelength limit  }
\label{Appendix:longwavelengthapproximationplasmon}
In this appendix, we derive the polarization function in the long-wavelength limit presented in the main text. Our starting point is the Lindhard formula in Eq.~\eqref{retpol}. Let us first calculate the real part of the polarization. The main contribution is from the intraband transition, when $\lambda=\lambda'$. It reads
\begin{eqnarray}
	\Re\Pi^+_{R}(\vec{p},\omega) &\approx& N\sum_{\lambda=\pm1} \int\frac{d\vec{q}}{(2\pi)^2} 
	\frac{-\vec{p}\cdot\vec{\nabla}_{\vec{q}}f_\lambda(\vec{q})}{\omega}\Big[1+\frac{\vec{p}\cdot\vec{\nabla}_{\vec{q}}\epsilon_\lambda(\vec{q})}{\omega}\Big],\nonumber\\
	&=&\frac{N}{\omega^2} \int\frac{d\vec{q}}{(2\pi)^2} \Big[ 
	f_+(\vec{q})(\vec{p}\cdot\vec{\nabla}_{\vec{q}})^2\epsilon_+(\vec{q})-\left(1-f_-(\vec{q})\right)(\vec{p}\cdot\vec{\nabla}_{\vec{q}})^2\epsilon_-(\vec{q})\Big], \nonumber\\
	&=& \frac{Np^2}{\omega^2} \int\frac{qdqd\theta}{(2\pi)^2} \Big[ 
	f_+(\vec{q})+\left(1-f_-(\vec{q})\right)\Big]\frac{\sin^2\theta}{q}\nonumber\\&=& \frac{Np^2}{4\pi\omega^2}\int dq\Big[f_+(\vec{q})+(1-f_-(\vec{q}))\Big]\nonumber\\&=& -\frac{Np^2}{4\pi\omega^2}T\Big[\text{Li}_1\left(-e^{\mu/T}\right)+\text{Li}_1\left(-e^{-\mu/T}\right)\Big]\nonumber\\&=& \frac{Np^2}{4\pi\omega^2}T\Big[\log\left(1+e^{\mu/T}\right)+\log\left(1+e^{-\mu/T}\right)\Big]\nonumber\\&=&\frac{Np^2}{4\pi\omega^2}T\Big[\log\left(2+2\cosh\mu/T\right)\Big].
	\label{eq:realpol}
\end{eqnarray}

In contrast, the interband contribution, when $\lambda=-\lambda'$, gives a logarithmic correction which will be neglected in evaluating the plasmon energy dispersion. For the case of non-zero dopings, at zero temperature, this interband contribution reads
$\frac{Np^2}{16\pi\omega} \log \left(\Big|\frac{\omega-2\mu}{\omega+2\mu}\Big|\right)$.

Next, we consider the imaginary part of the polarization function. The main contribution to the imaginary part
is from the interband transition, when $\lambda=-\lambda'$. This gives
\begin{eqnarray}
	\Im\Pi^-_{R}(\vec{p},\omega) &\approx&-N \pi \sum_{\lambda=\pm1} \int\frac{d\vec{q}}{(2\pi)^2} \frac{1}{4}\left(\vec{p}\cdot\vec{\nabla}_{
		\vec{q}}\theta_{\vec{q}}\right)^2 \Big[ 
	f_\lambda(\vec{q})-f_{-\lambda}(\vec{q}) \Big]  \delta(\omega+\epsilon_\lambda(\vec{q})-\epsilon_{-\lambda}(\vec{q})),\nonumber\\
	&=&-N \pi  \int\frac{d\vec{q}}{(2\pi)^2}\frac{1}{4}\frac{p^2}{q^2}\cos^2(\theta) \Big[\Big(
	f_+(\vec{q})-f_{-}(\vec{q}) \Big) \delta(\omega+\epsilon_+(\vec{q})-\epsilon_{-}(\vec{q}))\\&&\hspace{5cm}+\Big(
	f_-(\vec{q})-f_{+}(\vec{q}) \Big) \delta(\omega+\epsilon_-(\vec{q})-\epsilon_{+}(\vec{q}))\Big],\nonumber\\
	&=&-\frac{N}{16}p^2 \int \frac{dq}{q} \Big[\Big(
	f_+(\vec{q})-f_{-}(\vec{q}) \Big) \delta(\omega+2q)+\Big(
	f_-(\vec{q})-f_{+}(\vec{q}) \Big) \delta(\omega-2q)\Big],
	\nonumber\\
	&=&-\frac{N}{16}p^2 \int \frac{dq}{q} \Big(
	f_+(\vec{q})-f_{-}(\vec{q}) \Big) \Big(  \delta(\omega+2q)- \delta(\omega-2q)\Big),\nonumber\\
	&=&-\frac{N}{16}p^2 \int \frac{dq}{q} \Big(
	f_+(\vec{q})-f_{-}(\vec{q}) \Big) \frac{1}{2} \Big( \Theta(-\omega)\delta(q+\omega/2)- \Theta(\omega)\delta(q-\omega/2)\Big),\nonumber\\
	&=&-\frac{N}{16}p^2 \int \frac{dq}{q} \Big(
	f_+(\vec{q})-f_{-}(\vec{q}) \Big) \frac{1}{2} \Big( \Theta(-\omega)\delta(q+\omega/2)- \Theta(\omega)\delta(q-\omega/2)\Big)
	\\&=& \frac{N}{16}\frac{q^2}{\omega} \left(\frac{1}{e^{\frac{|\omega|/2-\mu}{T}}+1}-\frac{1}{e^{\frac{-|\omega|/2-\mu}{T}}+1}\right)\;.
	\label{eq:impol}
\end{eqnarray}
In the limit of zero temperature, this becomes
\begin{equation}
	\Im\Pi^{R}(\vec{q},\omega) \approx -\frac{N}{16}\frac{q^2}{\omega} \Theta(|\omega|-2|\mu|),
\end{equation}
which was found previously in \cite{Wunsch2006}. It vanishes when $|\omega|<2|\mu|$, consequently, the long-lived plasmon mode exists in this region.
By substituting the real part in Eq.(\ref{eq:realpol}) and the imaginary part in Eq.(\ref{eq:impol}) into Eq.(\ref{eq:plasmonlifetime}), we find the decay of plasmon. It reads
\begin{equation}
	\gamma_{p}(\vec{q}) = -\frac{\pi\omega_p(\vec{q})^2}{8T\log(2+2\cosh\mu/T)}\left(\frac{1}{e^{\frac{|\omega_p(\vec{q})|/2-\mu}{T}}+1}-\frac{1}{e^{\frac{-|\omega_p(\vec{q})|/2-\mu}{T}}+1}\right).
\end{equation}

\section{Derivaltion of Eq.(\ref{eq:fermionboltzmannequationRPA})}
\label{Appendix:Derivationof fermionequationRPA}
 The GW diagram in Fig.(\ref{fig:fermionselfenergy}) is interpreted into the expression in Eq.(\ref{eq:fermionselfenergyfreefermionRPA}). From that, we obtain the retarded and Keldysh components.
\begin{eqnarray}
	\Sigma^{R}(\vec{x},\vec{x}',t,t')&=&i\Big[D^{K}(\vec{x},\vec{x}',t,t')G^{R}_{\phi}(\vec{x},\vec{x}',t,t')+D^{R}(\vec{x},\vec{x}',t,t')G_{\phi}^K(\vec{x},\vec{x}',t,t')\Big],\nonumber\\
\end{eqnarray}
and
\begin{eqnarray}
	\Sigma^{K}(\vec{x},\vec{x}',t,t')&=&i\Big[D^{K}(\vec{x},\vec{x}',t,t')G_\phi^{K}(\vec{x},\vec{x}',t,t')+\left(D^{R}(\vec{x},\vec{x}',t,t')-D^{A}(\vec{x},\vec{x}',t,t')\right)\left(G_\phi^{R}(\vec{x},\vec{x}',t,t')-G_\phi^{A}(\vec{x},\vec{x}',t,t')\right)\Big].\nonumber\\
\end{eqnarray}
In writing down the Keldysh component of the self-energy above, we use again the fact that a product of retarded and advanced Green functions vanishes. We proceed with a Wigner transformation followed by transforming the self-energy into the quasi-particle basis. Within such a basis, the off-diagonal elements are irrelevant. This gives
\begin{eqnarray}
		\sigma^R_{\lambda\lambda}(\vec{p},\omega)&=&\Big[\mathcal{U}_{\vec{p}}\Sigma^{R}(\vec{p},\omega)\mathcal{U}^\dagger_{\vec{p}}\Big]_{\lambda\lambda}\nonumber\\&=&i\int \frac{d\vec{q}}{(2\pi)^2}\frac{d\nu}{2\pi}\mathcal{F}_{\lambda\lambda'}(\vec{p}-\vec{q},\vec{p})\Big[D^{K}(\vec{q},\nu)g_{\phi,\lambda'}^{R}(\vec{p}-\vec{q},\omega-\nu)+D^{R}(\vec{q},\nu)g_{\phi,\lambda'}^{K}(\vec{p}-\vec{q},\omega-\nu)\Big],\nonumber\\
		\sigma^K_{\lambda\lambda}(\vec{p},\omega)&=&\Big[\mathcal{U}_{\vec{p}}\Sigma^{K}(\vec{p},\omega)\mathcal{U}^\dagger_{\vec{p}}\Big]_{\lambda\lambda}
		\nonumber\\&=&-4i\int \frac{d\vec{q}}{(2\pi)^2}\frac{d\nu}{2\pi}\mathcal{F}_{\lambda\lambda'}(\vec{p}-\vec{q},\vec{p})\Im D^{R}(\vec{q},\nu)  \Im g^R_ {\phi,\lambda'}(\vec{p}-\vec{q},\omega-\nu)\Big[\Big(1+2b(\vec{q},\nu)\Big)\Big(1-2f_{\lambda'}(\vec{p}-\vec{q},\omega-\nu)\Big)+1\Big].\nonumber\\
\end{eqnarray}
where the retarded component of the plasmon Green 's function, $D^R$, is determined by an inverse of Eq.(\ref{eq:plasmoninverseGreenfunction}).
\begin{equation}
	D^R(\vec{q},\nu)=\frac{\pi\alpha \nu^2}{q} \frac{1}{(\nu+i\delta)^2-\omega_p^2(\vec{q})}.
\end{equation}
The Keldysh component is calculated by the Wigner transform of Eq.(\ref{eq:KeldyshbosonGreenfunction}).
\begin{equation}
	D^K(\vec{q},\nu) = 2i\Im D^R(\vec{q},\nu)(1+2b(\vec{q})).
\end{equation}
By substituting the self-energy into the right-hand side of the Keldysh equation, we obtain the collision terms in Eq.(\ref{eq:fermionboltzmannequationRPA}).

Next, we consider the right-hand side of the equation on which there are two renormalization effects arises from the real part of the self-energy: (i) Fermi velocity renormalization given by $\lambda v_F \hat{p}+\partial_{\vec{p}}\Re\sigma^R_{\lambda\lambda}(\vec{x},t)$ and (ii) internal forces among  electrons themselves and from plasmons given by $-\partial_{\vec{x}}\Re\sigma^R_{\lambda\lambda}(\vec{x},t)$. We will evaluate the self-energy here using the same approximation as we use for plasmon. To this end, let us consider
\begin{eqnarray}
	\sigma^R_{\lambda\lambda}(\vec{p},\omega)&=&\Big[\mathcal{U}_{\vec{p}}\Sigma^{R}(\vec{p},\omega)\mathcal{U}^\dagger_{\vec{p}}\Big]_{\lambda\lambda}\nonumber\\&=&i\int \frac{d\vec{q}}{(2\pi)^2}\frac{d\nu}{2\pi}\mathcal{F}_{\lambda\lambda'}(\vec{p}-\vec{q},\vec{p})\Big[D^{K}(\vec{q},\nu)g_{\phi,\lambda'}^{R}(\vec{p}-\vec{q},\omega-\nu)+D^{R}(\vec{q},\nu)g_{\phi,\lambda'}^{K}(\vec{p}-\vec{q},\omega-\nu)\Big]
\end{eqnarray}
The real part of the self-energy is written as a sum of two terms.
\begin{equation}
	\Re\sigma^R_{\lambda\lambda}(\vec{p},\omega)= \sigma^{(1)}(\vec{p},\omega)+\sigma^{(2)}(\vec{p},\omega),
\end{equation}
where below we evaluate it on-shell at $\omega = \epsilon_\lambda(\vec{p})$.
\begin{eqnarray}
	\sigma^{(1)}(\vec{p},\epsilon_\lambda(\vec{p})) &=& 	\int \frac{d\vec{q}}{(2\pi)^2}\frac{d\nu}{2\pi}\mathcal{F}_{\lambda\lambda'}(\vec{p}-\vec{q},\vec{p})\Big[ \frac{\pi^2\alpha \nu}{q}\left(\delta(\nu+\omega_p(\vec{q}))+\delta(\nu-\omega_p(\vec{q}))\right)B(\vec{q},\nu)\frac{1}{\epsilon_\lambda(\vec{p})-\nu-\epsilon_{\lambda'}(\vec{p}-\vec{q})}\nonumber\\&=&	-\int \frac{d\vec{q}}{(2\pi)^2}\mathcal{F}_{\lambda\lambda'}(\vec{p}-\vec{q},\vec{p}) \frac{\pi\alpha \omega_p(\vec{q}) }{2q}B(\vec{q},-\omega_p(\vec{q}))\frac{1}{\epsilon_\lambda(\vec{p})+\omega_p(\vec{q})-\epsilon_{\lambda'}(\vec{p}-\vec{q})}\nonumber\\&&+\int \frac{d\vec{q}}{(2\pi)^2}\mathcal{F}_{\lambda\lambda'}(\vec{p}-\vec{q},\vec{p}) \frac{\pi\alpha \omega_p(\vec{q})}{2q}B(\vec{q},\omega_p(\vec{q}))\frac{1}{\epsilon_\lambda(\vec{p})-\omega_p(\vec{q})-\epsilon_{\lambda'}(\vec{p}-\vec{q})}
	\nonumber\\&=&	-\int \frac{d\vec{q}}{(2\pi)^2}\mathcal{F}_{\lambda\lambda'}(\vec{p}+\vec{q},\vec{p}) \frac{\pi\alpha \omega_p(\vec{q}) }{2q}B(-\vec{q},-\omega_p(\vec{q}))\frac{1}{\epsilon_\lambda(\vec{p})+\omega_p(\vec{q})-\epsilon_{\lambda'}(\vec{p}+\vec{q})}\nonumber\\&&+\int \frac{d\vec{q}}{(2\pi)^2}\mathcal{F}_{\lambda\lambda'}(\vec{p}-\vec{q},\vec{p}) \frac{\pi\alpha \omega_p(\vec{q})}{2q}B(\vec{q},\omega_p(\vec{q}))\frac{1}{\epsilon_\lambda(\vec{p})-\omega_p(\vec{q})-\epsilon_{\lambda'}(\vec{p}-\vec{q})}
	\nonumber\\&=&	\int \frac{d\vec{q}}{(2\pi)^2}\mathcal{F}_{\lambda\lambda'}(\vec{p}+\vec{q},\vec{p}) \frac{\pi\alpha \omega_p(\vec{q}) }{2q}B(\vec{q},\omega_p(\vec{q}))\frac{1}{\epsilon_\lambda(\vec{p})+\omega_p(\vec{q})-\epsilon_{\lambda'}(\vec{p}+\vec{q})}\nonumber\\&&+\int \frac{d\vec{q}}{(2\pi)^2}\mathcal{F}_{\lambda\lambda'}(\vec{p}-\vec{q},\vec{p}) \frac{\pi\alpha \omega_p(\vec{q})}{2q}B(\vec{q},\omega_p(\vec{q}))\frac{1}{\epsilon_\lambda(\vec{p})-\omega_p(\vec{q})-\epsilon_{\lambda'}(\vec{p}-\vec{q})}
\end{eqnarray}
Now we expand the denominators  in both terms and find that 
\begin{eqnarray}
	\sigma^{(1)}(\vec{p},\epsilon_\lambda(\vec{p})) &\approx& 	\int \frac{d\vec{q}}{(2\pi)^2} \frac{\pi\alpha \omega_p(\vec{q}) }{2q}B(\vec{q},\omega_p(\vec{q}))\frac{1}{\omega_p(\vec{q})}\left[1-\frac{\epsilon_\lambda(\vec{p})-\epsilon_{\lambda}(\vec{p}+\vec{q})}{\omega_p(\vec{q})}-1+\frac{-\epsilon_\lambda(\vec{p})+\epsilon_{\lambda}(\vec{p}-\vec{q})}{\omega_p(\vec{q})}\right]\nonumber\\&\approx&
	\int \frac{d\vec{q}}{(2\pi)^2} \frac{\pi\alpha \omega_p(\vec{q}) }{2q}B(\vec{q},\omega_p(\vec{q}))\frac{1}{\omega^2_p(\vec{q})}\left(\vec{q}\cdot\partial_{\vec{p}}\right)^2\epsilon_{\lambda}(\vec{p})\nonumber\\&=&\lambda\frac{\pi\alpha}{2p}\int\frac{d\vec{q}}{(2\pi)^2}\frac{q}{\omega_p(\vec{q})}B(\vec{q},\omega_p(\vec{q})) \sin^2\theta
\end{eqnarray}
where $\theta$ is an angle between $\vec{p}$ and $\vec{q}$. This is the result correct up to 
the lowest order in $q/\omega_p(\vec{q})$ where $q$ and $\omega_p(\vec{q})$ are plasmon momentum and energy.
\begin{equation}
	\sigma^{(2)}(\vec{p},\epsilon_\lambda(\vec{p}))=\int \frac{d\vec{q}}{(2\pi)^2}\frac{d\nu}{2\pi}\mathcal{F}_{\lambda\lambda'}(\vec{p}-\vec{q},\vec{p})\frac{\pi\alpha\nu}{2q}\left(\frac{1}{\nu+\omega_p(\vec{q})}+\frac{1}{\nu-\omega_p(\vec{q})}\right)\left(2\pi  \delta(\epsilon_{\lambda}(\vec{p})-\nu-\epsilon_{\lambda'}(\vec{p}-\vec{q}))(1-2f_{\lambda'}(\vec{p}-\vec{q}))\right)
\end{equation} 
The second term is of next order in $q/\omega_p(\vec{q})$, we therefore neglect it. 
\begin{eqnarray}
	\sigma^{(2)}(\vec{p},\epsilon_\lambda(\vec{p}))&\approx&\int \frac{d\vec{q}}{(2\pi)^2}\frac{\pi\alpha}{2q}\frac{\vec{q}\cdot\partial_{\vec{p}}\epsilon_{\lambda}(\vec{p})}{\omega_p(\vec{q})}\left(1-\frac{\vec{q}\cdot\partial_{\vec{p}}\epsilon_\lambda(\vec{p})}{\omega_p(\vec{q})}-1-\frac{\vec{q}\cdot\partial_{\vec{p}}\epsilon_\lambda(\vec{p})}{\omega_p(\vec{q})}\right)(1-2f_{\lambda}(\vec{p}-\vec{q}))\nonumber\\&\approx& -\int \frac{d\vec{q}}{(2\pi)^2}\frac{\pi\alpha}{q}\left(\frac{\vec{q}\cdot\partial_{\vec{p}}\epsilon_{\lambda}(\vec{p})}{\omega_p(\vec{q})}\right)^2(1-2f_{\lambda}(\vec{p}-\vec{q})).
\end{eqnarray}
\end{document}